\documentclass{aa}  

\usepackage{graphicx}
\usepackage[varg]{txfonts}
\usepackage[colorlinks,citecolor=blue]{hyperref}
\usepackage{float}
\usepackage{placeins,afterpage}
\usepackage[caption = false]{subfig}

\usepackage{booktabs}
\usepackage{pdflscape}
\usepackage{float}
\usepackage[normalem]{ulem}
\usepackage[dvipsnames]{xcolor}
\setcitestyle{notesep={ }}

\begin{document} 

   \title{The EDIBLES Survey}
   \subtitle{VIII. Band profile alignment of diffuse interstellar bands} 

   \author{A.~Ebenbichler\inst{1}
         \and
         J.~V.~Smoker\inst{2,3}
         \and
         R.~Lallement\inst{4}
         \and
         A.~Farhang\inst{5}
         \and
         N.~L.~J.~Cox\inst{6}
         \and
         C.~Joblin\inst{7}
         \and
         J.~Th.~van~Loon\inst{8}
         \and
         H.~Linnartz\inst{9,}\thanks{This paper is dedicated to the memory of Harold Linnartz, who passed away far too early on December 31, 2023.}
         \and
         N.~Przybilla\inst{1}
         \and
         P.~Ehrenfreund\inst{9}
         \and 
         J.~Cami\inst{10,11,12}
         \and
         M.~Cordiner\inst{13,14}
          }


   \institute{Universit\"at Innsbruck, Institut f\"ur Astro- und Teilchenphysik, 
              Technikerstr. 25/8, 6020 Innsbruck, Austria\\
              \email{Alexander.Ebenbichler@uibk.ac.at}
       \and
       European Southern Observatory, Alonso de Cordova 3107, Vitacura, Santiago, Chile
       \and  
       UK Astronomy Technology Centre, Royal Observatory, Blackford Hill, Edinburgh EH9 3HJ, UK
       \and
       GEPI, Observatoire de Paris, PSL Research University, CNRS, Universit\'e Paris-Diderot, Sorbonne Paris Cit\'e, Place Jules Janssen, 92195 Meudon, France
       \and
       School of Astronomy, Institute for Research in Fundamental Sciences, 19395-5531 Tehran, Iran
       \and
           Centre d’Etudes et de Recherche de Grasse, ACRI-ST, Av. Nicolas Copernic, Grasse 06130, France
	     \and
       Institut de Recherche en Astrophysique et Plan\'etologie (IRAP), Universit\'e Toulouse III - Paul Sabatier, CNRS, CNES, 9 Av. du Colonel Roche, 31028 Toulouse Cedex 04, France
       \and
       Lennard-Jones Laboratories, Keele University, ST5 5BG, UK
       \and
	         Laboratory for Astrophysics, Leiden Observatory, Leiden University, 
	        P.O.~Box 9513, 2300 RA Leiden, The Netherlands
       \and
       Department of Physics and Astronomy, The University of Western Ontario, London, ON N6A 3K7, Canada
       \and
       Institute for Earth and Space Exploration, The University of Western Ontario, London, ON N6A 3K7, Canada
       \and
       SETI Institute, 189 Bernardo Ave, Suite 100, Mountain View, CA 94043, USA
       \and
       Astrochemistry Laboratory, NASA Goddard Space Flight Center, Code 691, 8800 Greenbelt Road, Greenbelt, MD 20771, USA
       \and
       Department of Physics, The Catholic University of America, Washington, DC 20064, US
}
   \date{Received 7 December 2024; accepted 16 February 2024}

 
  \abstract
   {There have been many attempts to identify families of diffuse interstellar bands (DIBs) with perfectly correlating band strengths. Although major efforts have been made to classify broadly based DIB families and important insights have been gained, no family has been identified with sufficient accuracy or statistical significance to prove that a series of selected DIBs originates from the same carrier. This can be attributed in part to the exclusive use of equivalent widths to establish DIB families.}
   {In a change of strategy, we search for DIBs that are highly correlated in both band strength {\em and} profile shape. 
   This approach increases the chance of correlating DIBs being members of one family and originating from the same carrier molecule. 
   We also search for correlations between DIB profile families and atomic interstellar lines, with the goal of further chemically constraining possible DIB carriers.}
   {We adapted the well-known method of time-series alignment to perform a spectral alignment; that is, DIB alignment. In a second step, we analysed the alignment results using a clustering analysis. This method required a statistically significant data set of DIB sight lines. The ESO Diffuse Interstellar Bands Large Exploration Survey (EDIBLES) data were perfectly suited for this application.}
   {We report eight DIB families with correlating strengths and profiles, as well as four previously unreported DIBs in the visual range, found using DIB alignment. 
   All profile family members show Pearson correlation coefficients in band strength higher than 0.9.
   In particular, we report the 6614 -- 6521\,\AA~DIB pair, in which both DIBs show the same triple-peak substructure and an unprecedented band strength Pearson correlation coefficient of 0.9935. 
   The presented approach opens up new perspectives that can guide the laboratory search for DIB carriers.
   }
   {}

   
   \titlerunning{The EDIBLES Survey. VIII. DIB profile alignment}
   \authorrunning{Ebenbichler et al.}

   \maketitle
%

\section{Introduction}
At the time of writing, diffuse interstellar bands (DIBs) have been known for a little more than 100 years, since their first detection by \citet{Heger22}. 
They originate from material, most probably large molecules in the gas phase, populating the interstellar medium (ISM) and show a good but not perfect correlation with interstellar extinction \citep{1995ARA&A..33...19H}.
Diffuse interstellar bands have been observed mostly in the visual wavelength range \citep[e.g.][]{hobbs08,Hobbsetal09}, while more recently discovered DIBs were identified in the near-infrared \citep{Geballeetal11,2014A&A...569A.117C, ebenbichler22, 2022hamano}.
Close to 600 DIBs are known at this time \citep{2019ApJ...878..151F}, but only four have an unambiguously identified carrier molecule, the Buckminster fullerene cation C$_{60}^+$ \citep{1994Natur.369..296F,2015Natur.523..322C,2017ApJ...846..168S,2019ApJ...875L..28C}.
Over the last three decades, several DIB surveys have been reported \citep{herbig1991,jenniskens1994,Tuairisg2000, Hobbsetal09, 2014A&A...569A.117C}, systematically extending the number of DIBs and updating band profile parameters. 
These show strong variations, both in band intensity and band width.
The full width at half maximum (FWHM) of DIBs has a large range, from 0.4~\AA\ or 1~cm$^{-1}$ for the 6196~\AA~DIB to 176.6~\AA\ or 300~cm$^{-1}$ for the 7700~\AA~DIB \citep{apellaniz19}.
DIBs are also used as a diagnostic tool.
\citet{Zasowski15} searched the Apache Point Observatory Galactic Evolution Experiment (APOGEE) data set of about 60\,000 sight lines in the Milky Way in order to map large-scale structures and distributions using the 15272~\AA~DIB.

There have been numerous correlation studies on DIBs with interesting results, 
such as a recent one by \citet{Fanetal22}, who performed a data-driven correlation analysis and even found anti-correlating DIBs.
However, these correlation studies have only established rough correlations of DIBs, dividing them into general families, that is $\sigma$, $\zeta$ \citep{1987Krelowski}, and C$_2$ DIBs.
Even for the best correlating DIB pair, the 6614 and 6196~\AA~DIBs, their profiles are vastly different in width and shape \citep{McCall_2009}, raising the question of how one molecule can form bands with such different profiles.
Since several DIBs show substructures, many studies have concentrated on interpreting band profiles, to extract information about their possible carriers \citep[see the review by][]{2014IAUS..297...34S}.
Several studies analysed the profile variations of the strongest narrow DIBs in different sight lines \citep[e.g.][]{1997ApJ...477..209K,2008ApJ...682.1076G}.
Additionally, other work explored the strongest DIBs by fitting synthetic models and extracting information about a likely carrier molecule this way. Specifically, 
\citet{cami2004} and \citet{edibles5} measured the sub-peak separation of the 5797, 6379, and 6614~\AA~DIBs, respectively, assuming P(Q)R branches of linear, spherical, prolate, or oblate molecules to derive  rotational constants of potential carrier molecules.
The 6614\,\AA~DIB, most known for its substructure, has been investigated by multiple groups. 
\citet{bernstein15} examined the behaviour of the profile in diverse sight lines including Herschel~36, where the profile shows a broad red wing. They modelled the triple-peak profile and the broad wing with two DIBs, assuming two separate carriers.
The presence of an extended red tail in several strong DIBs towards Herschel~36 was interpreted by \citet{oka2013} as the result of different rotational constants in ground and electronically excited state and radiative pumping, increasing populations of higher rotational levels.
\citet{huang2015} modelled the 5797\,\AA~DIB assuming polar molecules, which implies a very low rotational temperature of $T_\mathrm{r}$\,=\,2.73~K of the carrier molecule due to rotational cooling, down to the cosmic microwave background temperature.
They also provided a very detailed discussion of the symmetry and size of possible carrier molecules.

Looking at the bigger picture, correlation studies have mostly been performed using one parameter: the equivalent width ($EW$); some studies have also used the central depth ($A_\mathrm{c}$).  It has to be noted that both quantities are sensitive to the choice of continuum and to noise. Such a parameterisation is well suited to strong DIBs, but for weaker DIBs the contribution of noise and other interstellar, stellar, or telluric spectral features to the flux makes the perceived continuum rather variable, leading to a potentially large scatter of $EW$ measurements depending on subjective choices. 
Correlation studies based on $EW$ showed that the strong DIBs do not originate from the same carrier, but for weaker DIBs no conclusive findings have been made.
There also exist a large number of studies on DIB profiles, but these are typically limited to a small set of DIBs, because the analysis of a larger data set requires new, more efficient measurement techniques.

\begin{figure}
\centering
\includegraphics[width = .78\linewidth]{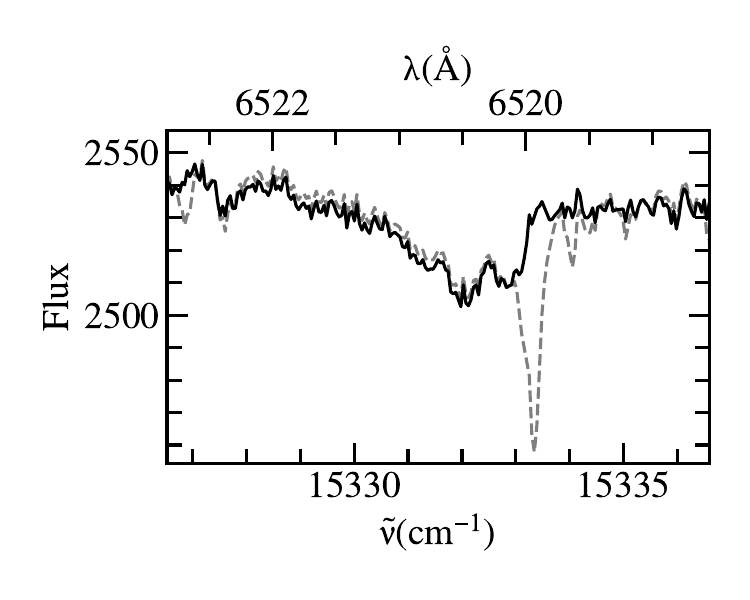}\\[-6mm]
\caption{Telluric blends of the 6521 \AA \, DIB towards \object{HD 166937}. The spectrum before telluric correction is displayed as a dashed grey line, whereas the continuous black line shows it afterwards. The water vapour during these observations ranges from 1.92 to 1.99~mm at the zenith. 
Wavelengths (air) and wave numbers (vacuum) are given in the barycentric rest frame.}
\label{fig:6521_tell}
\end{figure}

In the present study, we extend on previous work using a different approach. 
We developed an algorithm to detect DIBs with similar profiles, in order to establish strength correlations without using the $EW$. 
This method turns out to be more robust, in particular for weaker DIBs. 
Moreover, the application of this algorithm made it possible to identify several new weak DIBs in the optical spectral range.
It should be noted that this approach comes with the assumption that upon electronic excitation the molecular geometry does not change substantially; that is, the difference in rotational parameters in ground and excited states only varies marginally, maintaining band profiles.
This restricts our survey, as we will miss correlations to which this assumption does not apply, but where it does, a strong constraint is introduced.

A brief overview on the observational data employed for the present work is given in Sect.~\ref{sec:obs}. 
We describe the algorithm for DIB alignment in Sect.~\ref{sec:alignment}. 
Results from the application of this new approach are discussed in Sect.~\ref{sec:results}. 
This includes DIB families with matching profiles, DIB profile families with weaker strength correlations, and the first mention of previously unreported DIBs identified using DIB alignment. 
Finally, we discuss the band profiles in relation to previous data in Sect.~\ref{sec:discussion} and conclusions are drawn in Sect.~\ref{sec:conclusions}.

\begin{figure*}[!ht]
\centering
\subfloat[]{
\includegraphics[width = .33\linewidth]{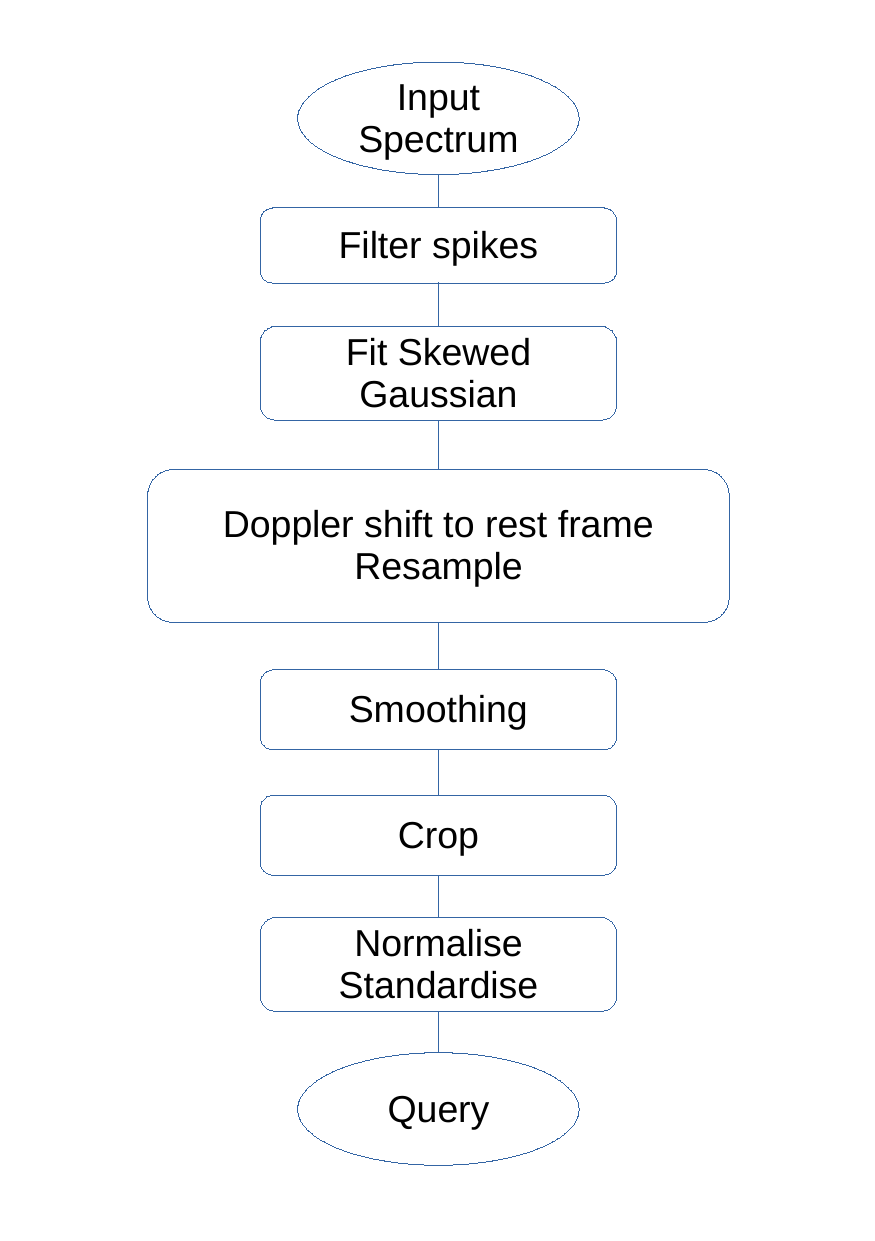}
\label{fig:flow_query}
}
\subfloat[]{
\includegraphics[width = .23\linewidth]{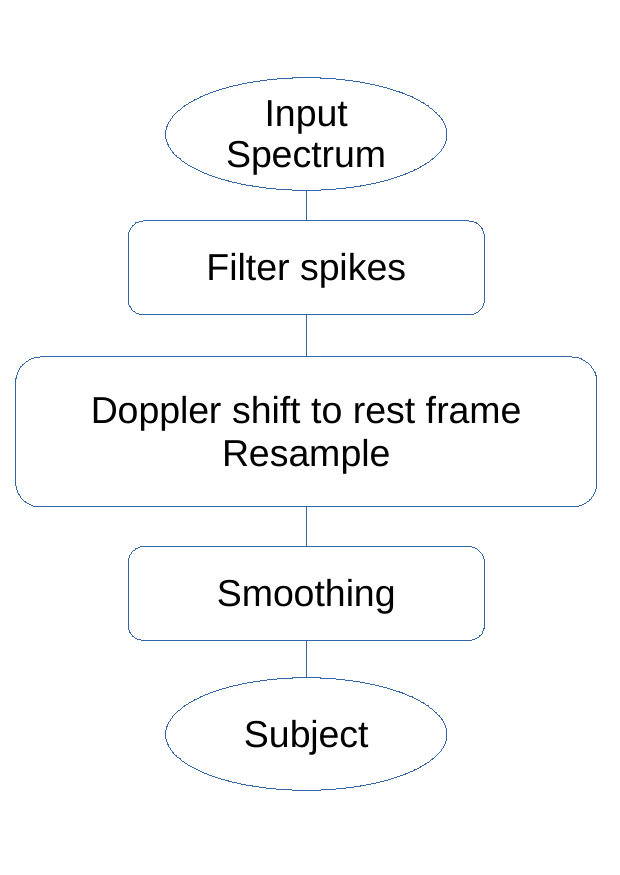}
\label{fig:flow_subject}
}
\subfloat[]{
\includegraphics[width = .33\linewidth]{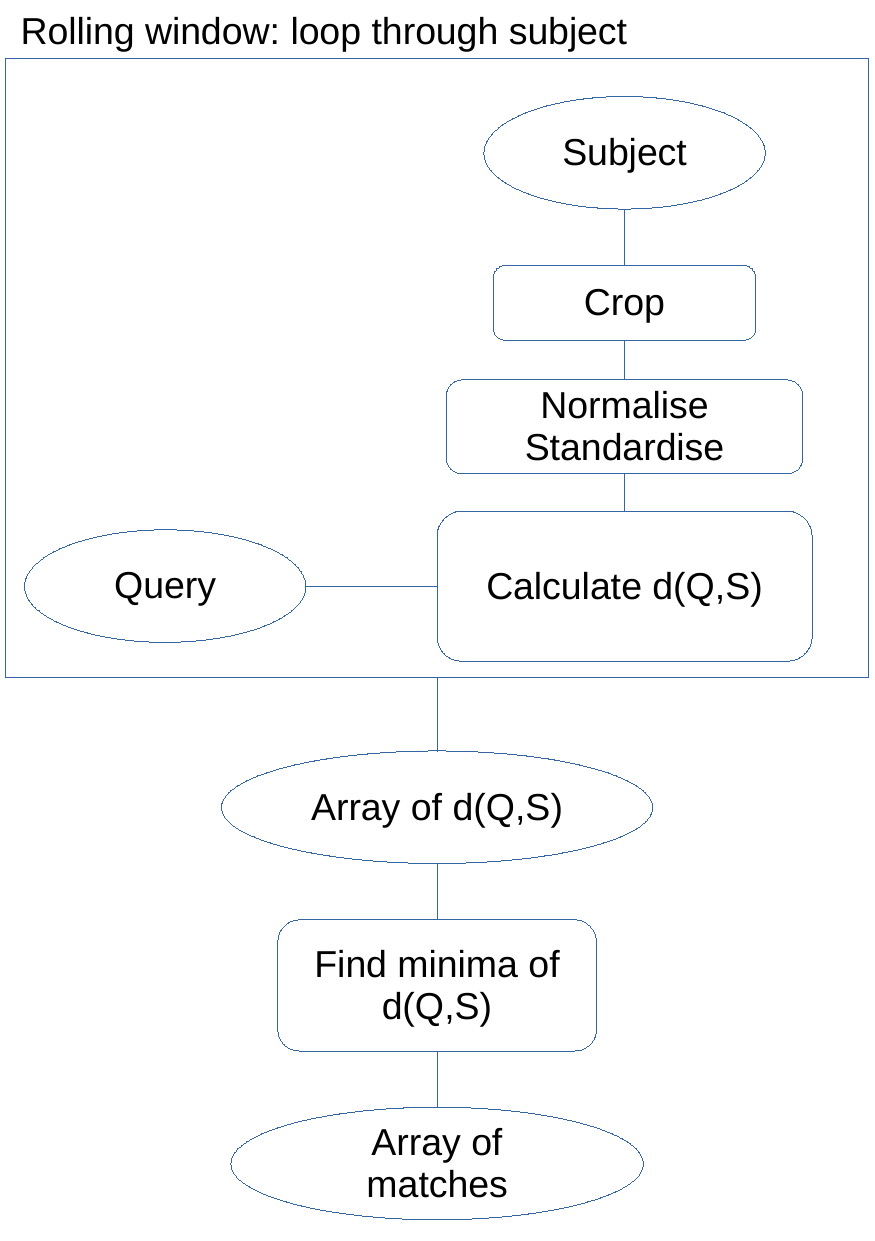}
\label{fig:flow_alignment}
}
\caption{Flow charts of the major processes.
Panel \ref{fig:flow_query}: Flow chart of the query preparation. Panel \ref{fig:flow_subject}: Flow chart of the subject preparation. Panel \ref{fig:flow_alignment}: Flow chart of the DIB alignment process. An array of matches was calculated per sight line. For the final analysis, all matches of all sight lines were inspected to find clusters~of~matches.}
\end{figure*}

\section{Observational data}\label{sec:obs}
We adopted the homogeneous, high-quality observational data set of the ESO Diffuse Interstellar Bands Large Exploration Survey \citep[EDIBLES,][]{edibles1} for the present work. 
The data consist of spectra obtained with the  Ultraviolet and Visual Echelle Spectrograph \citep[UVES,][]{Dekkeretal00} on the Very Large Telescope UT-2 at the European Southern Observatory (ESO) at Paranal in Chile for 123 sight lines towards Galactic O- and B-type stars. 
Four instrumental setups using the spectrograph's blue and red arms in combination with a dichroic mirror were employed to achieve near-complete wavelength coverage between $\sim$3050 and 10420\,{\AA} at a high spectral resolving power, $R$\,=\,$\lambda/\Delta\lambda$\,$\approx$\,70\,000--110\,000, using a 0.4\arcsec \, (blue) and 0.3\arcsec \, (red) slit.  The EDIBLES survey employed customised procedures based on the UVES pipeline to  fully reduce and calibrate the data, such that very high signal-to-noise (S/N) ratios of $S/N$\,$\approx$\,1000 were reached (in the red part) for the final data products (see \citet{edibles1} for details on the target selection, observing strategy, and data reduction). 
We repeated the telluric correction for the fourteenth order of the w564u setting using {\sc molecfit} \citep{molecfit,molecfit2} in order to optimise the observational data for the 6521~\AA~DIB.
An accurate and precise telluric correction is necessary, because even small residuals can change the band profile.
The contribution of telluric water vapour lines to this DIB is visualised in Fig.~\ref{fig:6521_tell}.
To enhance the S/N ratio, we co-added the spectral orders around the 6614 and 6521\,{\AA}~DIBs.

\section{Methods: Diffuse interstellar band alignment \label{sec:alignment}}
We searched for DIBs that have very similar band profiles and a strong correlation in their band strength.
To automatise this procedure, we adapted a common technique from time-series analysis: time-series alignment.
We took a part of our spectrum, which is a strong, unblended DIB, and used it as a ‘query.’
Then we used all spectra from the same sight line as ‘subjects.’
In the subjects, we searched for patterns that were similar to the query shape but possibly different in band strength. 
If we found good matches, we saved the properties of query and subject.
If there was a cluster of matches around the same wavelength for multiple sight lines, we used the result for cluster analysis.

The most important step is choosing the spectral space to compare in: wavelength, $\lambda$, radial velocity (RV), or wave number, $\tilde\nu$.
The intrinsic band structure of a DIB is probably caused by unresolved ro-vibronic transitions.
We searched for DIBs that exhibit the same (sub)structure and that stretch over the same wave number range, and that hence exhibit the same or a rather similar profile.
The spacing of those unresolved transitions is constant in energy, but not in wavelength.
So, we compared our spectra in the $\tilde\nu$ space and not in the wavelength or radial velocity spaces, as is commonly done in the literature.
Hence, we started with spectra in wave numbers as the input for the algorithm.
The spectra were transformed into the wave number space using the formula of \citet{1991ApJS...77..119M} for the air-to-vacuum transformation of the observed wavelengths.

We resampled the spectra to an equidistant $\tilde\nu$ grid. 
With this uniform spacing of data points, we could compare a query with any part of a subject. 
The comparison between query and subject was done in a rolling window approach.
We began at the start of the subject spectrum and took a spectral cut-out of the length of the query and made our comparison. 
Then we took a cut-out shifted by one grid point and made the same comparison. 
This way, our comparison window rolled through the subject and found the features that fitted the query.

\subsection{Query selection}
We used the DIBs towards \object{HD 183143} and \object{HD 204927} listed by \citet{hobbs08,Hobbsetal09} as queries. 
We started with the DIBs that have the highest $EW$/$E(B-V)$/FWHM values, until the queries became too weak to get reliable results.
Very broad DIBs with FWHM > 10\,cm$^{-1}$, like the 4430\,\AA~DIB, were omitted, because UVES orders are narrow enough, such that broad profile shapes may be distorted by the blaze curve of the spectrograph.
A flow chart on the preparation of a query is displayed in Fig.~\ref{fig:flow_query}.
The query should always be a DIB without stellar line blends or telluric residuals.
In order to know which range of the spectrum is part of the query DIB, we fitted an analytic function to the absorption profile,
choosing a skewed Gaussian function,
\begin{equation}
    f_\mathrm{gs}(\tilde\nu) = a \exp{\left[-\left(\frac{\tilde\nu - m}{2 w}\right)^2\right]} \exp{\left[-s \arctan{\frac{\tilde\nu - m}{w}}\right]}\,.
    \label{skew_gauss}
\end{equation}

\begin{figure}[ht]
\centering
\includegraphics[width = .8\linewidth]{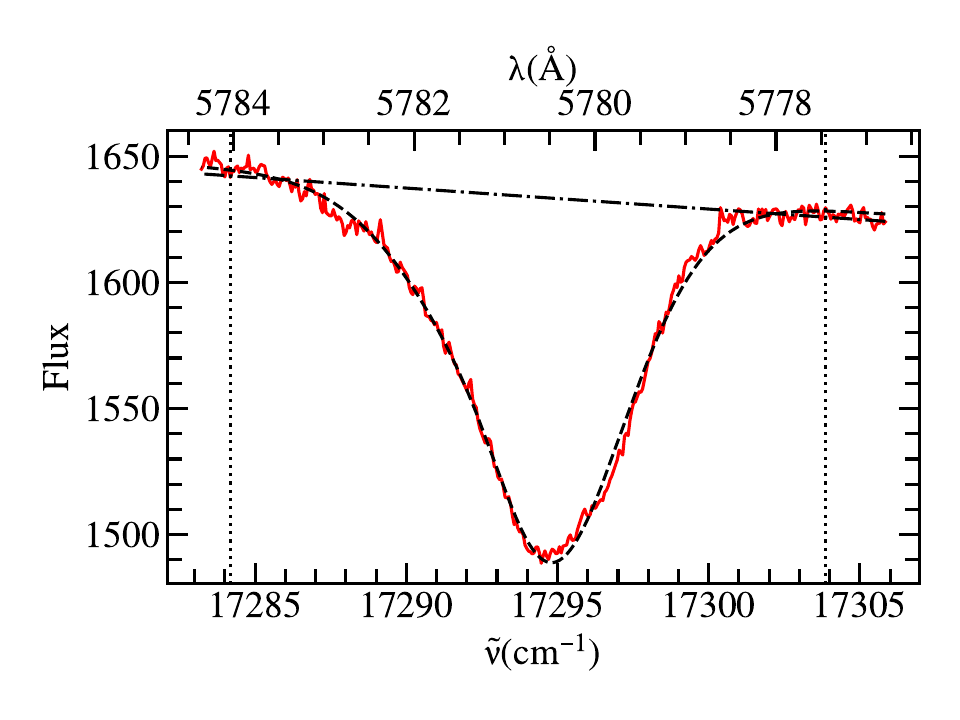}
\caption{Fitted curve (dashed black line) of the query DIB spectrum for the 5780\,\AA~DIB (red line) of \object{HD 103779}. The limits of the query range are marked by dotted lines. 
The intercept points of the dashed and dash-dotted black lines aid in the limit determination (see the text for further details).}
\label{fig:query_fitting_fit}
\end{figure}

The first exponential is a Gaussian function with a centre wave number, $m$, a width, $w$, and an amplitude, $a$.
The second exponential skews the Gaussian and is derived from the Pearson~IV function.
The skewness is proportional to the parameter, $s$. For $s$\,=\,0, $f_\mathrm{gs}$ is a normal Gaussian.
It is important to note that $f_\mathrm{gs}$ becomes narrower and its peak becomes stronger and gets wavelength-shifted as $|s|$ increases.
In our case, this does not matter, because we primarily wanted to fit an analytical function to a DIB.
The peak wavelength and DIB width were determined numerically after fitting.
The final fit function then incorporated $f_\mathrm{gs}$ in a linear continuum model,
\begin{equation}
    f_\mathrm{fit}(\tilde\nu) = (\alpha + \beta \tilde\nu) (1 - f_\mathrm{gs}(\tilde\nu))\,,
    \label{fit_model}
\end{equation}
where $\alpha$ denotes a constant flux value and $\beta$ the slope of the continuum flux.

The width of the fit profile represents the DIB width.
The wave number range of the query was defined as the range in which the fit profile is stronger than 2\% of its maximum depression plus a 10\% padding on each side (see Fig.~\ref{fig:query_fitting_fit}) to account for the effects of smoothing afterwards.
The points where the fit profile is as strong as 2\% of its maximum depression are indicated by the intercept points of the dashed and dash-dotted lines in Fig.~\ref{fig:query_fitting_fit}.
This way, we ensured that the continuum is part of the query spectrum most of the time.
The extent of continuum points was minimised in order to reduce the chances of other spectral lines being included.
After fitting, the spectrum was cropped using the fitted query range and shifted to the rest frame of the query DIB.
Then, the spectrum got smoothed using a boxcar profile with a width of 10\% of the query range. 
This was needed to suppress noise effects during DIB alignment. 
The subject spectra were smoothed in the same way.

In the first run, we used wavelength values from \cite{hobbs08,Hobbsetal09} as approximate laboratory wavelengths. 
In a second run, we calibrated the wavelengths of promising queries using atomic ISM lines.
Then, the spectrum was resampled into an equidistant wave number grid to make the comparison with the subject spectra more efficient.

\subsection{Normalisation and standardisation}
\begin{figure}[t!]
\centering
\includegraphics[width=0.8\linewidth]{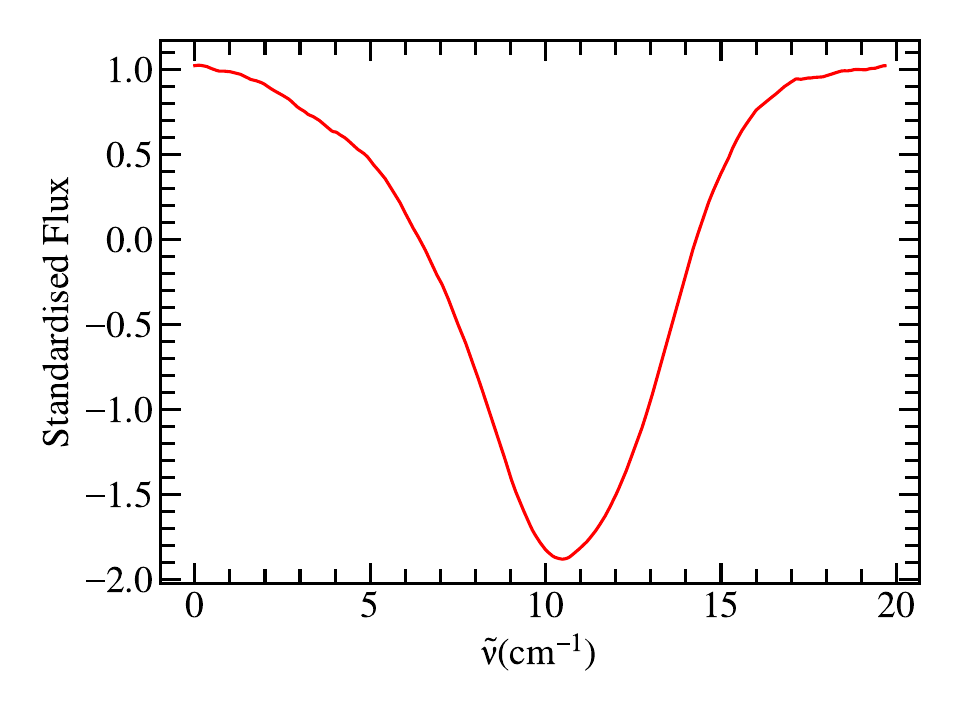}
\caption{Smoothed, continuum-corrected, and standardised query derived from the spectrum in Fig.~\ref{fig:query_fitting_fit} (see the text for further discussion).}
\label{fig:query_fitting_query}
\end{figure}

In order to make DIBs of different strengths comparable, they had to be standardised in terms of their flux.
A slope in the continuum could become a problem, so we took the first and last points of the smoothed query and approximated a linear continuum through them.
As a first standardisation step, we divided the spectrum by this continuum, 
resulting in a normalised spectrum with a constant continuum of 1.
Then, we standardised the flux,
\begin{equation}
    f_\mathrm{st} = \frac{f - \mu_f}{\sigma_f}\,,
\end{equation}
where $f$ is the normalised flux, $\sigma_f$ the standard deviation, and $\mu_f$ the mean flux of the query.
We note that this standardisation step results in negative flux values for the standardised flux at some wave numbers, which is a pure mathematical consequence of the definition.
An example of the resulting standardised query is given in Fig.~\ref{fig:query_fitting_query}.
It has to be noted that $f_\mathrm{st}$ is not continuum-normalised, so if the continuum is at unity value, like in Fig.~\ref{fig:query_fitting_query}, this is by coincidence.

\subsection{Rolling window comparison}
Once our standardised query was realised, we compared it to the subjects, which are all spectra from the sight line of the query.
Comparisons were always made for DIBs recorded along the same line of sight. 
The subject was filtered for spikes, transformed into the query DIB rest frame, smoothed, and transformed into an equidistant wave number grid with the same step size as the query (Fig.~\ref{fig:flow_subject}). 
Then, query and subject were aligned (Fig.~\ref{fig:flow_alignment}). 
We performed a rolling window comparison between the query and cut-outs of the subject. 
The subject cut-out is always the same size as the query. 
This is the comparison window. 
This window was rolled through the spectrum grid point by grid point, and we compared the query to the subject. 
In each comparison, the subject cut-out got normalised and standardised in the same way as the query. 
Then, we calculated the Euclidean distance between the standardised query and subject and divided the result by the square root of the number of query pixels,
\begin{equation}
    d(Q,S) = \lVert Q - S \rVert_2 / \sqrt{N} = \left[\sum_{i=1}^{N} (q(i) - s(i))^2\right]^{1/2} \bigg/ \sqrt{N}\,,
    \label{eq:euc_dist}
\end{equation}
where $q(i)$ and $s(i)$ are the query and subject flux at index $i$ and $N$ is the size of the query in resampled grid points.
The rolling window comparison produces a spectrum of distances, $d(Q,S)$. 
If $d(Q,S)$ is larger than 0.5, two DIBs generally do not look similar upon visual inspection.
Hence, we selected the minima with $d(Q,S)$\,$<$\,0.5 and counted those occurrences as matches, saving the match wave number, $\tilde\nu_\mathrm{m}$, and the standard deviations used for standardisation, $\sigma_\mathrm{Q}$ and $\sigma_\mathrm{S}$.

\begin{table}[t!]
\caption{Wavelength and FWHM measurements of the query DIBs for the DIB profile family investigations.} \label{tab:query_list}
\centering
{\small
\begin{tabular}{ccccc}
\hline\hline
Query &  $\nu$       & $\lambda_\mathrm{air}$ & FWHM        &   FWHM  \\
DIB   &  (cm$^{-1}$) & (\AA)                  & (cm$^{-1}$) &   (\AA) \\
\hline
4963 &  20140.0$\pm$0.4 &             4963.85$\pm$0.10 &    2.23$\pm$0.27 &  0.55$\pm$0.07 \\
5512 &  18135.1$\pm$0.2 &             5512.64$\pm$0.07 &    1.60$\pm$0.11 &  0.49$\pm$0.03 \\
5780 &  17294.5$\pm$0.3 &             5780.57$\pm$0.10 &    5.83$\pm$0.23 &  1.95$\pm$0.08 \\
5797 &  17245.4$\pm$0.3 &             5797.05$\pm$0.10 &    1.98$\pm$0.12 &  0.66$\pm$0.04 \\
6196 &  16135.1$\pm$0.2 &             6195.96$\pm$0.07 &    0.99$\pm$0.15 &  0.38$\pm$0.06 \\
6203 &  16116.4$\pm$0.2 &             6203.14$\pm$0.09 &    2.89$\pm$0.43 &  1.11$\pm$0.17 \\
6379 &  15671.4$\pm$0.2 &             6379.28$\pm$0.09 &    1.33$\pm$0.08 &  0.54$\pm$0.03 \\
6614 &  15116.0$\pm$0.3 &             6613.68$\pm$0.11 &    2.06$\pm$0.15 &  0.90$\pm$0.07 \\
\hline
\end{tabular}
\tablefoot{All measurements are mean values from our single cloud sight lines. The errors are derived from the standard deviation of those measurements (see Sect.~\ref{sec:query_calib} for further information).}}
\end{table}

\begin{figure*}[ht]
\centering
\includegraphics[width=.995\linewidth]{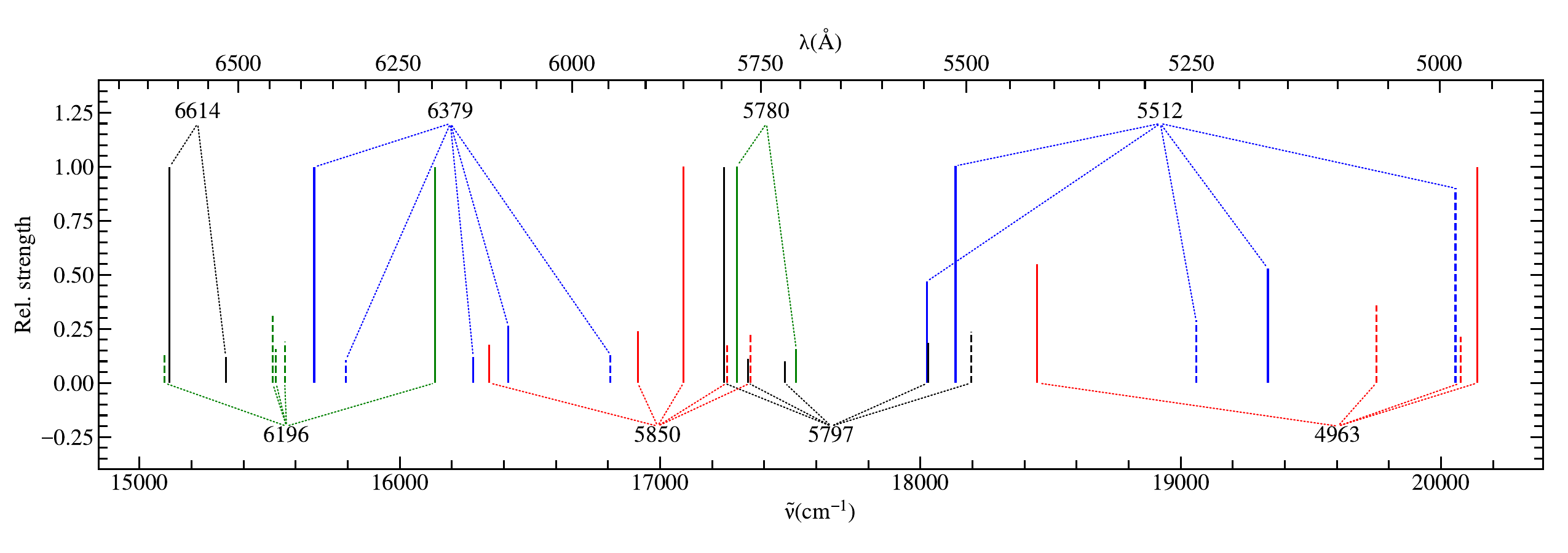}
\caption{Spectrum of the identified DIB families. The strengths are relative to the respective query DIB, which is set to a strength of 1.0. Solid lines represent quality 1 members, while dashed lines indicate quality 2 members.}
\label{fig:dib_fam_spectrum}
\end{figure*}

\begin{figure*}
\centering
\includegraphics[width = .90\linewidth]{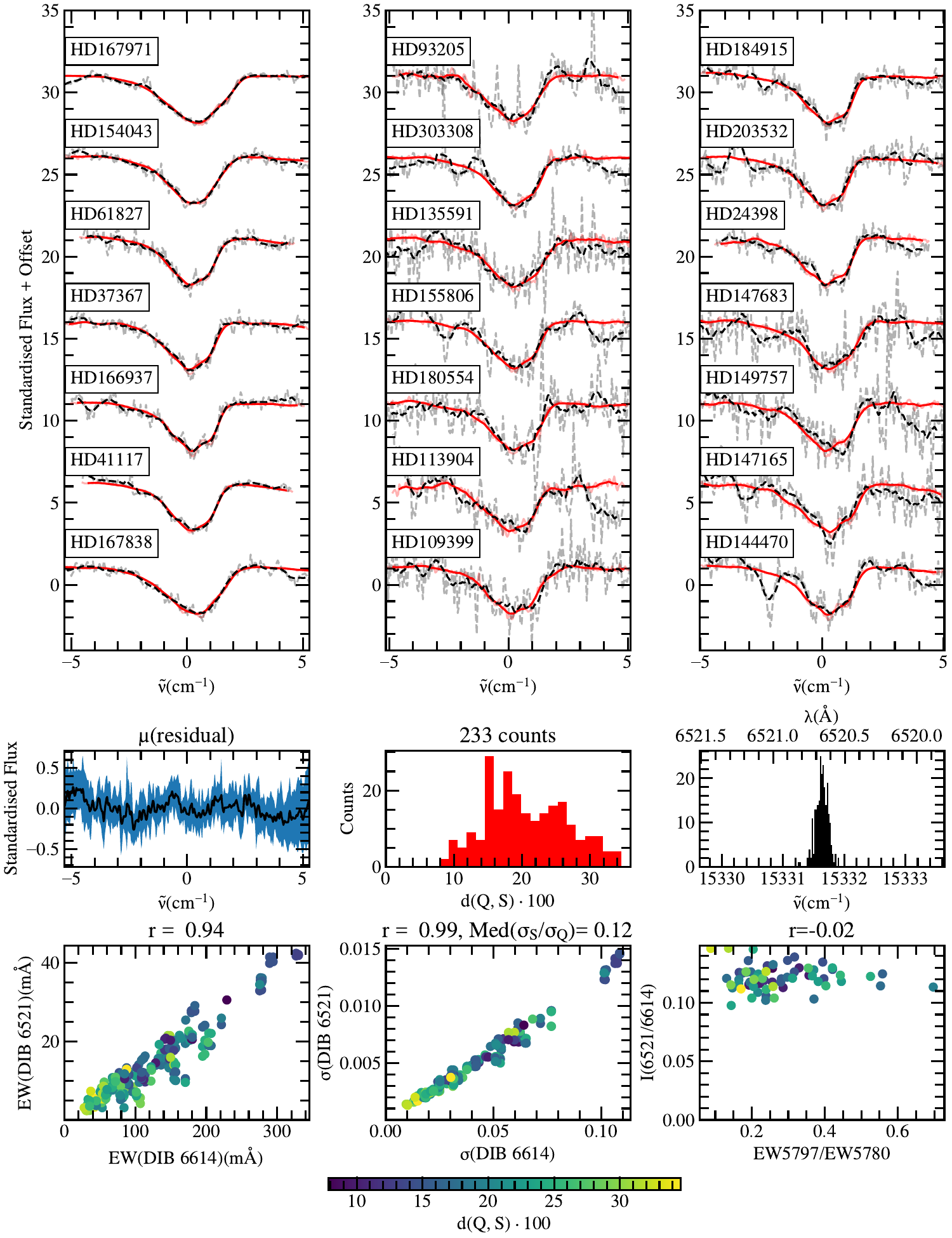}
\caption{Comparison between the 6614 (solid red line) and 6521~\AA~ (dashed black line) DIBs. 
The top left column shows the spectra of the seven best-matching sight lines, with the best at the top. 
The top central panel displays the seven worst-matching sight lines, with the worst at the bottom. 
The top right panel displays the seven best-matching single cloud sight lines, with the best at the top. 
Each DIB is shifted to its central wave number.
The smoothed spectra used by the algorithm are displayed opaquely, while the un-smoothed spectra are depicted as semi-transparent. 
In the middle left panel, the mean residuals between the standardised query and subject for the 20 best-matching pairs are visualised.
The blue area represents the 1$\sigma$ confidence interval of the mean residual.
In the middle central panel, a histogram of the matched distances of all pairs is shown. 
In the middle right panel, a histogram of the wave numbers where the matches were found is displayed.
In the bottom left panel, the $EW$ correlation of the two investigated DIBs is indicated, with their $\sigma$ correlation to its right. 
In the bottom right panel, the intensity ratio, $I$, is plotted against $EW_{5797}/EW_{5780}$.}
\label{fig:fam_6614_6521}
\end{figure*}

\begin{table*}[h]
\centering
\caption{DIB profile families derived from DIB alignment.}\label{tab:dib_families}
{\small
\begin{tabular}{cccrcccc}
\hline\hline
Query & Member &  $\tilde\nu$\ & $E_\mathrm{diff}$ & $\lambda_\mathrm{air}$ & $I$ &     $r$ & Quality \\
DIB   & DIBs   & (cm$^{-1}$) & (cm$^{-1}$) & (\AA) & & & Flag\\
\hline
4963 & 4963 & 20139.9$\pm$0.2 & 0.0$\pm$0.2 & 4963.89$\pm$0.04 & 1.00$\pm$0.16 & 0.994 & 1 \\
 & 4979 & 20076.3$\pm$0.2 & $-$63.6$\pm$0.2 & 4979.61$\pm$0.04 & 0.21$\pm$0.04 & 0.927 & 2 \\
 & 5062 & 19751.4$\pm$0.2 & $-$388.5$\pm$0.2 & 5061.52$\pm$0.05 & 0.36$\pm$0.06 & 0.961 & 2 \\
 & 5418 & 18448.9$\pm$0.2 & $-$1691.0$\pm$0.2 & 5418.87$\pm$0.04 & 0.55$\pm$0.13 & 0.964 & 1 \\
5512 & 4984 & 20055.5$\pm$0.1 & 1920.5$\pm$0.2 & 4984.77$\pm$0.03 & 0.90$\pm$0.21 & 0.945 & 2 \\
 & 5170 & 19335.2$\pm$0.1 & 1200.1$\pm$0.2 & 5170.47$\pm$0.03 & 0.53$\pm$0.07 & 0.942 & 1 \\
 & 5245 & 19059.0$\pm$0.2 & 923.9$\pm$0.2 & 5245.40$\pm$0.04 & 0.29$\pm$0.08 & 0.943 & 2 \\
 & 5512 & 18135.1$\pm$0.1 & 0.0$\pm$0.2 & 5512.65$\pm$0.04 & 1.00$\pm$0.09 & 0.990 & 1 \\
 & 5546 & 18024.6$\pm$0.1 & $-$110.5$\pm$0.1 & 5546.44$\pm$0.02 & 0.47$\pm$0.08 & 0.981 & 1 \\
5780 & 5705 & 17523.0$\pm$0.4 & 228.6$\pm$0.5 & 5705.22$\pm$0.13 & 0.16$\pm$0.02 & 0.979 & 1 \\
 & 5780 & 17294.4$\pm$0.2 & 0.0$\pm$0.3 & 5780.62$\pm$0.08 & 1.00$\pm$0.03 & 0.999 & 1 \\
5797 & 5494 & 18196.3$\pm$0.1 & 951.0$\pm$0.2 & 5494.11$\pm$0.04 & 0.24$\pm$0.04 & 0.975 & 2 \\
 & 5545 & 18029.2$\pm$0.1 & 784.0$\pm$0.1 & 5545.01$\pm$0.03 & 0.19$\pm$0.02 & 0.983 & 1 \\
 & 5720 & 17479.1$\pm$0.1 & 233.9$\pm$0.1 & 5719.52$\pm$0.04 & 0.10$\pm$0.02 & 0.982 & 1 \\
 & 5766 & 17337.7$\pm$0.1 & 92.4$\pm$0.2 & 5766.19$\pm$0.04 & 0.11$\pm$0.02 & 0.990 & 1 \\
 & 5797 & 17245.3$\pm$0.1 & 0.0$\pm$0.1 & 5797.09$\pm$0.03 & 1.00$\pm$0.05 & 0.999 & 1 \\
5850 & 5762 & 17348.2$\pm$0.2 & 258.3$\pm$0.3 & 5762.69$\pm$0.07 & 0.23$\pm$0.09 & 0.953 & 2 \\
 & 5793 & 17257.1$\pm$0.2 & 167.1$\pm$0.3 & 5793.12$\pm$0.05 & 0.18$\pm$0.04 & 0.948 & 2 \\
 & 5850 & 17089.9$\pm$0.2 & 0.0$\pm$0.3 & 5849.77$\pm$0.07 & 1.00$\pm$0.07 & 0.998 & 1 \\
 & 5911 & 16914.2$\pm$0.2 & $-$175.8$\pm$0.3 & 5910.57$\pm$0.07 & 0.24$\pm$0.06 & 0.946 & 1 \\
 & 6117 & 16343.9$\pm$0.1 & $-$746.0$\pm$0.2 & 6116.79$\pm$0.04 & 0.18$\pm$0.06 & 0.970 & 1 \\
6196 & 6196 & 16135.2$\pm$0.1 & 0.0$\pm$0.1 & 6195.93$\pm$0.03 & 1.00$\pm$0.06 & 0.998 & 1 \\
 & 6426 & 15558.3$\pm$0.1 & $-$576.9$\pm$0.1 & 6425.67$\pm$0.04 & 0.19$\pm$0.03 & 0.973 & 2 \\
 & 6440 & 15524.8$\pm$0.1 & $-$610.4$\pm$0.2 & 6439.54$\pm$0.06 & 0.16$\pm$0.04 & 0.919 & 2 \\
 & 6445 & 15511.1$\pm$0.1 & $-$624.0$\pm$0.1 & 6445.21$\pm$0.05 & 0.31$\pm$0.06 & 0.956 & 2 \\
 & 6623 & 15095.2$\pm$0.1 & $-$1039.9$\pm$0.1 & 6622.77$\pm$0.04 & 0.13$\pm$0.02 & 0.935 & 2 \\
6379 & 5947 & 16809.8$\pm$0.1 & 1138.2$\pm$0.1 & 5947.27$\pm$0.04 & 0.13$\pm$0.07 & 0.946 & 2 \\
 & 6090 & 16416.3$\pm$0.1 & 744.8$\pm$0.1 & 6089.80$\pm$0.03 & 0.27$\pm$0.05 & 0.983 & 1 \\
 & 6140 & 16282.3$\pm$0.1 & 610.8$\pm$0.2 & 6139.93$\pm$0.04 & 0.12$\pm$0.02 & 0.959 & 1 \\
 & 6330 & 15793.6$\pm$0.2 & 122.1$\pm$0.2 & 6329.94$\pm$0.07 & 0.11$\pm$0.03 & 0.938 & 2 \\
 & 6379 & 15671.5$\pm$0.1 & 0.0$\pm$0.1 & 6379.23$\pm$0.04 & 1.00$\pm$0.06 & 1.000 & 1 \\
6614 & 6521 & 15331.6$\pm$0.1 & 215.6$\pm$0.1 & 6520.66$\pm$0.04 & 0.12$\pm$0.01 & 0.989 & 1 \\
 & 6614 & 15116.1$\pm$0.1 & 0.0$\pm$0.1 & 6613.65$\pm$0.03 & 1.00$\pm$0.04 & 0.999 & 1 \\
\hline
\end{tabular}
\tablefoot{Wavelengths and wave numbers are mean values of the clusters and the errors are standard deviations of the clusters.
Energy differences with respect to the query are also given as $E_\mathrm{diff}$. The vacuum-to-air-conversion formula of \citet{1991ApJS...77..119M} was used to calculate the air wavelengths,~$\lambda_\mathrm{air}$. 
The strength ratios, $I$, are the median values of $\sigma_\mathrm{S}/\sigma_\mathrm{Q}$ with the standard deviation as errors, and $r$ denotes the Pearson correlation coefficient. 
Family members are divided in two quality groups: 1 and 2. 
Quality 1 implies that membership is certain because of tight strength correlations and strong profile correlations. 
Quality 2 indicates a weaker strength correlation, but still strong profile similarities. 
The quality flags are a subjective rating of the correlation. 
If, for example, two DIBs have the same distinct substructure, they can be rated quality 1, even if the Pearson coefficient is not as high as for some quality 2 DIBs.}}
\end{table*}

After performing this analysis, we searched for clusters of matches from all sight lines using the Python Sklearn package DBSCAN \citep{dbscan}.
If there was a cluster of match-wave numbers, we had a closer look and compared queries and subjects at the matches by eye, using comparison plots (see Sect.~\ref{sec:dib_families}). 
If the spectra matched well, the query and subject DIBs were considered members of the same DIB profile family. 
If additionally for two DIBs in this family, $\sigma_\mathrm{Q}$ and $\sigma_\mathrm{S}$ correlated well and $\tilde\nu_\mathrm{m}$ showed a very narrow histogram, we concluded that both DIBs belong to the same DIB family, which increases the chance that they share a common carrier or are chemically linked.

\subsection{Query wavelength calibration\label{sec:query_calib}}
The algorithm requires two values of the query DIBs to be extracted from the input spectrum: $\tilde\nu$ and FWHM.
For the initial alignment runs, we employed the values derived by \citet{hobbs08,Hobbsetal09} for HD~183143 and HD~204827, respectively.
To obtain accurate DIB rest-wavelengths, we analysed as many DIBs as possible for EDIBLES single-cloud sight lines \citep{edibles5}.
We used RVs from that work to shift the spectra into the cloud rest frame, and used skewed Gaussian fits to determine the central wave number and FWHM.
The fit results are shown in Table \ref{tab:query_list}.
The derived wave numbers are the positions of the central absorption of the skewed Gaussian fit.
For the query DIBs, between ten and thirty spectra were fitted successfully.
From the results we calculated mean values and used the standard deviations as error estimates.
The resulting wave numbers and FWHMs were used as starting values for fitting the skewed Gaussian by the algorithm.
It is important to note that the algorithm returns slightly different DIB wave numbers (of the order of 0.1\,cm$^{-1}$) than the skewed Gaussian, because it saves the mean wave number of the matches as opposed to the wave number of the central absorption.
We tried to correct for this shift, but this led to larger scatters in the wave numbers of each match.
As a consequence, small shifts exist when comparing the wave numbers of the query DIBs in Tables \ref{tab:query_list} and \ref{tab:dib_families}.
It is possible to calibrate the wave numbers in Table \ref{tab:dib_families} to get the wave numbers of the maximal absorption, but in our case there is a focus on using the standard deviation of those wave numbers as a clustering argument.
The smaller the standard deviation, the more likely it becomes that two DIBs share the same carrier molecule.

\section{Results\label{sec:results}}

\subsection{Diffuse interstellar band profile families\label{sec:dib_families}}
We found eight DIB families that are also DIB profile families -- their members have both matching profiles {\em and} high correlations in band strength.
We consider such family members to have either the same carrier or at least carriers that are chemically closely related and that have very similar molecular properties.
The DIB profile families are listed in Table~\ref{tab:dib_families}.
For each family, the used query is also part of the members.
For each member, the mean wave number, $\tilde\nu$, energy difference, $E_\mathrm{diff}$, and intensity ratio, $I$, relative to the query, band strength Pearson correlation coefficient, $r$, and quality class are given.
The errors are the standard deviations of each value.
The DIB profile families are also visualised in Fig.~\ref{fig:dib_fam_spectrum} in different colours.
Each DIB family in the figure is normalised to the strength of the query DIB, which is always the strongest member.
Solid lines represent quality 1 members, while dashed lines indicate quality 2 members.
The quality classes are explained below.

Comparison plots for each relation can be found in the appendix,
with a representative example shown in Fig.~\ref{fig:fam_6614_6521}.
The plots show three columns with profile comparisons for the 6614 (query) and 6521\,{\AA} DIBs. 
The top left column shows the seven best-matching sight lines and the next column shows the seven worst matches.
The top right panel displays the seven best-matching single cloud sight lines, with the best at the top. 
The query is depicted as a red line and the subject match as a dashed black line.
The smoothed spectra used by the algorithm are plotted as opaque lines and the un-smoothed spectra are semi-transparent in the background.
The mean residuals between the DIB pair are shown in the middle left panel.
We cropped the matches with a $d(Q,S)$-cutoff of 0.35, such that the worst matches look acceptable.
This cutoff can be seen in the histogram in the middle column of Fig.~\ref{fig:fam_6614_6521}, second from the top.
The middle right panel shows a histogram of the wave numbers where matches were found using the 6379\,{\AA}~DIB as a query.
In the bottom left panel, there is a classic $EW$ correlation diagram of the DIB pair.
The bottom middle panel shows the band strength correlation using the flux standard deviations, $\sigma$, of the two DIBs.
In the bottom right panel, the intensity ratio, $I$, is plotted against $EW_{5797}/EW_{5780}$. Further details of this plot are discussed in Sect.~\ref{sec:i_variation_sigma_zeta}.
We performed a 3-$\sigma$ clipping on $I$ to exclude outliers.

One sees that the correlations between some DIB pairs are not as tight as between others but that they are still considered family members.
When the respective DIBs are weaker, the criteria necessarily have to be relaxed, because of the higher contamination by other spectral features and effects of noise in the spectra.
We decided to divide family members in two quality groups: 1 and 2.
Quality~1 means that membership is certain because of very tight band strength correlations and strong profile correlations.
Quality~2 means a weaker strength correlation, but still strong profile similarities.
Out of the eight DIB profile families found, three are linked to C$_2$ DIBs, and all have prominent strong DIBs as queries.
None of the queries correlate sufficiently well with each other either in band strength or profile shape for them to be considered members of the same profile family.

Auto-correlations of the individual queries provide an insight into what a perfect match should look like, while there are still some errors introduced by calibration and different observation conditions. An example is shown in Fig.~\ref{fig:fam_6379_6379} for the 6379\,{\AA} DIB.
For the query-query comparison we only compared different observations of the same sight line, because for the same observation there will always be a perfect match.

\begin{table*}[ht]
\centering
\caption{Pearson coefficients of the strength correlations between all queries and matches in the present work.}\label{tab:corr_df}
{\small
\begin{tabular}{lrrrrrrrr}
\hline\hline
{} &   4963 &  5512 &  5780 &  5797 &  5850 &   6196 &   6379 & 6614 \\
\hline
4963 & 0.994 & 0.906 & ... & 0.828 & 0.879 & ... & 0.637 & 0.659 \\
4979 & 0.927 & 0.890 & ... & 0.659 & 0.783 & ... & ... & 0.542 \\
4984 & 0.980 & 0.947 & ... & 0.834 & 0.906 & 0.321 & 0.756 & 0.660 \\
5062 & 0.963 & 0.845 & ... & 0.619 & 0.705 & ... & 0.685 & 0.453 \\
5170 & 0.905 & 0.981 & ... & 0.714 & 0.848 & 0.459 & 0.650 & 0.561 \\
5245 & 0.878 & 0.967 & ... & 0.848 & 0.918 & ... & ... & 0.752 \\
5418 & 0.965 & 0.872 & ... & 0.600 & 0.745 & ... & 0.542 & 0.397 \\
5494 & 0.829 & 0.799 & ... & 0.975 & 0.939 & 0.909 & 0.925 & 0.931 \\
5512 & 0.930 & 0.990 & ... & 0.848 & 0.903 & 0.144 & 0.769 & 0.654 \\
5545 & 0.773 & 0.507 & ... & 0.983 & 0.963 & ... & ... & 0.927 \\
5546 & ... & 0.982 & ... & ... & ... & ... & 0.787 & ... \\
5705 & ... & ... & 0.981 & ... & ... & ... & ... & ... \\
5720 & 0.784 & 0.779 & ... & 0.985 & 0.902 & ... & ... & 0.910 \\
5762 & 0.688 & 0.847 & ... & 0.927 & 0.953 & 0.763 & 0.933 & 0.777 \\
5766 & 0.774 & 0.811 & ... & 0.991 & 0.966 & ... & 0.966 & 0.917 \\
5780 & ... & ... & 1.000 & ... & ... & ... & ... & ... \\
5793 & 0.806 & ... & ... & 0.932 & 0.948 & ... & ... & 0.817 \\
5797 & 0.811 & 0.799 & ... & 0.999 & 0.977 & ... & 0.978 & 0.940 \\
5850 & 0.885 & 0.919 & ... & 0.980 & 0.998 & ... & 0.925 & 0.863 \\
5911 & 0.794 & 0.863 & ... & 0.960 & 0.950 & ... & 0.941 & 0.823 \\
5947 & ... & 0.811 & ... & ... & ... & 0.802 & 0.966 & ... \\
6090 & ... & 0.805 & ... & 0.954 & ... & 0.834 & 0.984 & 0.808 \\
6117 & 0.844 & 0.888 & ... & 0.969 & 0.972 & ... & ... & 0.891 \\
6140 & ... & 0.841 & ... & 0.977 & 0.975 & 0.822 & 0.975 & 0.848 \\
6196 & ... & 0.359 & ... & ... & ... & 0.999 & 0.906 & ... \\
6330 & ... & 0.672 & ... & 0.878 & ... & 0.911 & 0.940 & 0.947 \\
6379 & ... & 0.749 & ... & 0.930 & 0.821 & 0.900 & 1.000 & 0.879 \\
6426 & ... & 0.673 & ... & ... & ... & 0.973 & 0.905 & ... \\
6440 & ... & 0.829 & ... & ... & ... & 0.956 & 0.959 & 0.963 \\
6445 & 0.405 & 0.537 & ... & 0.925 & 0.831 & 0.975 & 0.898 & 0.979 \\
6521 & 0.632 & 0.527 & ... & 0.897 & 0.802 & 0.969 & 0.775 & 0.989 \\
6614 & 0.662 & 0.334 & ... & 0.939 & 0.851 & ... & 0.872 & 0.999 \\
6623 & ... & ... & ... & ... & ... & 0.955 & 0.831 & ... \\
\hline
\end{tabular}
\tablefoot{The displayed correlations are calculated from matches with distances smaller than 0.35.
If there are less than 50 matches using that cut, no values are displayed.
The same maximum distance is used for all correlations to make them comparable.
We note that some DIB profile families are very similar, in particular those with queries 5797 and 5850~{\AA}. But upon closer inspection -- including the profile comparison -- these families show subtle but distinct differences, in particular when comparing the queries directly. 
The proposed DIB blend at 6440\,\AA\ shows high band strength correlations with the 6379\,\AA\ DIB, but on visual inspection the profiles are different.
}}
\end{table*}

The first test for DIB profile family members is band strength correlation.
Diffuse interstellar bands are considered for further investigation as family members if their band strength correlation coefficient is higher than 0.9.
As a second test, we made sure that ambiguous matches are not part of our DIB profile families by comparing the Pearson correlation coefficients for all combinations of queries and members in Table \ref{tab:corr_df}.
For this table, all matches were cropped by a maximal matching distance of $d(Q,S)=0.35$ to make them comparable.
If a DIB had a correlation coefficient of $r>0.9$ for multiple queries, we inspected all those matches more closely in terms of profile correlation and the relative band strengths versus $EW_{5797}/EW_{5780}$, as is described in Sect.~\ref{sec:i_variation_sigma_zeta}.
An optional test is the histogram of measured distances.
If there is a steep rise in counts, it implies that the profiles intrinsically match very well (Fig.~\ref{fig:profiles_6379_6090}, right column, second panel from the top).
If the rise is slow, this implies that the best matches are only coincidences, for example by overlapping cloud components (Fig.~\ref{fig:5797_5850}), but intrinsically the two DIBs have different profile shapes.
This test is susceptible to noise if one of the compared DIBs is very weak.

\begin{figure}[ht]
    \centering
    \includegraphics[width = .95\linewidth]{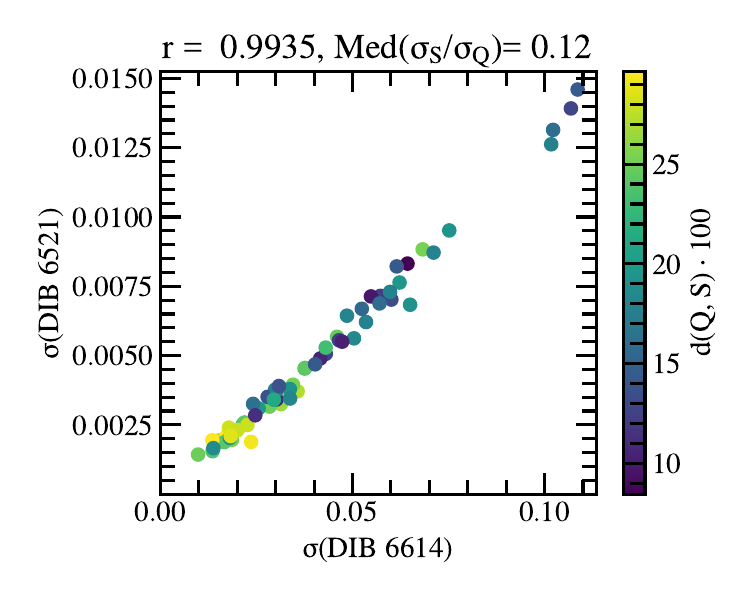}
    \caption{Band strength correlation of the 6614-6521\,{\AA}~DIB pair. The $\sigma$ values are derived from co-added spectra.}
    \label{fig:sigma_corr}
\end{figure}

\begin{figure}[ht]
\includegraphics[width = .85\linewidth]{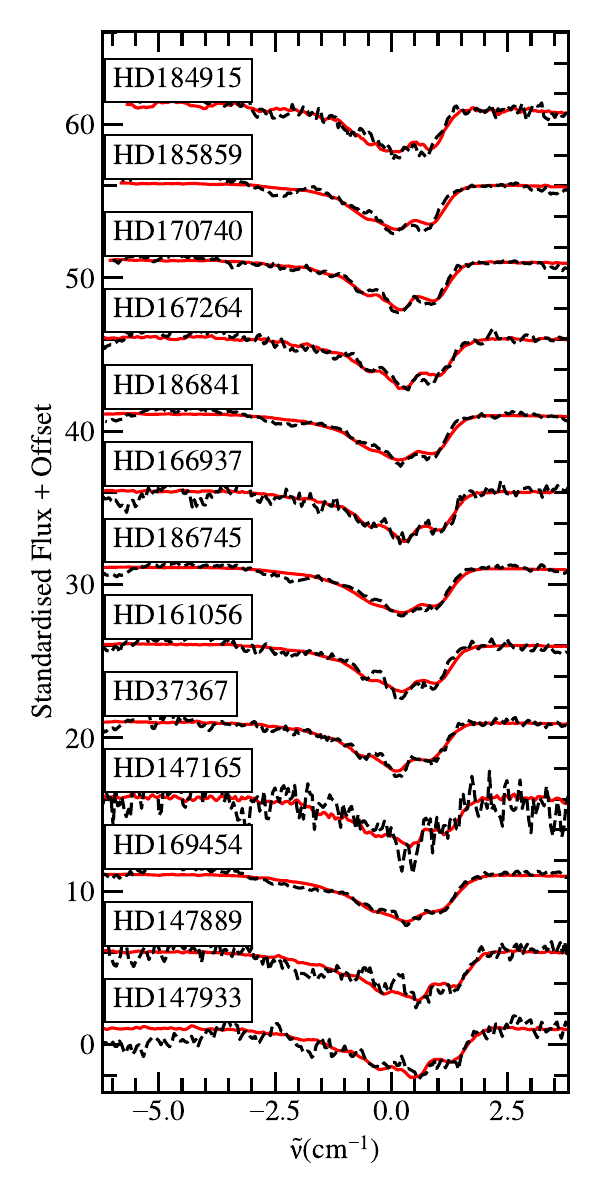}
\caption{Comparison of DIB 6614 (red) and DIB 6521 (dashed black) in sight lines with three sub-peaks, using co-added spectra.
The sight lines are sorted by the FWHM of the 6614\,{\AA}~DIB.}
\label{fig:6614_6521_sub_peaks}
\end{figure}

\begin{figure*}[ht]
\centering
\subfloat[]{
\includegraphics[width = .395\linewidth]{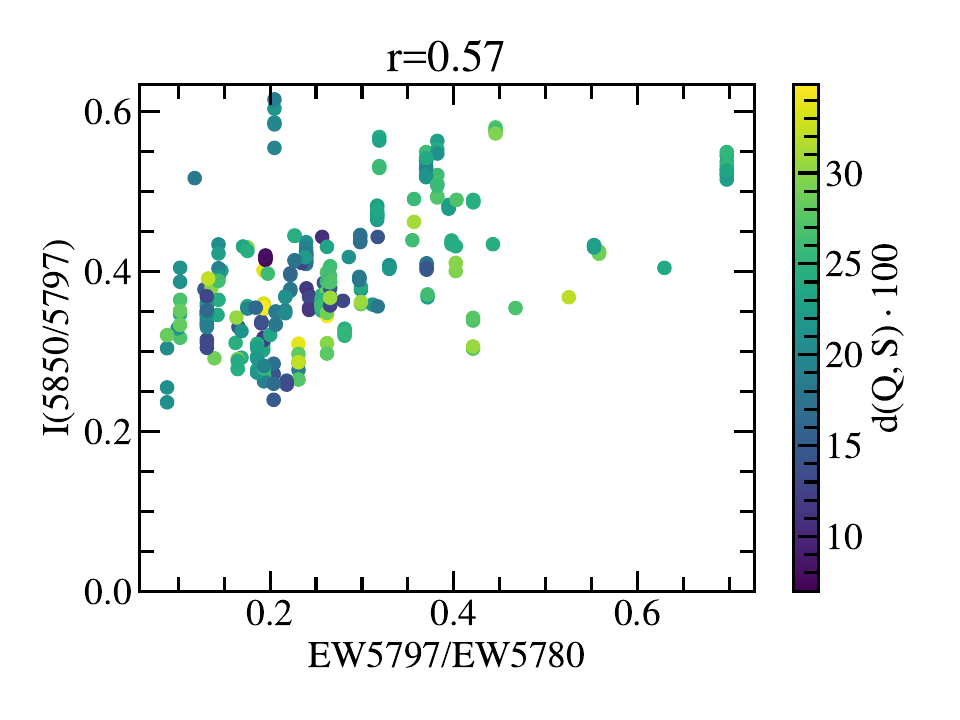}
\label{fig:sigma_zeta_5797_5850_q1}}
\subfloat[]{
\includegraphics[width = .395\linewidth]{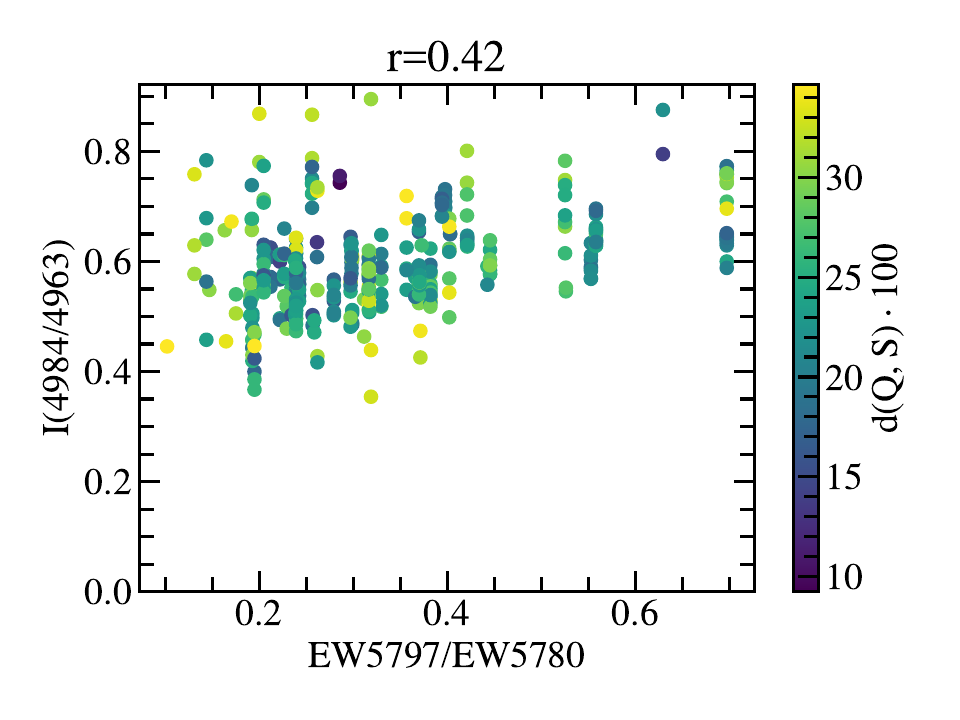}
\label{fig:sigma_zeta_4963_4984_q1}}

\subfloat[]{
\includegraphics[width = .395\linewidth]{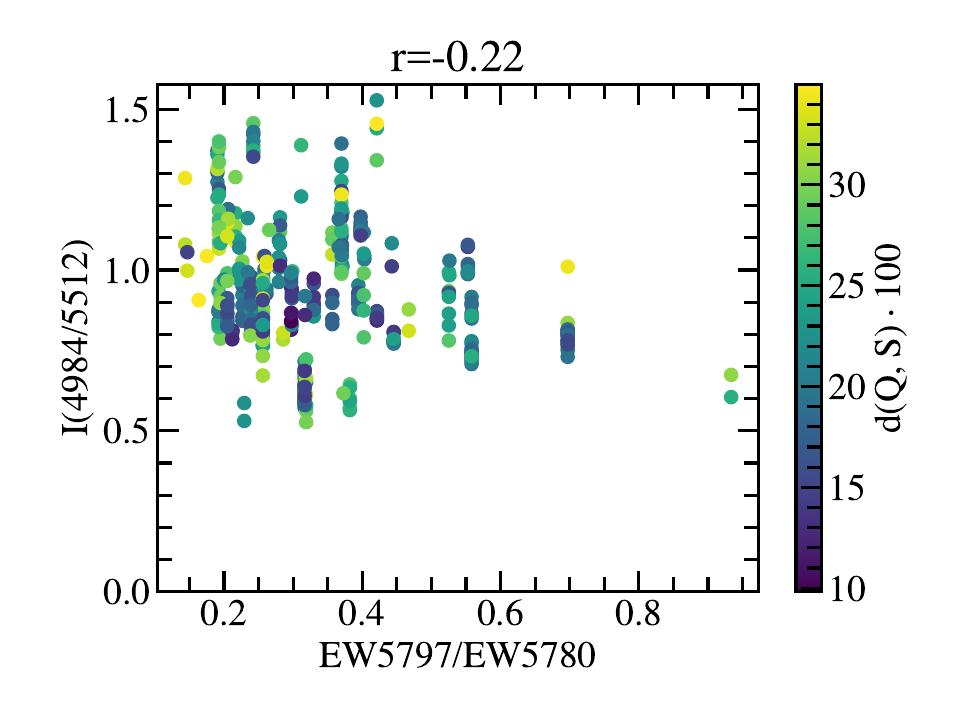}
\label{fig:sigma_zeta_5512_4984_q1}}
\subfloat[]{
\includegraphics[width = .395\linewidth]{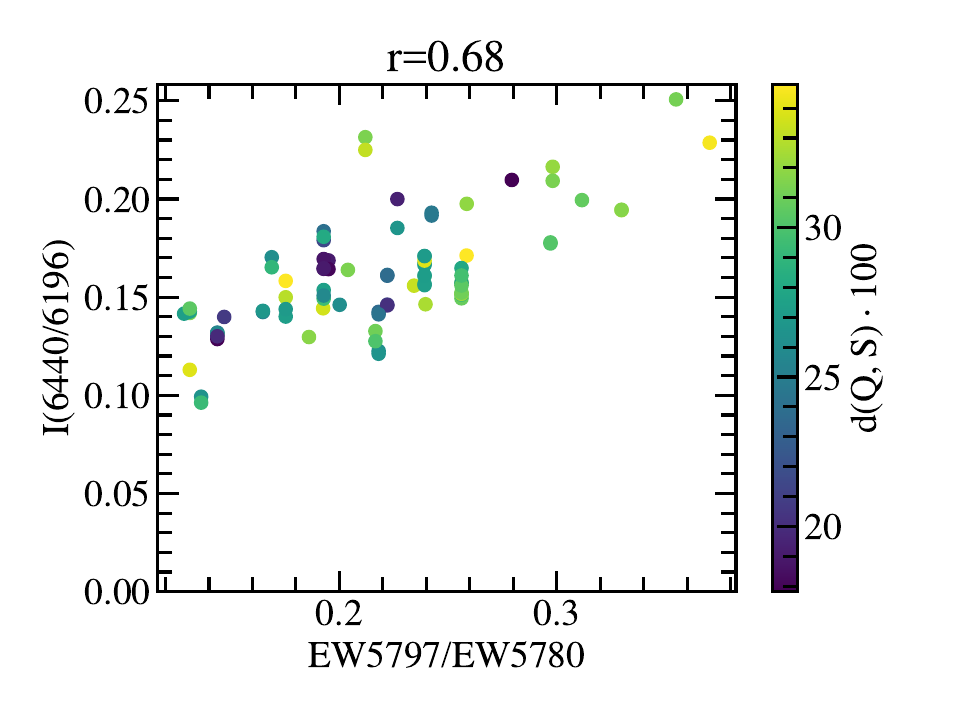}
\label{fig:sigma_zeta_6196_6440_q1}}
\caption{Band strength ratios, $I$, versus $EW_{5797}/EW_{5780}$ for four different DIB pairs. 
In panels~\ref{fig:sigma_zeta_5797_5850_q1} and \ref{fig:sigma_zeta_4963_4984_q1}, there is a positive correlation, but $d(Q,S)$ remains similar for $\zeta$ and $\sigma$ sight lines.
Panel~\ref{fig:sigma_zeta_5512_4984_q1} shows variation in $I$, but weaker correlation with $EW_{5797}/EW_{5780}$.
Consequently, the 4984\,\AA~DIB more likely shares a common carrier with the 5512\,\AA~DIB than the 4963\,\AA~one.
In panel~\ref{fig:sigma_zeta_6196_6440_q1}, $d(Q,S)$ decreases for $\zeta$ sight lines, contrary to all other pairs.
Measurement errors of $I$ can be estimated via the spread of measurements for a single sight line; that is, for one single $EW_{5797}/EW_{5780}$ value.}
\label{fig:sigma_zeta_i}
\end{figure*}

\begin{figure}[ht]
\includegraphics[width = .95\linewidth]{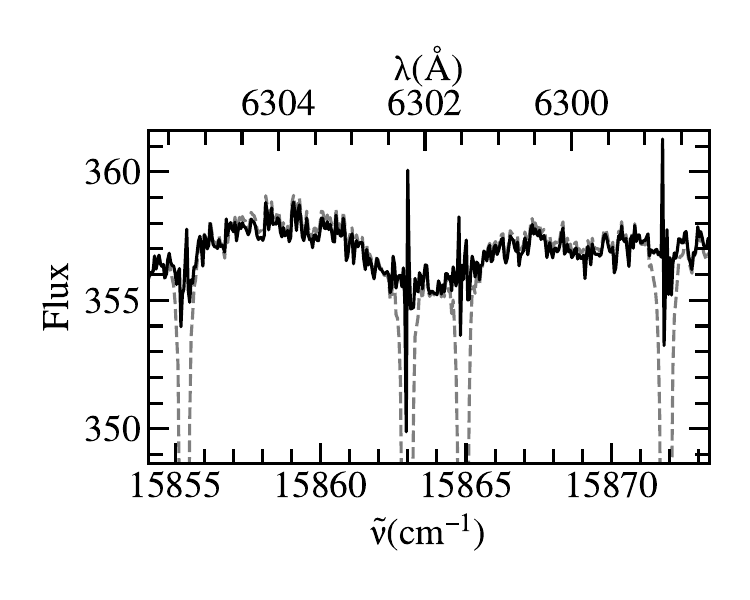}
\caption{Telluric blends of the previously unreported 6302~\AA~DIB towards \object{HD 148937}. The spectrum before telluric correction is displayed as a dashed grey line in the cases where a correction was applied. Wave numbers and air wavelengths are given in the barycentric rest frame.}
\label{fig:6302_tell}
\end{figure}

\subsection{The 6614-6521\,{\AA}~diffuse interstellar band pair}
The 6614-6521\,{\AA}~DIB pair is certainly the most intriguing one, because of its high band strength correlation and the triple-peak structure, which is very similar in both DIBs.
In order to get a better comparison between the two DIB profiles, we decided to co-add all available spectra for each sight line and to repeat the DIB alignment algorithm.
The correlation coefficient of the band strengths is higher than for any other DIB pair so far.
A plot of the correlation is visualised in Fig.~\ref{fig:sigma_corr}.
The 6521\,\AA~DIB is blended with a telluric H$_2$O line at 6520\,\AA\ and a stellar \ion{Si}{iii} line at 6521.5\,\AA. 
Both the stellar line, typical of B0-type stars, and the telluric line are weak and can be corrected with high accuracy.
Still, both features can affect the band strength correlation and profile comparison to a certain extent.
A series of spectra with this structure is shown in Fig.~\ref{fig:6614_6521_sub_peaks}.

\subsection{I variation versus EW$_{5797}$/EW$_{5780}$}\label{sec:i_variation_sigma_zeta}
While a Pearson correlation coefficient of $r=0.98$ might look like an almost perfect correlation, we often find significant variation in $I=\sigma_S/\sigma_Q$ for highly correlating pairs of DIBs.
This variation could be due to systematic errors, but a physical explanation is also possible, so it is helpful to look for correlations between $I$ and physical variations along the sight lines.
One of the most established physical parameters that can be determined using DIBs is the division in $\sigma$ and $\zeta$ sight lines \citep{1987Krelowski}, which is determined by the $EW$ ratio between the 5797 and 5780\,{\AA}~DIBs, $EW_{5797}/EW_{5780}$.
Sight lines with a low $EW$ ratio are $\sigma$ sight lines, while $\zeta$ sight lines show a high $EW$ ratio.
So if $I$ of a DIB pair shows a high correlation with $EW_{5797}/EW_{5780}$, this is an indication that those DIBs originate from slightly different carrier molecules that react differently to environmental conditions.
In the case of no or a weak correlation with $EW_{5797}/EW_{5780}$, this could mean that the carriers of the two DIBs, which are in the same DIB family, are more likely to be due to the same carrier, as they respond the same to changing conditions.
This correlation is useful if a DIB is not discernible between DIB profile families based on profile and Pearson correlation alone.
We plot the correlations between $I$ and $EW_{5797}/EW_{5780}$ for DIB pairs in Fig. \ref{fig:sigma_zeta_i} and also Fig.~\ref{fig:fam_6614_6521} to verify any correlation.

\subsection{New diffuse interstellar bands}
Interestingly, the algorithm introduced here to correlate DIBs also turned out to be successful in identifying so-far unknown and rather weak DIBs. 
These new DIBs are listed in Table~\ref{tab:new_dibs}, along with the queries used, intensity ratios, and band strength correlation coefficients. 
The new DIBs are in part also blended with other spectral features such as stellar or telluric lines. 
Despite these limitations, our approach  shows the consistent presence of these features along different lines of sight. 
In future works, using higher S/N data sets, not limited by the intrinsic disadvantages of a filler program like EDIBLES, more new DIBs may be found in this way.\\[-8mm]

\begin{table}[t]
\centering
\caption{New DIBs found using DIB alignment.}\label{tab:new_dibs}
\setlength{\tabcolsep}{1mm}
{\small
\begin{tabular}{ccccccc}
\hline\hline
Query & New &  $\tilde\nu$\ & $\lambda_\mathrm{air}$ & $I$ &     $r$ & Qual.\\
DIB   & DIB    & (cm$^{-1}$) & (\AA) & & & Flag \\
\hline
5762 & 5522 & 18105.7$\pm$0.2 & 5521.59$\pm$0.05 & 0.17$\pm$0.06 & 0.855 & 2 \\
5797 & 5620 & 17787.2$\pm$0.5 & 5620.46$\pm$0.14 & 0.07$\pm$0.02 & 0.932 & 2 \\
6203 & 6302 & 15862.4$\pm$0.4 & 6302.49$\pm$0.16 & 0.13$\pm$0.10 & 0.859 & 2 \\
6379 & 5522 & 18105.8$\pm$0.2 & 5521.56$\pm$0.06 & 0.02$\pm$0.05 & 0.497 & 2 \\
$\ldots$ & 6439 &  15525.7$\pm$0.4 &             6439.17$\pm$0.17 &            $\ldots$ &   $\ldots$  & ...\\
\hline
\end{tabular}
\tablefoot{The wavelengths and wave numbers are mean values of the clusters and the errors are standard deviations of the clusters.
The strength ratios, $I$, are the median values of $\sigma_\mathrm{S}/\sigma_\mathrm{Q}$.
The 6439~\AA~DIB was not found using a query, and hence has no query, $I$, $r$, or quality. The 5522~\AA~DIB was found using two different queries, but at the same wavelengths within the errors.}}
\end{table}

\begin{figure}[ht]
\centering
\includegraphics[width = .8\linewidth]{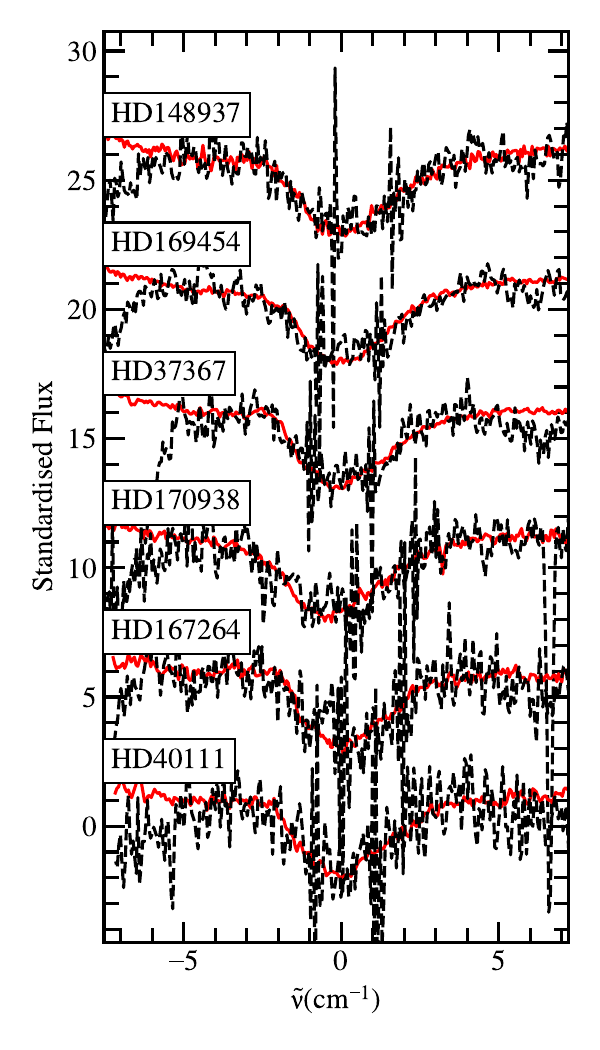}
\caption{Profile comparison between the 6203 and 6302\,\AA~DIBs for the six best-matching sight lines. The high noise in the 6302\,\AA~DIB region is caused by the correction for two strong telluric O$_2$ absorption lines. Over-correction of the telluric line cores can lead to emission features in the 6302~\AA~DIB and severely affect the $EW$ measurement.}
\label{fig:new_6203_6302}
\end{figure}

\paragraph{5522\,\AA~DIB.} This DIB is found with two queries independently.
Finding a new DIB with a query does not automatically mean that it belongs to the same DIB profile family or DIB family.
The extreme weakness introduces large uncertainties in profile shape and strength.\\[-8mm]

\paragraph{5620\,\AA~DIB.} This DIB is found in a number of sight lines. However, it can be affected by a stellar \ion{S}{ii} line at 5616.6\,{\AA}, especially if this line appears in emission, as is the case for HD~183143.\\[-8mm]

\paragraph{6302\,\AA~DIB.} This DIB is blended by two telluric O$_2$ lines, which is shown in Fig.~\ref{fig:6302_tell}, but is detected in several sight lines using the 6203\,\AA~DIB as a query.
The profile comparison for six sight lines is shown in Fig.~\ref{fig:new_6203_6302}.
The telluric lines can be over-corrected, which results in an emission feature at the centre of the 6302\,\AA~DIB. 
The correct choice of continuum is very difficult in this region because of multiple spectral features, which makes the $EW$ very unreliable.\\[-8mm]

\begin{figure}[ht!]
\centering
\includegraphics[width = .8\linewidth]{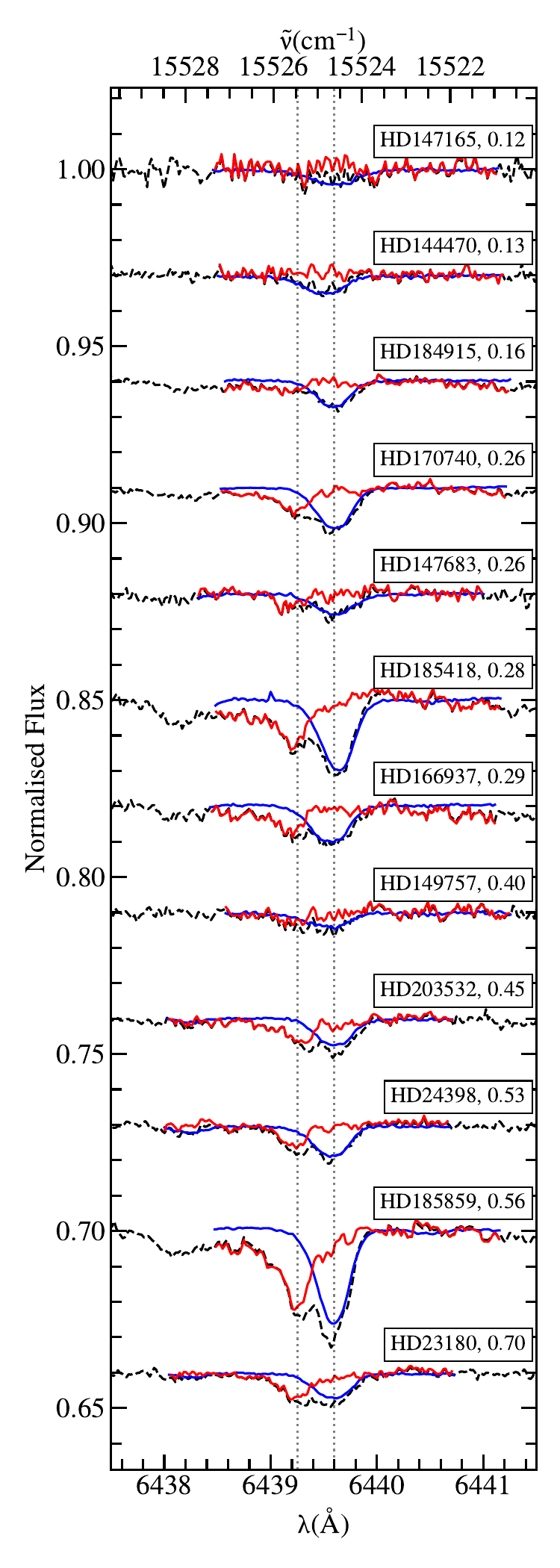}
\caption{Decomposition of the DIB blend at 6440~\AA\ for our single cloud sight lines. 
Besides each star name, the ratio $EW_{5797}/EW_{5780}$ is given. The original, co-added spectrum is shown in dashed black. 
The blue spectrum is from the 6196\,\AA~DIB, which was re-scaled and shifted in the wave number space. 
The red spectrum is the dashed black spectrum divided by the blue one and represents the blending 6439\,\AA~DIB.
The positions of the two blending DIBs are marked by dotted grey lines.
Wavelengths are given in the ISM rest frame.}
\label{fig:6196_6440_div_scl}
\end{figure}

\paragraph{6439\,\AA~DIB.} This DIB is blended with the 6440\,\AA~DIB and can be de-blended using DIB profile families.
The result of this process can be seen in Fig.~\ref{fig:6196_6440_div_scl} and further details are discussed in Sect.~\ref{sec:dib_blend}.

\section{Discussion}\label{sec:discussion}
\subsection{DIB profile families}
To confirm the identified DIB profile families, we compared our results with DIB families known from the literature.
The most recent correlation study on a large sample of 54 DIBs was performed by \citet{Fanetal22}.
We searched specifically for the smallest clusters shown in their Fig. 3 to affirm our DIB families.
Some of our best correlating pairs, pair 6379 - 6090\,\AA \, and pair 5797 - 5545\,\AA\ (Fig.~\ref{fig:profiles_5797_5545}), are also cluster pairs identified in \citet{Fanetal22}, fully in line with expectations.
We note that some DIB pairs can have very tight band strength correlations but slightly different profile shapes.
However, slight changes in the band profiles are to be expected for different transitions, even in the same molecule, as upon electronic excitation the molecular structure is expected to (slightly) change.

The 6614 -- 6521\,\AA~DIB pair is the most striking due to the common triple-peak structure, and therefore we give further attention to its discussion.
It is very likely that both DIBs are caused by the same or at least a very similar molecule, because of their common triple-peak structure and their very high band strength correlation.
A relatively high correlation coefficient of $r=0.88$ was found by \citet{xiang2012} for this DIB pair, using $EW$ correlations.
Their measurements of the 6521\,{\AA}~DIB $EW$ were probably affected by the stellar and telluric blend, leading to a lower correlation coefficient.
We see in Fig.~\ref{fig:6614_6521_sub_peaks} that their profiles correlate very well, including in sight lines with increased peak separation.
In accordance with \citet{edibles5}, we found enhanced broadening for both DIBs for the sight line of \object{HD 147889}, and additionally for \object{HD 147933}, with both sight lines penetrating the \object{Ophiuchus molecular cloud}.
From the uniformity of the variation in the sub-peak separation, we can conclude that the rotational constants of the respective excited states are identical within the margins of error.
However, in particular when looking at some of the single cloud sight lines, the 6521\,\AA~DIB appears to have a weaker P and a stronger R branch than the 6614\,\AA~DIB.
It is hard to determine whether those differences in intensity are intrinsic or whether they arise from unknown blends. This should be addressed in a detailed study focused on this DIB pair.
The detection of the relation between the 6521 and 6614\,\AA~DIBs supports the hypothesis that DIB 6614 is not a blend of two separate DIBs, caused by different carriers \citep{bernstein15}. It would be highly unlikely for the exact same blend to occur at two wavelengths.

Some DIB profile families are very similar at first sight.
The best example of this is the 5797 and 5850\,\AA~DIBs, which belong to different families.
They show a very high band strength correlation, but their profiles are systematically slightly different (Fig.~\ref{fig:5797_5850}), and their band strength ratios systematically change from $\sigma$ to $\zeta$ sight lines.
Figure \ref{fig:sigma_zeta_5797_5850_q1} helps to visualise these differences.
While the band strength correlation is very high (Fig. \ref{fig:5797_5850}), we see a trend of the 5850\,\AA~DIB getting stronger in $\zeta$ sight lines. 
In reference to the literature, this makes sense, because the 5850\,\AA~DIB has been proposed as a C$_2$ DIB before \citep{2014IAUS..297..121K}.
Hence, those DIBs are not in the same DIB family, but their carriers are very similar; that is, they may also share a chemical history.
We find more evidence of the distinct difference between the profiles in \cite{oka2013} in the sight line towards \object{Herschel 36}.
This sight line is known for its vastly broadened DIBs.
In Fig.~3 of \cite{oka2013}, both DIBs are shown in the velocity space.
The 5797\,\AA~DIB has an extended red wing, and the 5850\,\AA~DIB also has a red wing, but the latter's is far weaker.

The 4984\,\AA~DIB shows a high profile and band strength correlation with the 4963 and 5512\,\AA~DIBs (Table \ref{tab:corr_df}), but the two queries are clearly different (Fig.~\ref{fig:profiles_4963_5512}).
In Fig.~\ref{fig:sigma_zeta_4963_4984_q1} and \ref{fig:sigma_zeta_5512_4984_q1}, we show that the relative band strengths of the 5512 -- 4984\,\AA~DIB pair shows a weak correlation with $EW_{5797}/EW_{5780}$, while the 4963 -- 4984\,\AA~DIB pair increases with $EW_{5797}/EW_{5780}$.
Based on these correlations, we conclude that the 4984\,\AA~DIB is more likely to belong to the 5512\,\AA~DIB profile family.

However, the 5512 -- 4984\,\AA~DIB pair still shows a large $I$ variation of $\sim$25\,\%.
The obvious explanation is that they have different carriers that are closely chemically related. 
This could be by sharing a common chemical history or being sequential products in a reaction chain, and their ratios depending on the amount of ionising radiation in the interstellar radiation field (ISRF) or other unknown factors.
An alternative explanation would be that both DIBs have the same carrier, but the Franck-Condon factors of each transition can change depending on the ISRF or other environmental factors.
This possibility was already mentioned for C$_{60}^+$ by \citet{lykhin19}, who speculated that the intensity ratio of the two strongest C$_{60}^+$ DIBs can depend on the population of the rotational energy levels.
The variation in $I$ for the 5512 -- 4984\,\AA~DIB pair is hard to interpret, because of the higher effect of noise on those very weak DIBs, but there is seemingly only a weak correlation with $EW_{5797}/EW_{5780}$.

The $EW_{5797}/EW_{5780}$ $I$ correlations may have an advantage compared to traditional correlation studies, because they are not affected by the correlation with $E(B-V)$, as only strength ratios are used.
Furthermore, correlation coefficients are always decreased by noise, while this measurement error does not prevent us from calculating an $EW_{5797}/EW_{5780}$ $I$ correlation.
So far, we have only used this method as a tie-breaker, as in the case of the 4984\,\AA~DIB, and as a tool to identify DIB blends, but we note the potential of this method.

\subsection{C$_2$ diffuse interstellar bands\label{sec:c2_discussion}}

We were able to split the well-known C$_2$ DIBs into three DIB profile families.
The respective query of each family is the 4963, 5850, and 5512\,{\AA}~DIB.
We find that four DIBs previously not considered to be C$_2$ DIBs correlate well with known C$_2$ DIBs: the 5062, 5245, 5911, and 5850\,\AA~DIBs.
Hence, we propose to consider those DIBs as C$_2$ DIBs, or at least as C$_2$-related DIBs.
It was already suggested by \citet{edibles3} that there is no separate group of C$_2$ DIBs, but rather a gradual transition from non-C$_2$ to C$_2$ DIBs.
The 5762, 5850, and 5911\,\AA~DIBs fall in the intermediate range between the two qualifications.
We can confirm this hypothesis, thanks to the high band strength correlation that we find between the 5850 and 5797\,\AA~DIBs.

\begin{figure}[ht]
\includegraphics[width = .9\linewidth]{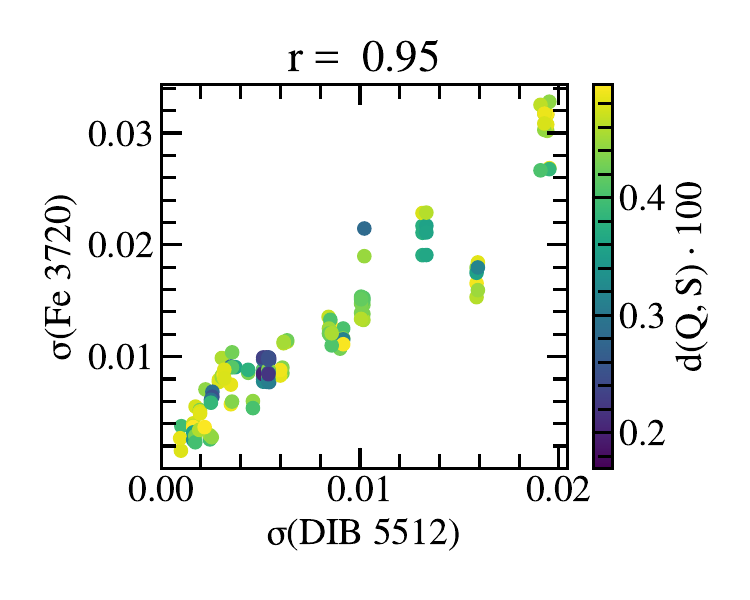}
\caption{Strength correlation between the 5512~\AA~DIB and the interstellar \ion{Fe}{i} line at 3720~\AA.
The very strong correlation indicates a strong connection between the DIB carrier and iron in the gas phase.
This result drew our attention to this relation, but because of the different profiles of the two features this method is not reliable and has to be confirmed using $EW$ correlations.}
\label{fig:corr_5513_3720}
\end{figure}

\begin{figure}[ht!]
\centering
\includegraphics[width = .8\linewidth]{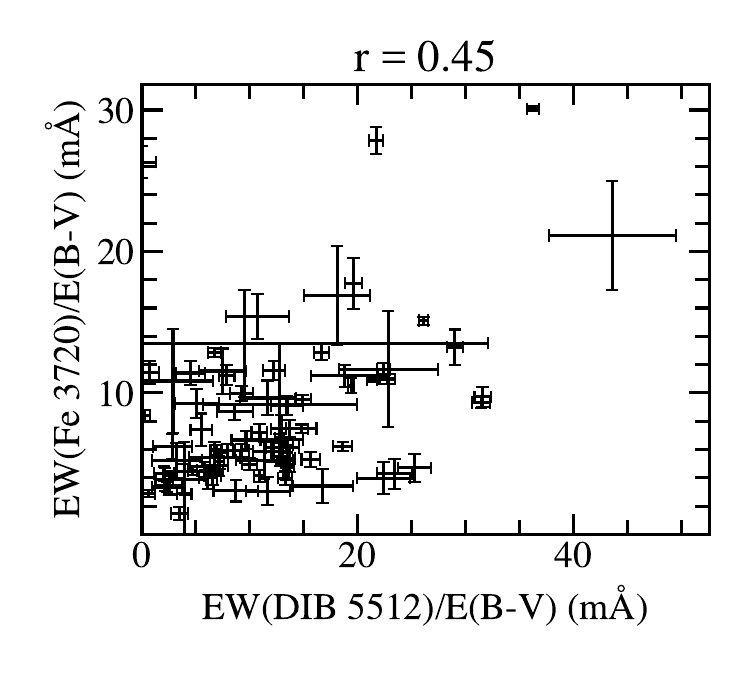}
\caption{Correlation plot of manual $EW$ measurements of the 5512~{\AA} DIB and the interstellar \ion{Fe}{i} line at 3720~\AA , normalised by $E(B-V)$. Single outliers like that at the top left (HD~37041) decrease the correlation coefficient significantly, but a linear trend is visible.}
\label{fig:ew_5513_3720}
\end{figure}

The 5512\,\AA~DIB is known to be a C$_2$ DIB \citep{edibles3}. 
It strongly correlates with other C$_2$ DIBs, and also to an interstellar feature, as is found in the present work: the interstellar \ion{Fe}{i} in the gas phase.
A strong correlation with iron is also very prominent in our data and may add extra information on the molecular origin of this feature.
The $\sigma$ correlation found by our algorithm is shown in Fig.~\ref{fig:corr_5513_3720}.
However, given the fact that the \ion{Fe}{i} line and the 5512\,\AA~DIB generally do not have the same intrinsic profile, our method might be prone to error; hence, we performed manual $EW$ measurements of both features to get a reliable correlation (see Fig.~\ref{fig:ew_5513_3720}). 
We normalised the $EW$s by $E(B-V)$ to mitigate the effect of the overall column density on both values.
These measurements were done on co-added spectra and show a $EW$ correlation of $r=0.45$, which supports the findings of the algorithm.
There are some outliers that decrease the correlation coefficient significantly, like the sight line to \object{HD 37041}, which decreases $r$ from 0.55 to 0.45. This sight line penetrates an extremely ionised region of the \object{Orion Bar}. Those conditions could prevent the DIB carrier from forming.
The manually measured $EW$s tend to be smaller than the automated values, because the \ion{Fe}{i} line can be blended with a stellar \ion{He}{i} line, which is easily discernible by eye because of its larger width.
We calculated the $EW$ errors using the following formula from \citet{2011A&A...533A.129V}:
\begin{equation}
    \sigma_{EW} = \sqrt{2 \Delta\lambda\delta\lambda}/(S/N)\,,
\end{equation}
where $\Delta\lambda$ is the wavelength range of the $EW$ measurement, $\delta\lambda$ 
is the spectral dispersion, and $S/N$ is measured per pixel. 

We know from previous studies that iron should be rather abundant in the ISM, but only a small fraction has been detected in the gas phase, \citep{Cartledgeetal06,Jenkins09},
while the remainder is thought to be bound to dust grains \citep[e.g.][]{Przybillaetal08b,NiPr12}.
In this context, the $EW$ correlation between the 5512\,\AA~DIB and \ion{Fe}{i} indicates that there is a reservoir of \ion{Fe}{i} that can be released into the gas phase in environments where C$_2$ DIBs are present. Laboratory experiments carried out on ionised iron-PAH clusters show that the dissociation of these clusters is possible under mild ultraviolet irradiation and that it releases \ion{Fe}{i} into the gas phase \citep{simon2009}. Depending on the number of Fe atoms and PAH units involved in these clusters, photodissociation can either release PAH$^+$ or [Fe-PAH]$^+$ species. While it is difficult to come to a conclusion as to the nature of the 5512\,\AA~DIB carrier, this laboratory study suggests that the 5512\,\AA~DIB carrier is a product of the photodissociation of iron-PAH clusters, which may explain the correlation observed with the \ion{Fe}{i} abundance.

\subsection{The 6440 -- 6439\,{\AA}~diffuse interstellar band blend\label{sec:dib_blend}}

The 6196 - 6440\,\AA~DIB pair is a peculiar case.
In the $\sigma$ correlation, one can see a linear trend of strength ratios, but there are also many outliers where the $\sigma$ values for the 6440\,\AA~DIB are too high (see Fig.~\ref{fig:6196_6440}).
This indicates that another feature is blended with the 6440\,\AA~DIB.
Looking at the single cloud sight lines, one can see a possible additional DIB component at 6440~\AA\ that is not correlated with the profile or strength of the 6196\,\AA~DIB and that we refer to as the 6439\,\AA~DIB. 
This component increases $\sigma_{6440}$ for many sight lines.
We investigated whether the component is a $\sigma$ or $\zeta$ DIB by plotting $\sigma_{6440}/\sigma_{6196}$ with $EW_{5797}/EW_{5780}$, and find a clear correlation (Fig.~\ref{fig:sigma_zeta_6196_6440_q1}).
This means that the 6440\,\AA~DIB is different from the 6196\,\AA~DIB, or that it could be a blend of two DIBs: the 6440 and 6439\,\AA~DIBs. 
The 6440\,\AA~DIB correlates with the 6196\,\AA~DIB, which is a $\sigma$ DIB \citep{Fanetal22}, and the 6439\,\AA~DIB, which is a $\zeta$-type DIB, blending with the other feature.
We tested this hypothesis by dividing the blend by the re-scaled and shifted 6196\,\AA~DIB. 
A visualisation of this de-blending process is shown in Fig.~\ref{fig:6196_6440_div_scl}.
The 6196\,\AA~DIB was re-scaled in strength by using an intensity ratio, $I$ = 0.18$\pm$0.06, and it was shifted by $\Delta \tilde\nu$ = $-$610.5$\pm$0.2\,cm$^{-1}$.
After division, the blending 6439\,\AA~DIB is visible with an increasing strength relative to the 6440\,\AA~DIB for $\zeta$ sight lines.
This second component cannot be of stellar origin because there are cases in which it is much narrower than the stellar lines; in particular, it shows a constant wavelength shift with respect to the 6440\,\AA~DIB in the investigated sight lines.
It is still uncertain if the blend is real, but the blue-ward sub-peak definitely increases relative to the red-ward sub-peak with increasing $EW_{5797}/EW_{5780}$ values.

\subsection{Vibrational progressions in diffuse interstellar band profile families\label{sec:Vibrational_progressions}}
The identified DIB families show energy differences of several hundredths of wave numbers (see Table~\ref{tab:dib_families}), which is in the energy regime of vibrational modes.
It may be possible to link those energy differences to vibrational modes of chemical bonds that are part of the carrier molecules by comparison with the  Vibrational and Electronic Energy Levels of Small Polyatomic Transient Molecules Database \citep[VEEL,][]{veel, jacox1998vibrational}.
Care has to be taken that quite often only matrix data exist, which exhibit a clear shift with respect to unperturbed gas phase spectra. 
Moreover, one has to realise that the energy difference originates from the excited state, for which much less accurate data are available. 
Nevertheless, taking into account a collection of energy differences and assuming that all transitions take place in one carrier species may offer further insights into its nature.
For example, the 5512\,\AA~DIB may be caused by a pure electronic transition, while the 5170\,\AA~DIB ($E_\mathrm{diff}$ = 1200.1$\pm$0.2\,cm$^{-1}$) may arise from the same electronic transition plus a CH$_2$ scissor mode. As a potential example,
 $\pi$ complexes of ethylene with iron monoxide or iron have infrared bands measured at $\tilde{\nu}\approx$\,1200\,cm$^{-1}$\, in solid argon \citep{kafafi1987}. 
Such complexes, similarly to PAH-Fe complexes, might be present in interstellar clouds and contribute to the depletion and/or release of iron in the gas phase depending on the environmental conditions.
For the 6614-6521\,\AA~DIB pair, it makes sense to look for species that have vibrational modes around 215.6\,cm$^{-1}$ and that upon simulation reproduce the triple-peak spectrum.
Assigning possible DIB carriers on the basis of the energy differences derived here is beyond the scope of this study and will await dedicated spectroscopic studies.

\section{Conclusions}\label{sec:conclusions}
We introduced DIB alignment as a new tool with which to investigate large data sets such as our EDIBLES data set, which shows great potential in several applications.
We report eight DIB families with highly correlating strengths and profiles, as well as four previously unreported DIBs in the visual range using DIB alignment. In particular, we report the 6614 -- 6521\,\AA~DIB pair, where both DIBs show the same triple-peak substructure and an unprecedented band strength Pearson correlation coefficient of 0.9935.
Specifically, the measurement of DIB band strengths using the flux standard deviation proves to be much more precise than the traditional $EW$ measurement.
However, the fact that it can be only used to compare DIBs with similar band profiles is a caveat.
Using DIB alignment, we have been able to link different DIBs more accurately, specifically to define DIB profile families and to identify a few previously unreported weak DIBs. 
The method also allows us to study blended DIBs more accurately. 
Several C$_2$ DIBs could be sorted into three families, including some DIBs previously not assigned as C$_2$ DIBs.
However, they do not include all C$_2$ DIBs. 
This could be either because the remaining DIBs are too weak in our observational data or because there are more C$_2$ DIB families that were not detected by our analysis. 
Their correlation with \ion{Fe}{i} is interesting and should be further examined.
In particular, this draws attention to new types of carriers for DIBs. For example, complexes of Fe with aromatic hydrocarbons are expected to exhibit strong bands in the visible region \citep{simon2009,Lanza2015}.
DIB alignment will be particularly
helpful for future DIB surveys in the near-infrared, making it much easier to find new DIBs (if their profiles resemble one of the input query DIBs sufficiently closely).

Further improvements to the method are necessary.
For example, the query detection was not always successful.
It might be better to compile custom DIB profile templates for each DIB profile family, which should be more robust than the skewed Gaussian but which also will be more work-intensive.
With enough refinement, it should be possible to enhance our understanding of DIBs from an observational point of view and create more and more detailed descriptions of possible carriers.

The discovery of DIB profile families is an important step towards a better understanding of DIB carrier molecules.
Detailed studies of specific DIB profile families or DIB pairs involving contour fitting and vibrational progressions will enhance our understanding of their carriers and test the hypothesis of common carriers.
In particular, subtle profile shape differences between DIB profile family members can provide more information about their carrier molecules.

\begin{acknowledgements}
A.E. thanks Mina Soltangheis for discussing the DIB profile problem and suggesting time-series alignment as a possible analysis method.
A.E. and N.P. gratefully acknowledge support by the Austrian Science Fund FWF project DK-ALM, grant W1259-N27. J.S. sincerely thanks the Astronomy Technology Centre for support. H.L. thanks NOVA (the Dutch research school for Astronomy) and NWO (the Netherlands Organisation for Scientific Research). This work is based on observations collected at the European Organisation for Astronomical Research in the Southern Hemisphere under ESO programme 194.C-0833. The data are publically available via the ESO Science Archive Facility at \url{http://archive.eso.org/eso/eso_archive_main.html}.
The DIB alignment algorithm is available at \url{https://github.com/alex-ebi/astro_scripts_uibk}.
   
\end{acknowledgements}


%
   \bibliography{bibliography} 
   \bibliographystyle{aa} 
   \listofobjects

%
\begin{appendix}
\setcounter{section}{1}
\setcounter{figure}{0}

\begin{figure*}[]
\begin{flushleft}{\large\textsf{\textbf{Appendix A: Supplementary comparison plots}}}\end{flushleft}
\includegraphics[width = .95\linewidth]{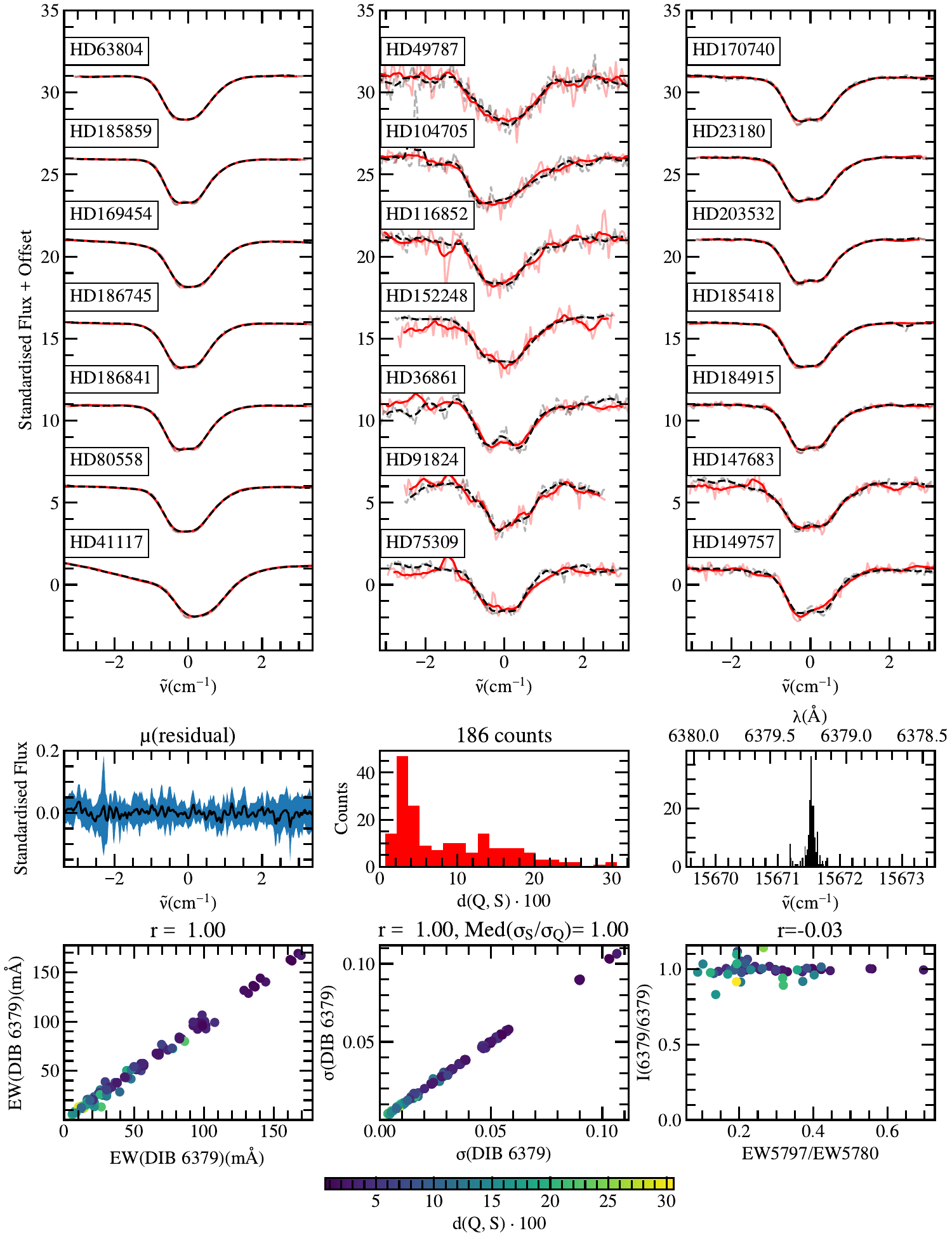}
\caption{Auto-correlation plot of the 6379~\AA~DIB. This is a very strong DIB, so the correlation is very high. The scatter of the matched wave numbers reflects the accuracy of the wavelength calibration.}
\label{fig:fam_6379_6379}
\end{figure*}

\begin{figure*}
\includegraphics[width = .95\linewidth]{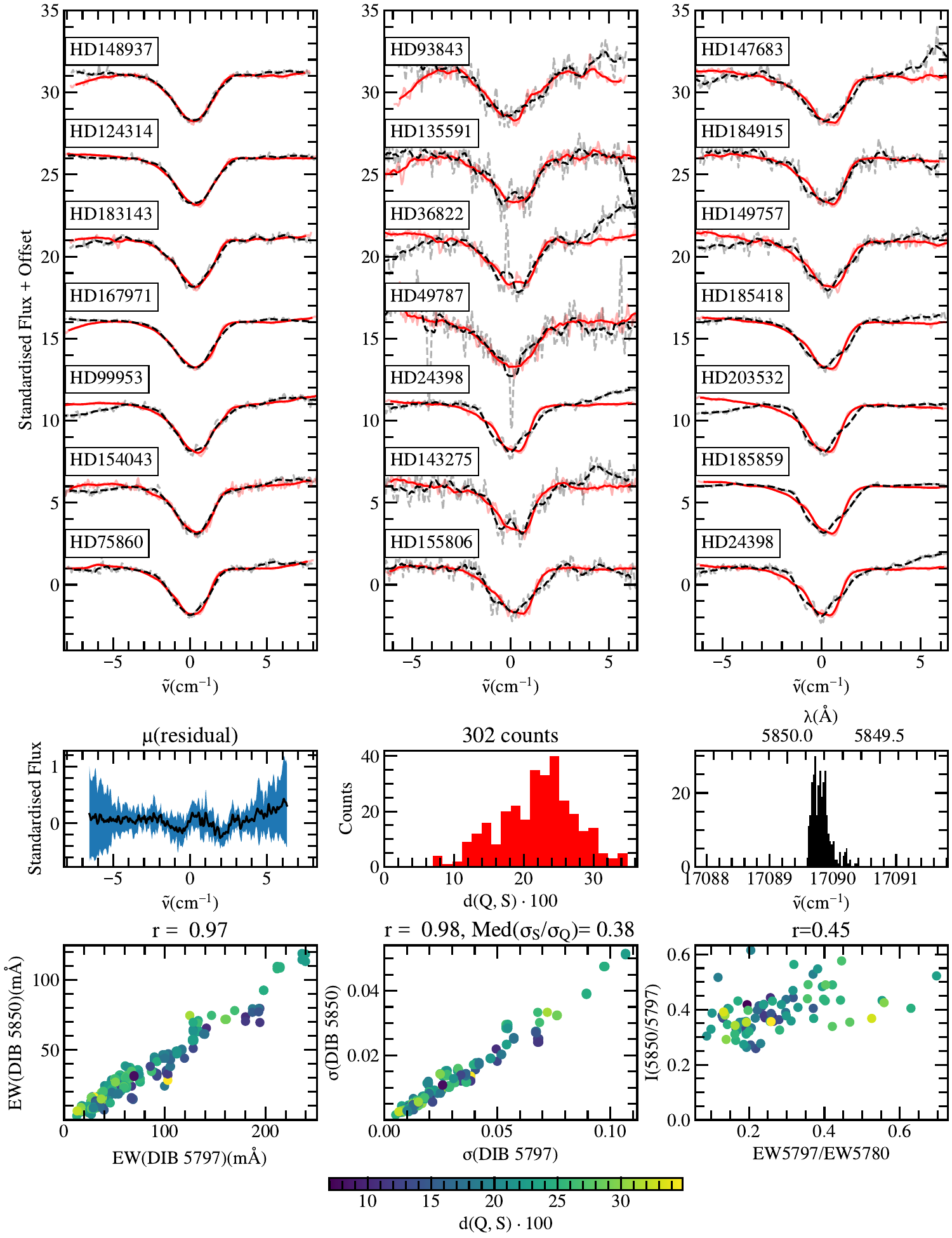}
\caption{Comparison between the 5797 (solid red line) and 5850\,\AA~ (dashed black line) DIBs. This pair has a very high strength correlation, but the 5797\,\AA~DIB always has its typical extended wing towards shorter wave numbers, while the 5850\,\AA~DIB tends to be more symmetric. For these small profile differences, the match distance histogram is very useful. The counts are slowly rising with increasing $d(Q,S)$. For matching profiles, there is a fast rise at the smallest distances (see Fig. \ref{fig:profiles_6379_6090}).}
\label{fig:5797_5850}
\end{figure*}

\begin{figure*}
\includegraphics[width = .99\linewidth]{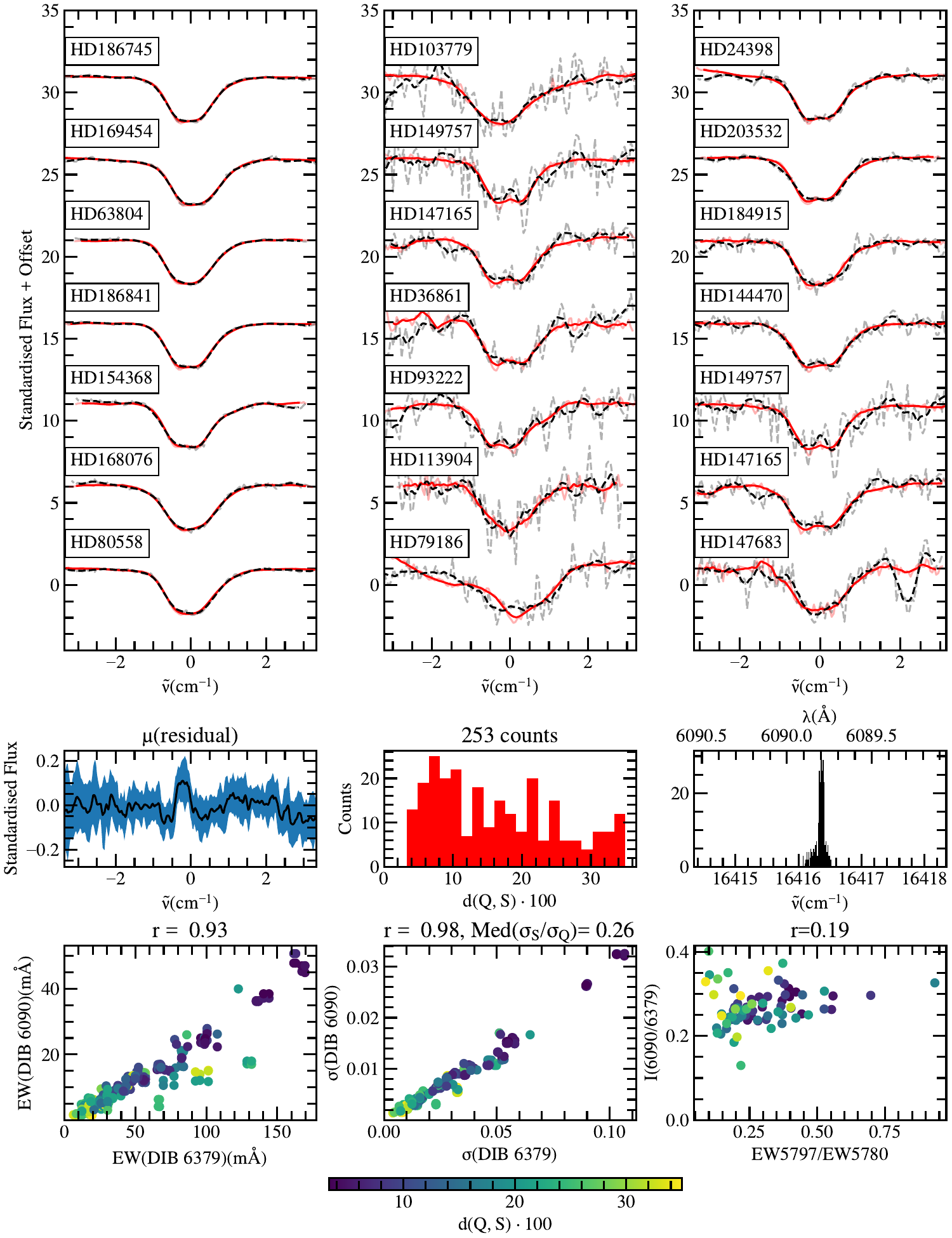}
\caption{Comparison between the 6379\,{\AA}~DIB and the 6090\,{\AA}~DIB.}
\label{fig:profiles_6379_6090}
\end{figure*}

\begin{figure*}
\includegraphics[width = .99\linewidth]{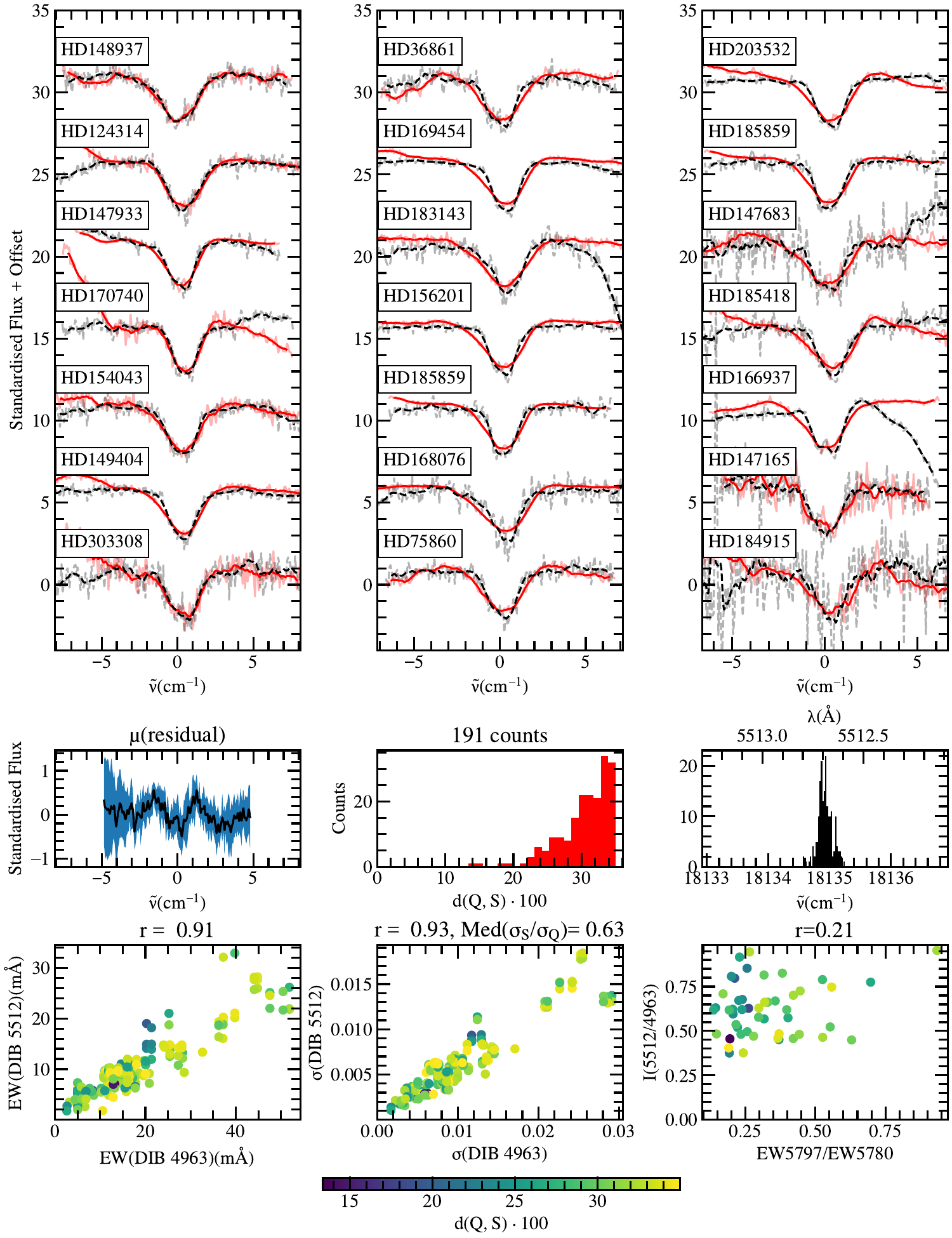}
\caption{Comparison between the 4963\,{\AA}~DIB and the 5512\,{\AA}~DIB.}
\label{fig:profiles_4963_5512}
\end{figure*}

\begin{figure*}
\includegraphics[width = .95\linewidth]{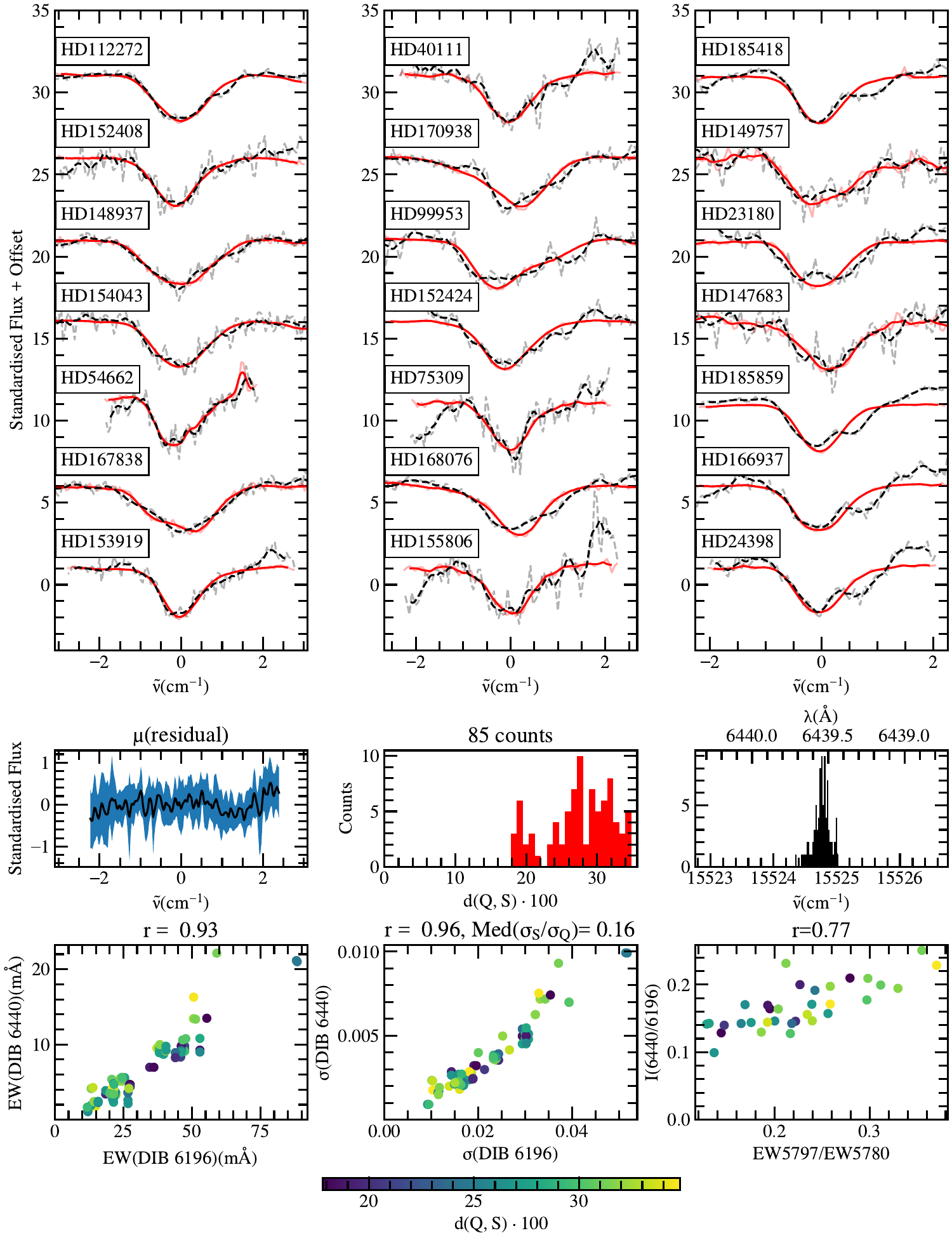}
\caption{Comparison between the 6196 (solid red line) and 6440\,\AA~ (dashed black line) DIBs. The blending 6439\,\AA~DIB can be seen in the residuals (middle left).}
\label{fig:6196_6440}
\end{figure*}

\setcounter{section}{2}
\setcounter{figure}{0}

\begin{figure*}
\begin{flushleft}{\large\textsf{\textbf{Appendix B: DIB profile families -- comparison plots}}}\end{flushleft}
\includegraphics[width = .97\linewidth]{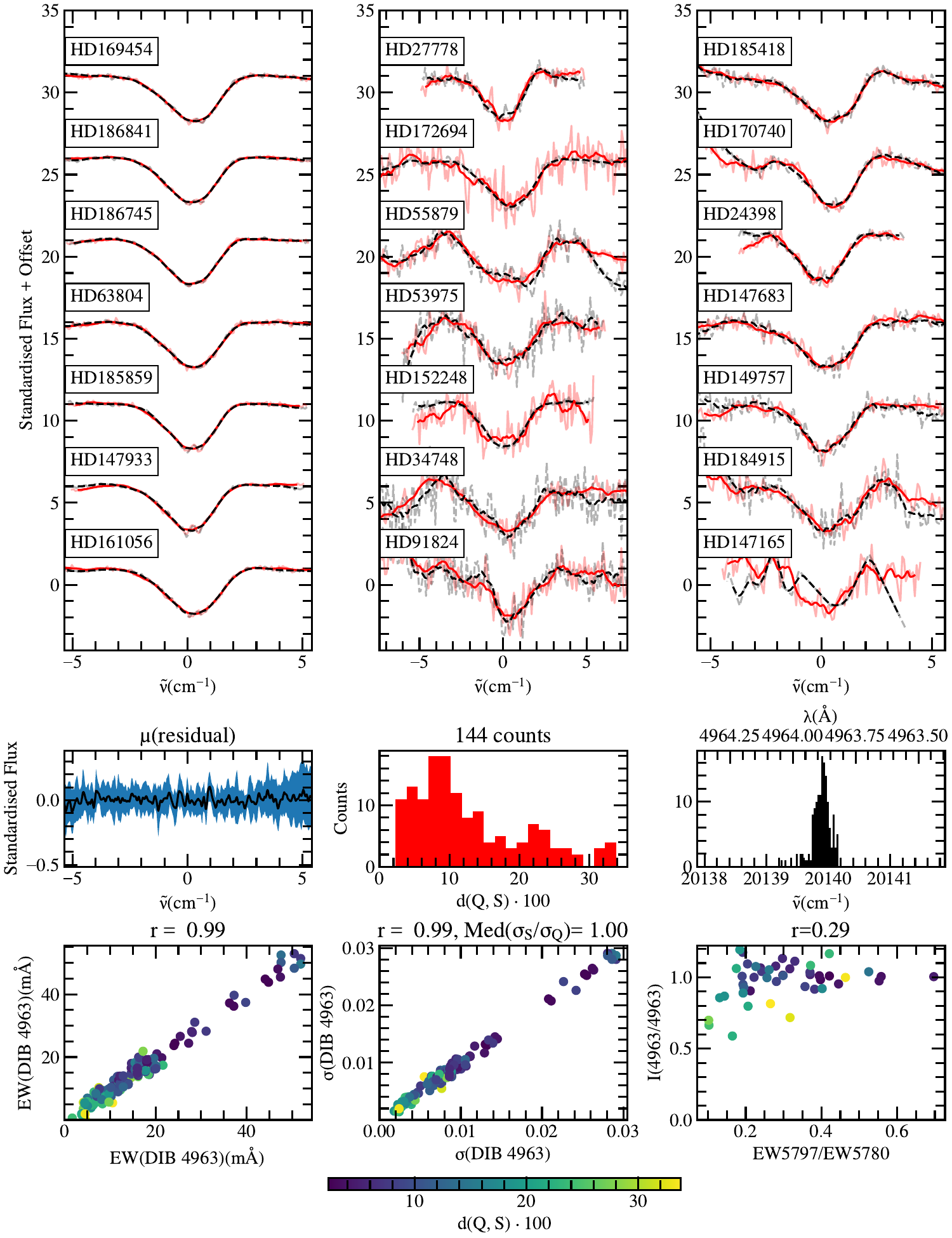}
\caption{Comparison between the 4963\,{\AA}~DIB and the 4963\,{\AA}~DIB.}
\label{fig:profiles_4963_4963}
\end{figure*}

\begin{figure*}
\includegraphics[width = .99\linewidth]{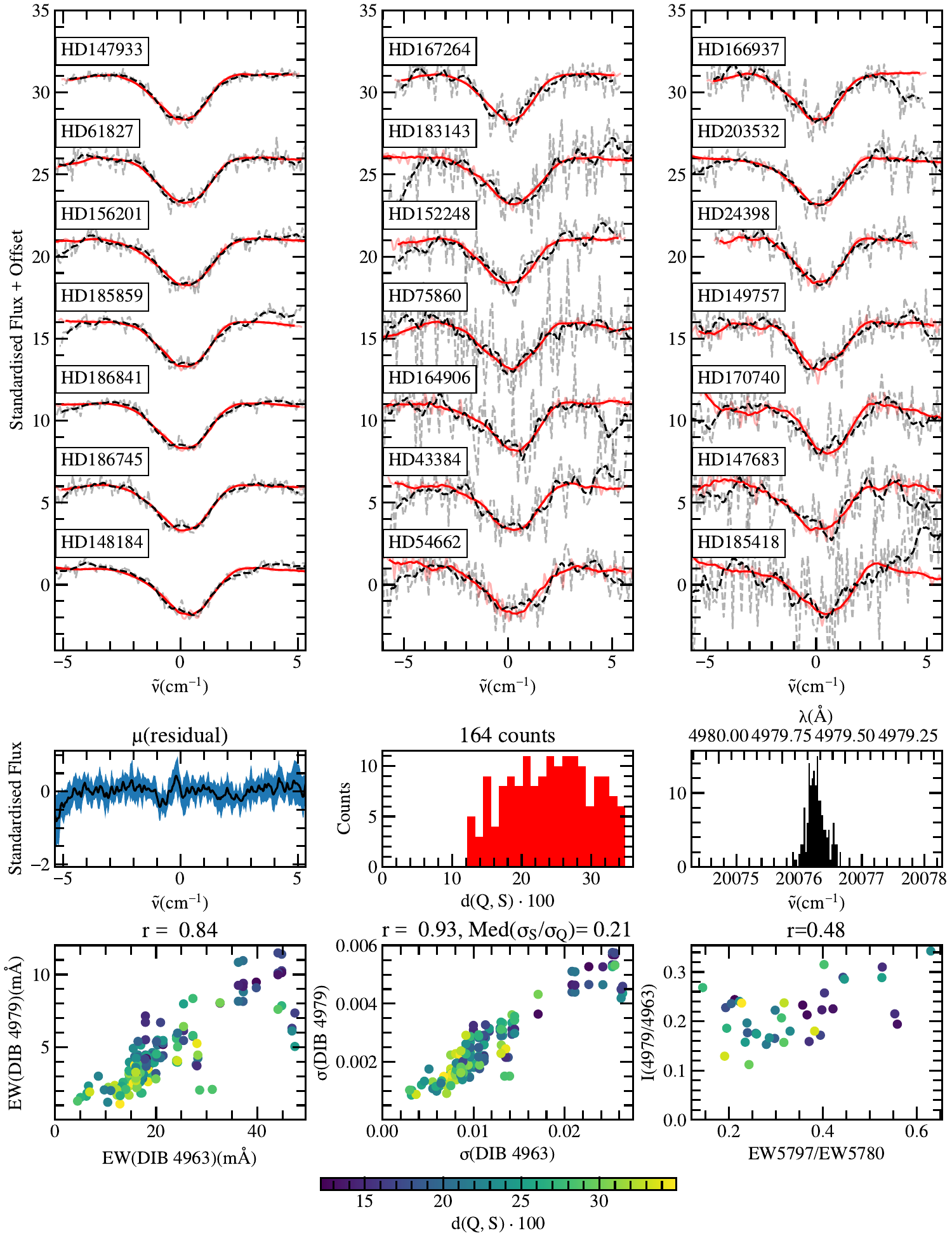}
\caption{Comparison between the 4963\,{\AA}~DIB and the 4979\,{\AA}~DIB.}
\label{fig:profiles_4963_4979}
\end{figure*}

\begin{figure*}
\includegraphics[width = .99\linewidth]{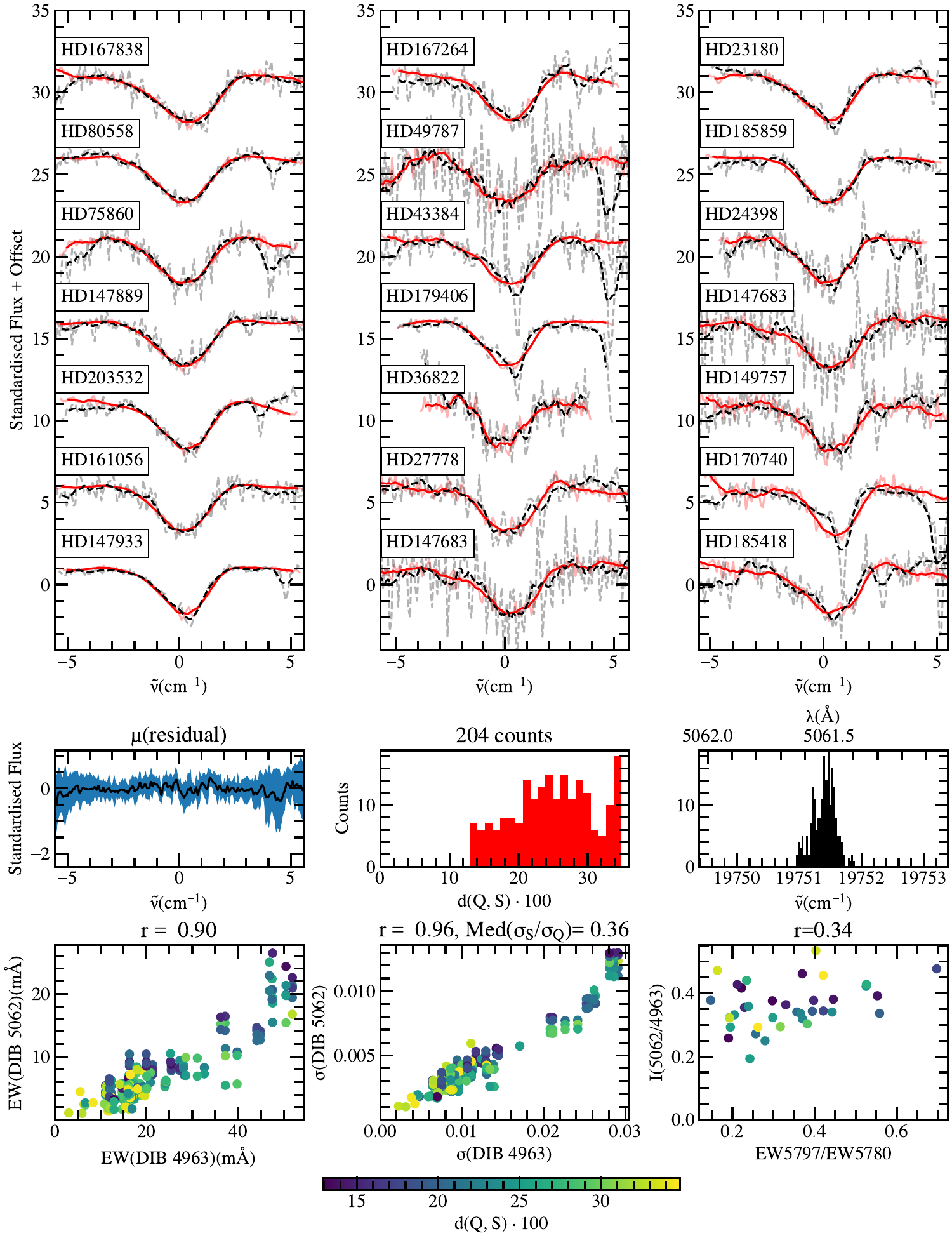}
\caption{Comparison between the 4963\,{\AA}~DIB and the 5062\,{\AA}~DIB.}
\label{fig:profiles_4963_5062}
\end{figure*}

\begin{figure*}
\includegraphics[width = .99\linewidth]{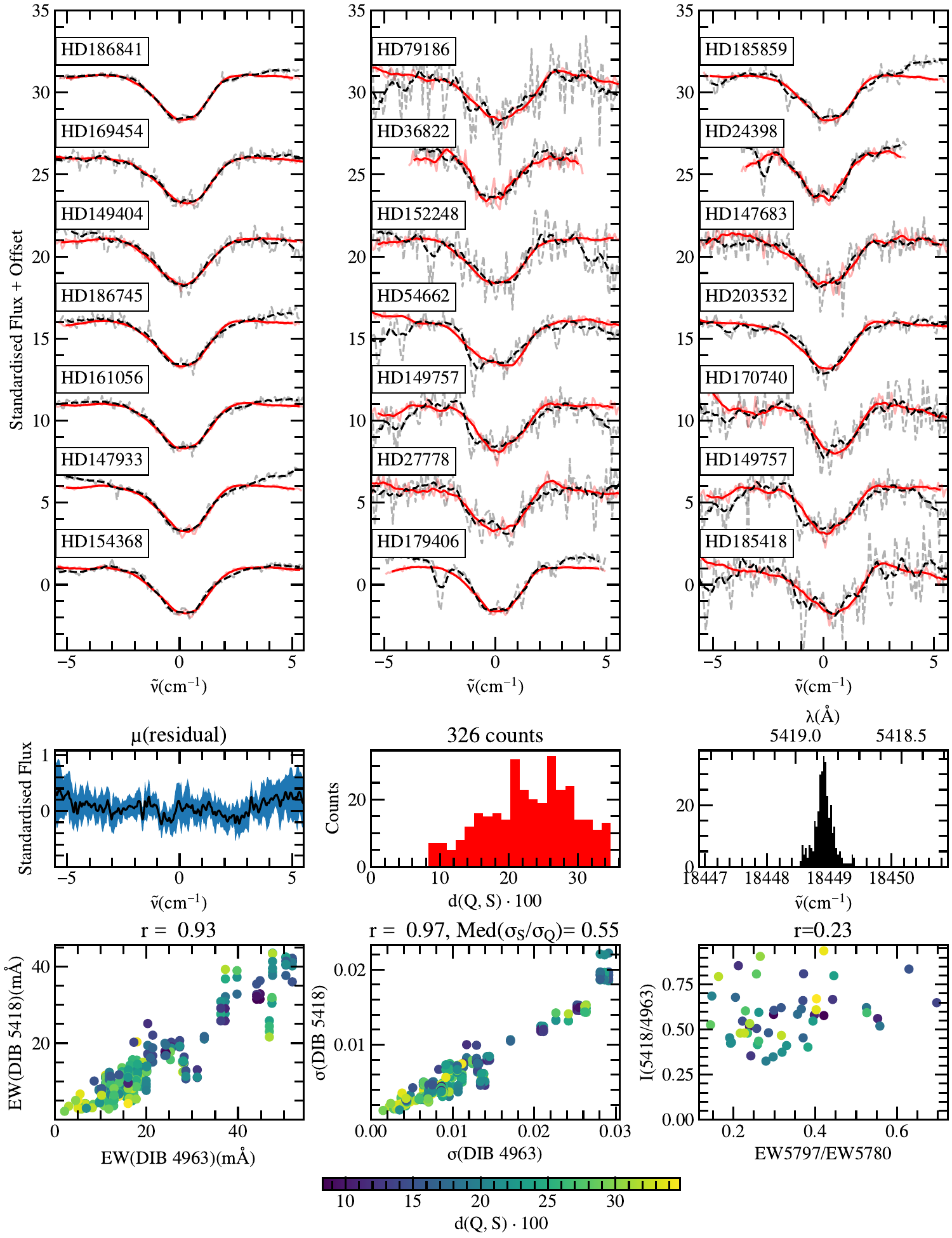}
\caption{Comparison between the 4963\,{\AA}~DIB and the 5418\,{\AA}~DIB.}
\label{fig:profiles_4963_5418}
\end{figure*}

\begin{figure*}
\includegraphics[width = .99\linewidth]{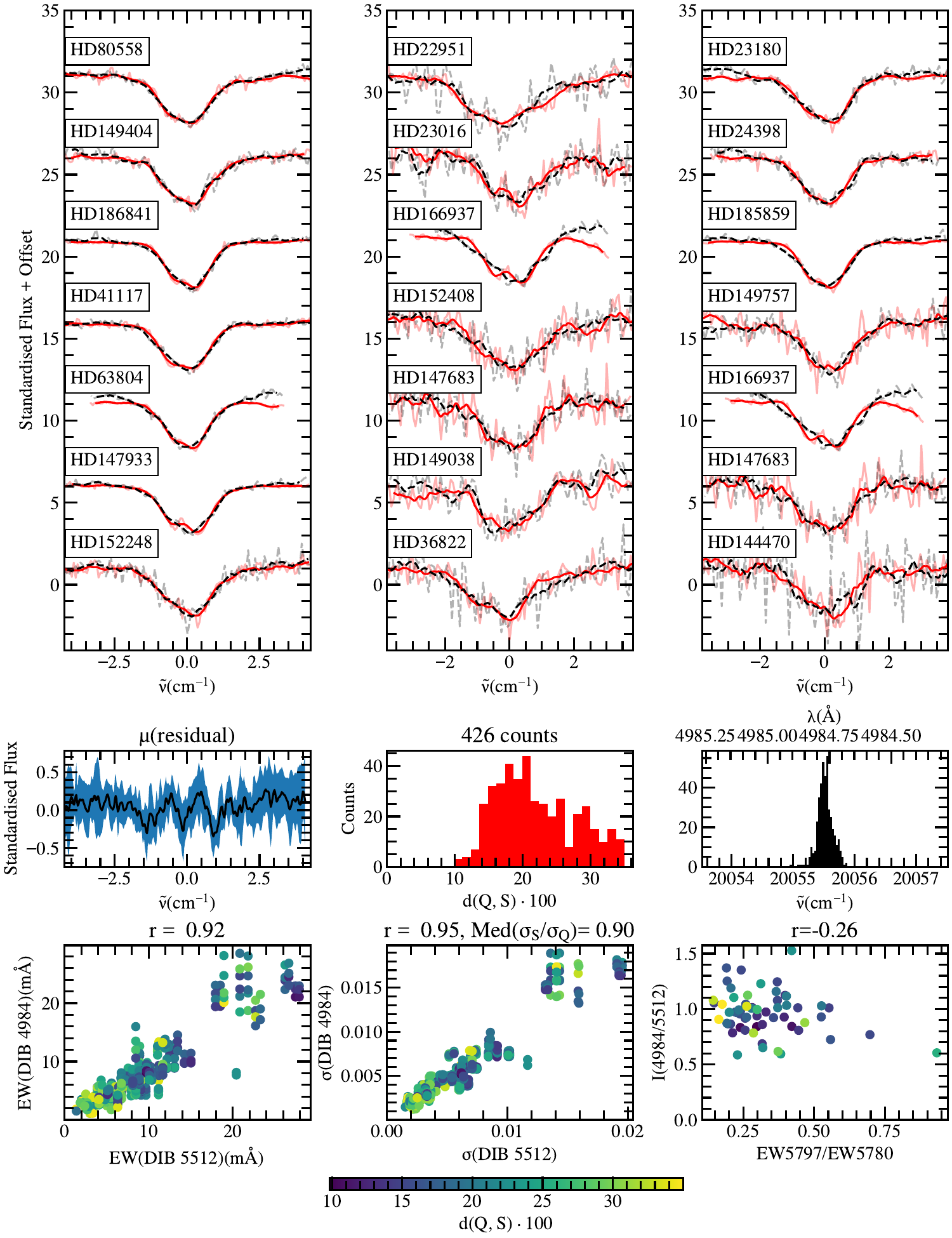}
\caption{Comparison between the 5512\,{\AA}~DIB and the 4984\,{\AA}~DIB.}
\label{fig:profiles_5512_4984}
\end{figure*}

\begin{figure*}
\includegraphics[width = .99\linewidth]{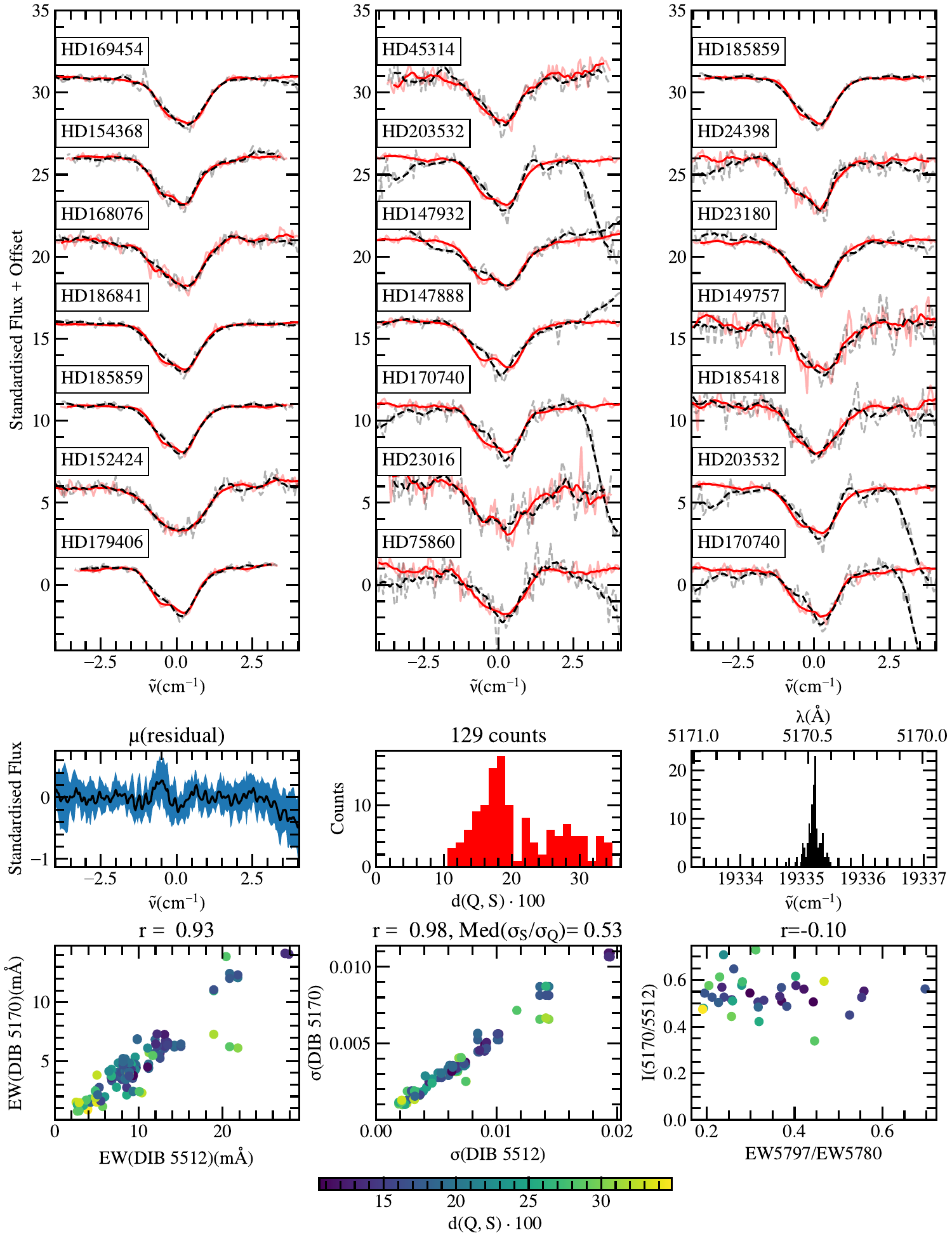}
\caption{Comparison between the 5512\,{\AA}~DIB and the 5170\,{\AA}~DIB.}
\label{fig:profiles_5512_5170}
\end{figure*}

\begin{figure*}
\includegraphics[width = .99\linewidth]{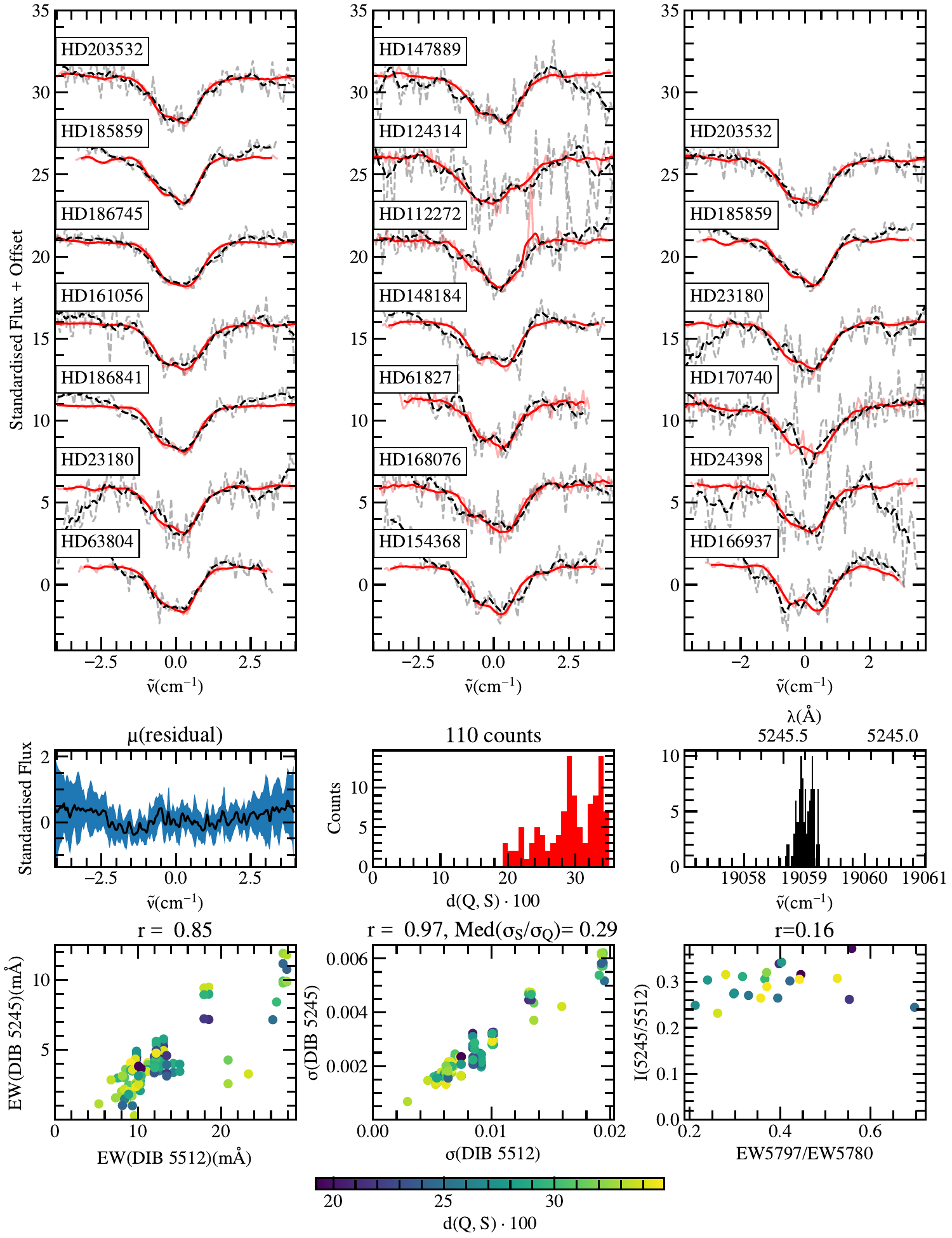}
\caption{Comparison between the 5512\,{\AA}~DIB and the 5245\,{\AA}~DIB.}
\label{fig:profiles_5512_5245}
\end{figure*}

\begin{figure*}
\includegraphics[width = .99\linewidth]{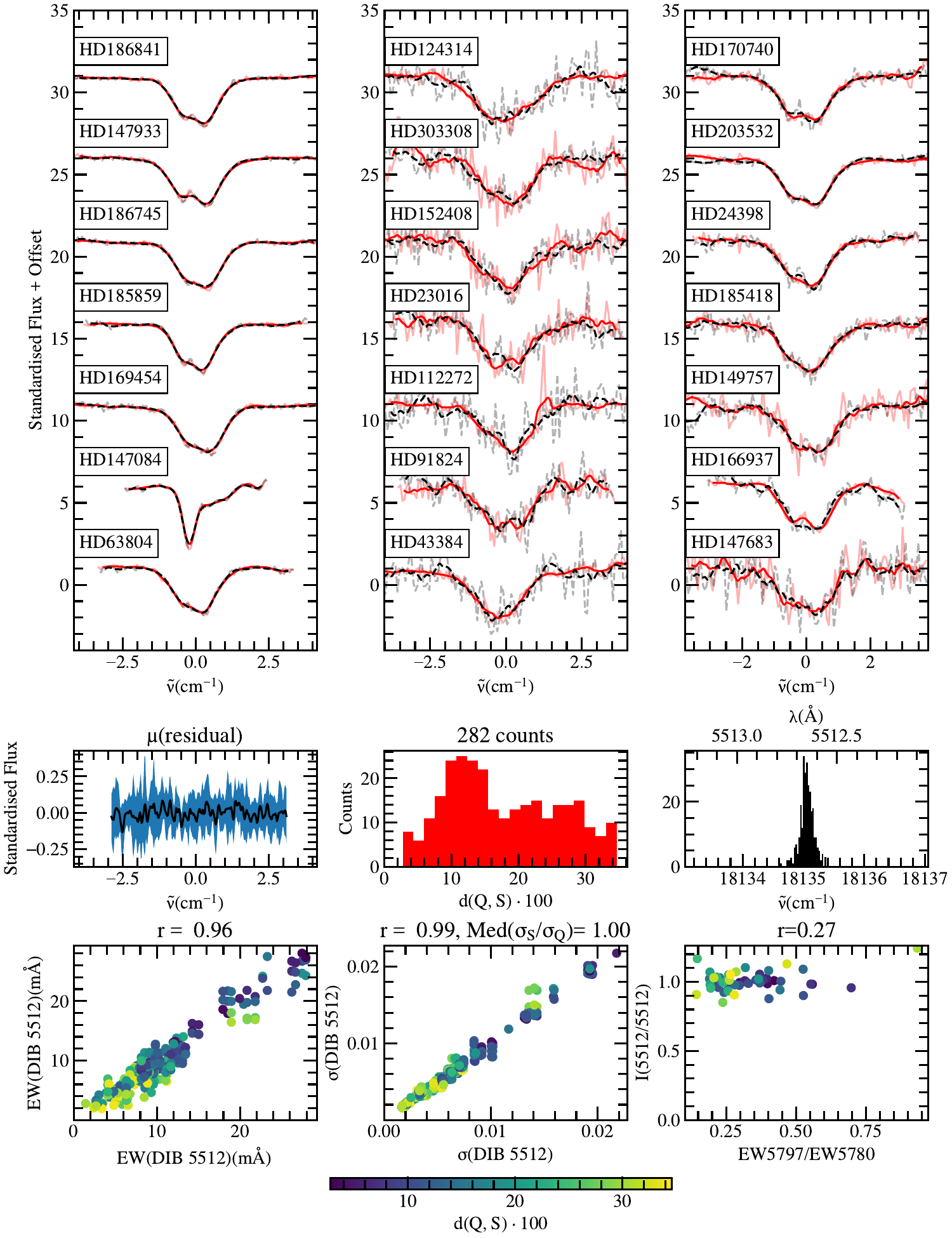}
\caption{Comparison between the 5512\,{\AA}~DIB and the 5512\,{\AA}~DIB.}
\label{fig:profiles_5512_5512}
\end{figure*}

\begin{figure*}
\includegraphics[width = .99\linewidth]{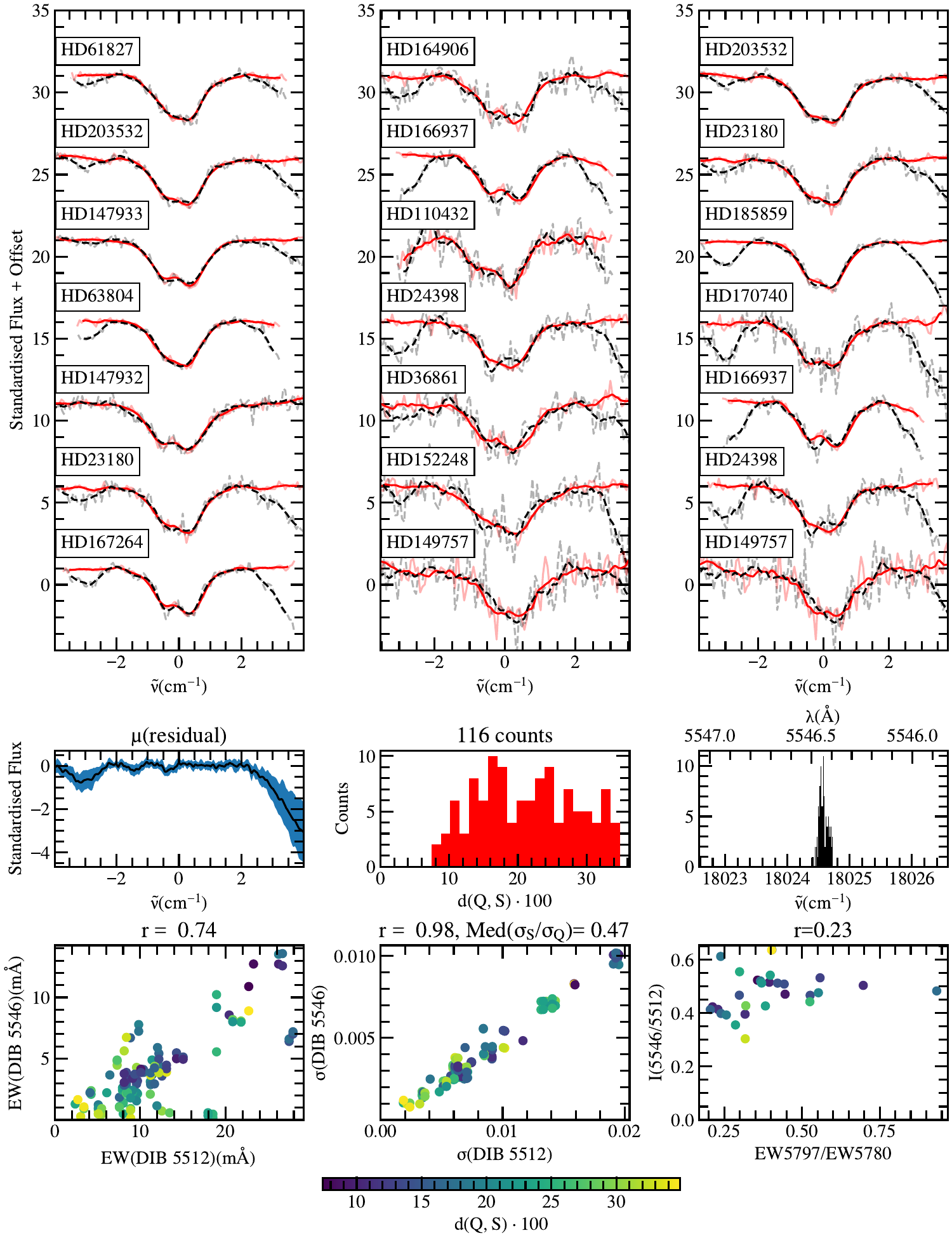}
\caption{Comparison between the 5512\,{\AA}~DIB and the 5546\,{\AA}~DIB.}
\label{fig:profiles_5512_5546}
\end{figure*}

\begin{figure*}
\includegraphics[width = .99\linewidth]{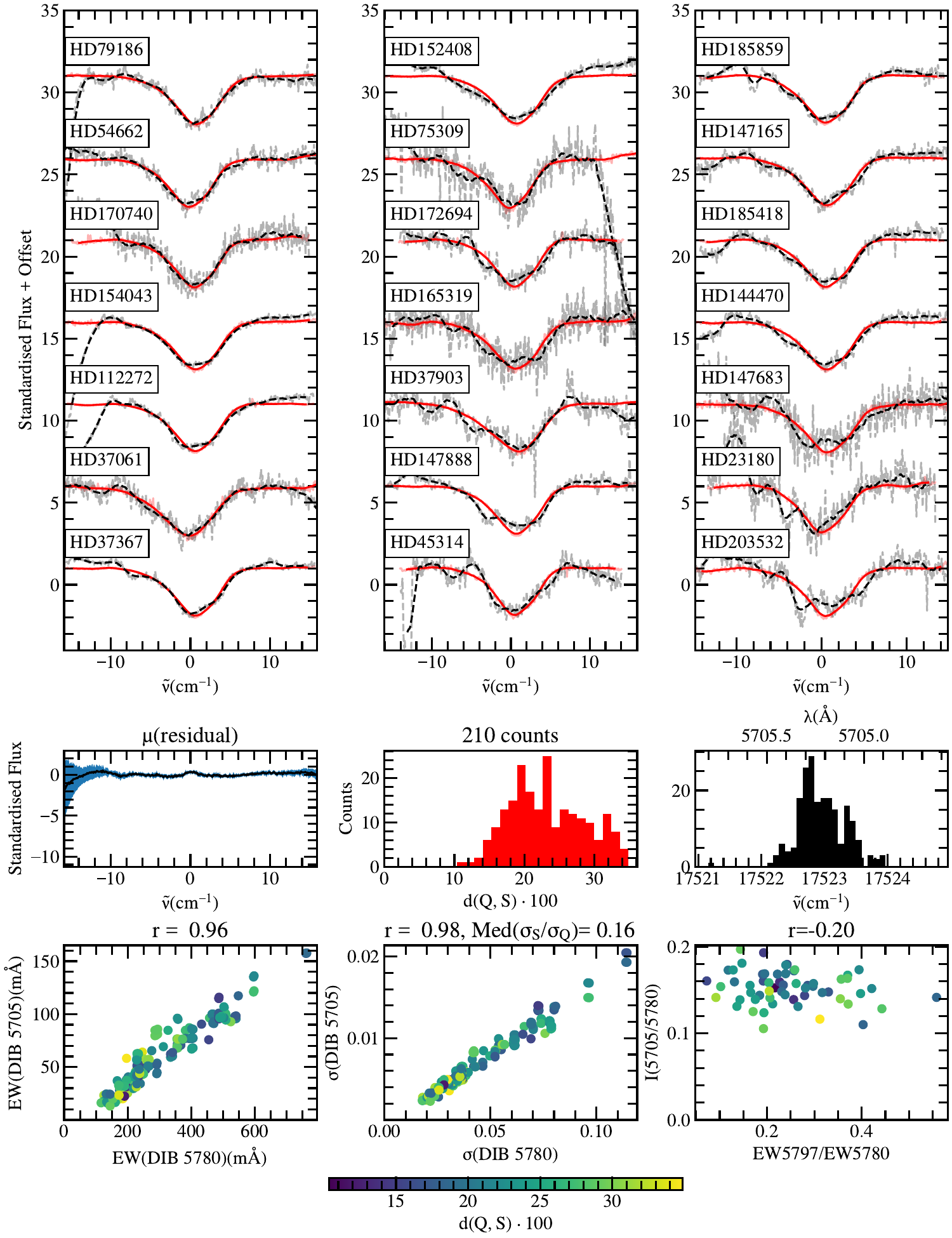}
\caption{Comparison between the 5780\,{\AA}~DIB and the 5705\,{\AA}~DIB.}
\label{fig:profiles_5780_5705}
\end{figure*}

\begin{figure*}
\includegraphics[width = .99\linewidth]{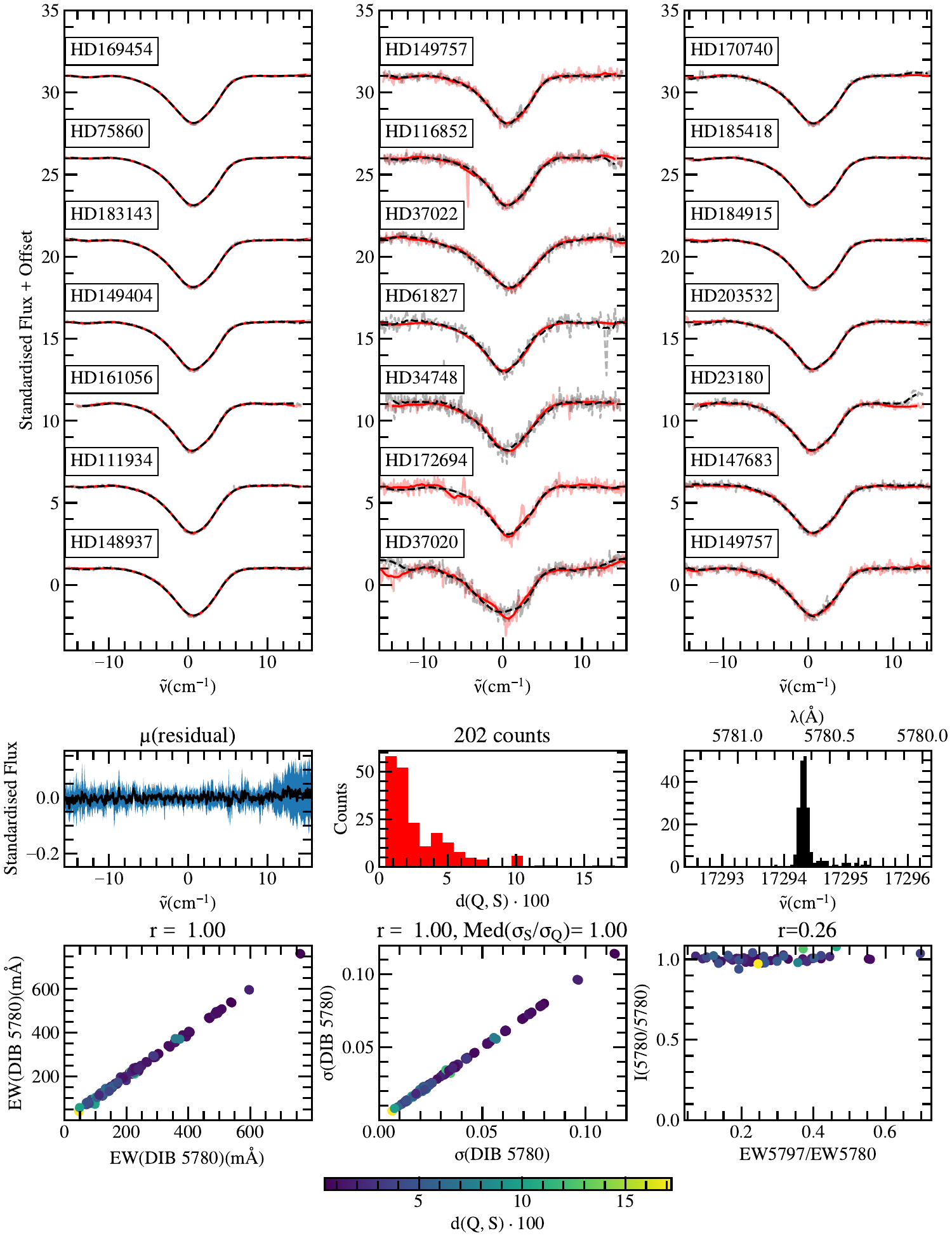}
\caption{Comparison between the 5780\,{\AA}~DIB and the 5780\,{\AA}~DIB.}
\label{fig:profiles_5780_5780}
\end{figure*}

\begin{figure*}
\includegraphics[width = .99\linewidth]{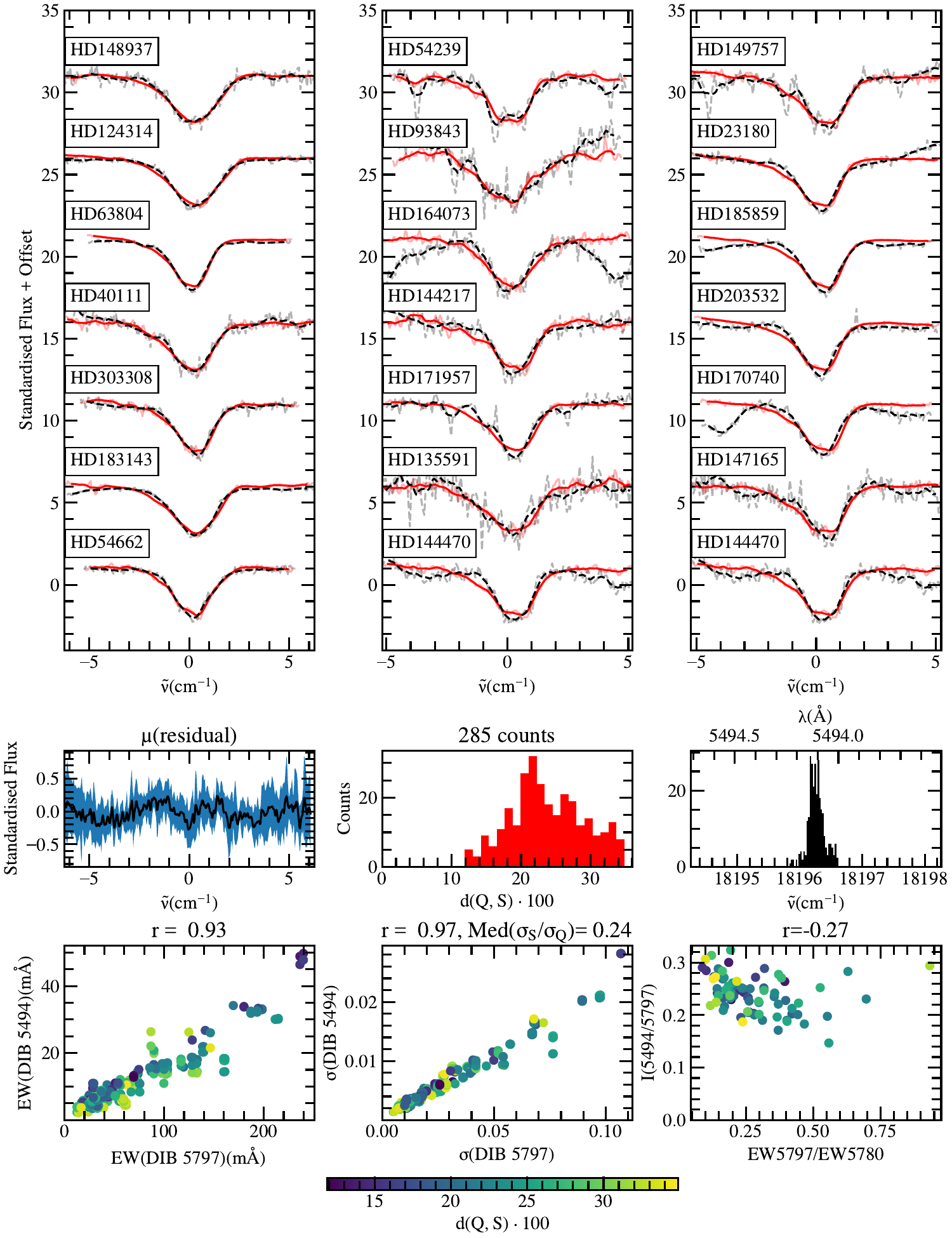}
\caption{Comparison between the 5797\,{\AA}~DIB and the 5494\,{\AA}~DIB.}
\label{fig:profiles_5797_5494}
\end{figure*}

\begin{figure*}
\includegraphics[width = .99\linewidth]{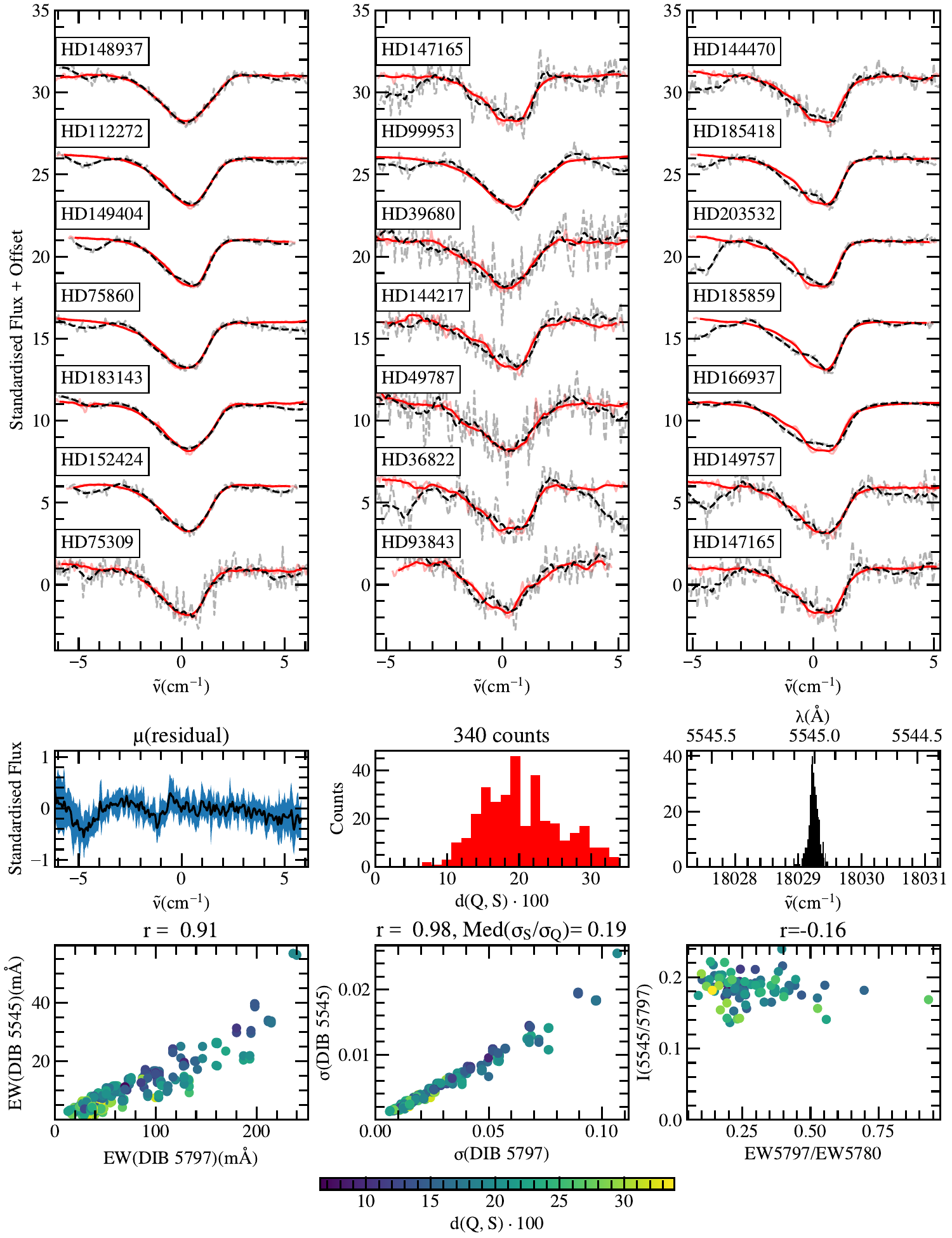}
\caption{Comparison between the 5797\,{\AA}~DIB and the 5545\,{\AA}~DIB.}
\label{fig:profiles_5797_5545}
\end{figure*}

\begin{figure*}
\includegraphics[width = .99\linewidth]{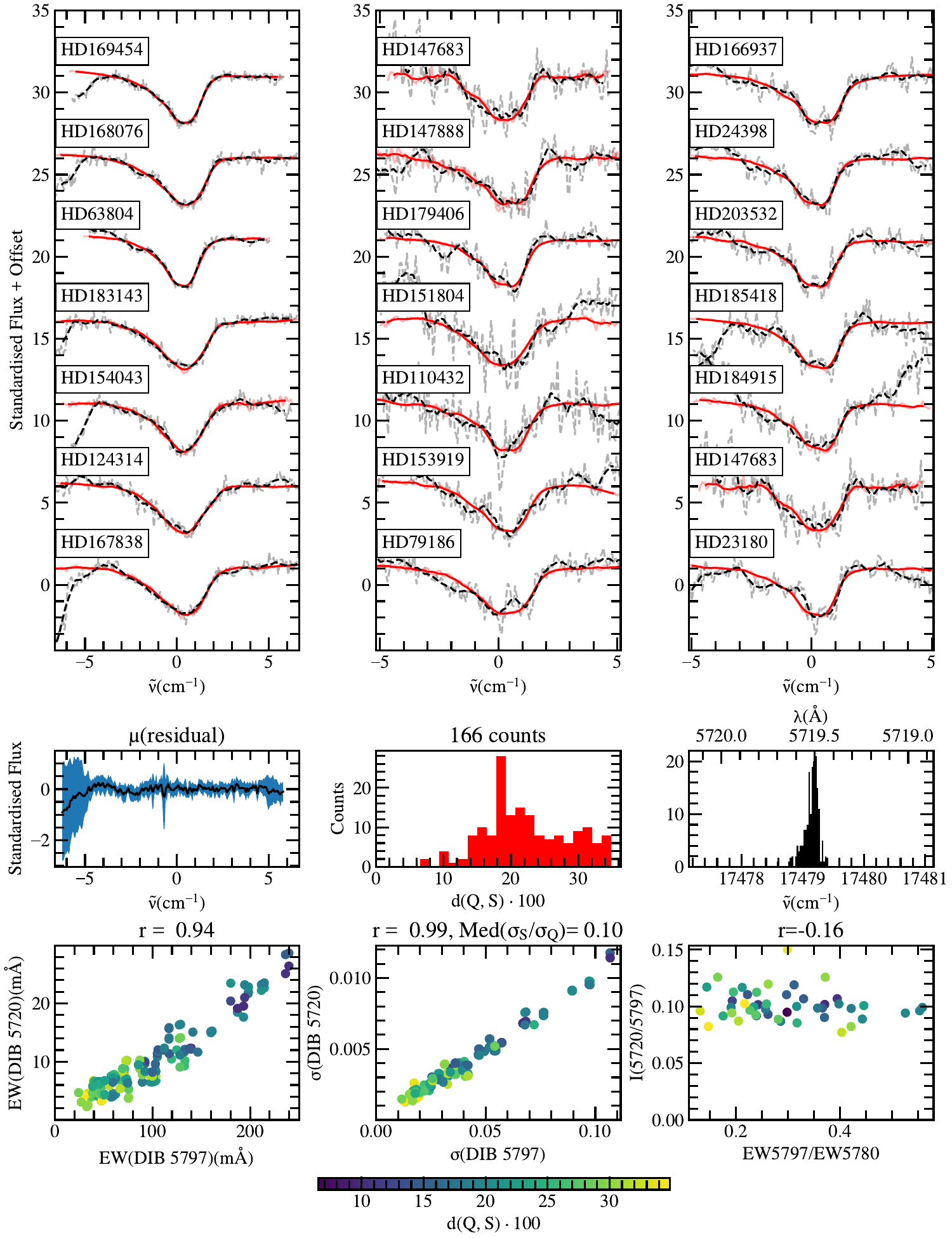}
\caption{Comparison between the 5797\,{\AA}~DIB and the 5720\,{\AA}~DIB.}
\label{fig:profiles_5797_5720}
\end{figure*}

\begin{figure*}
\includegraphics[width = .99\linewidth]{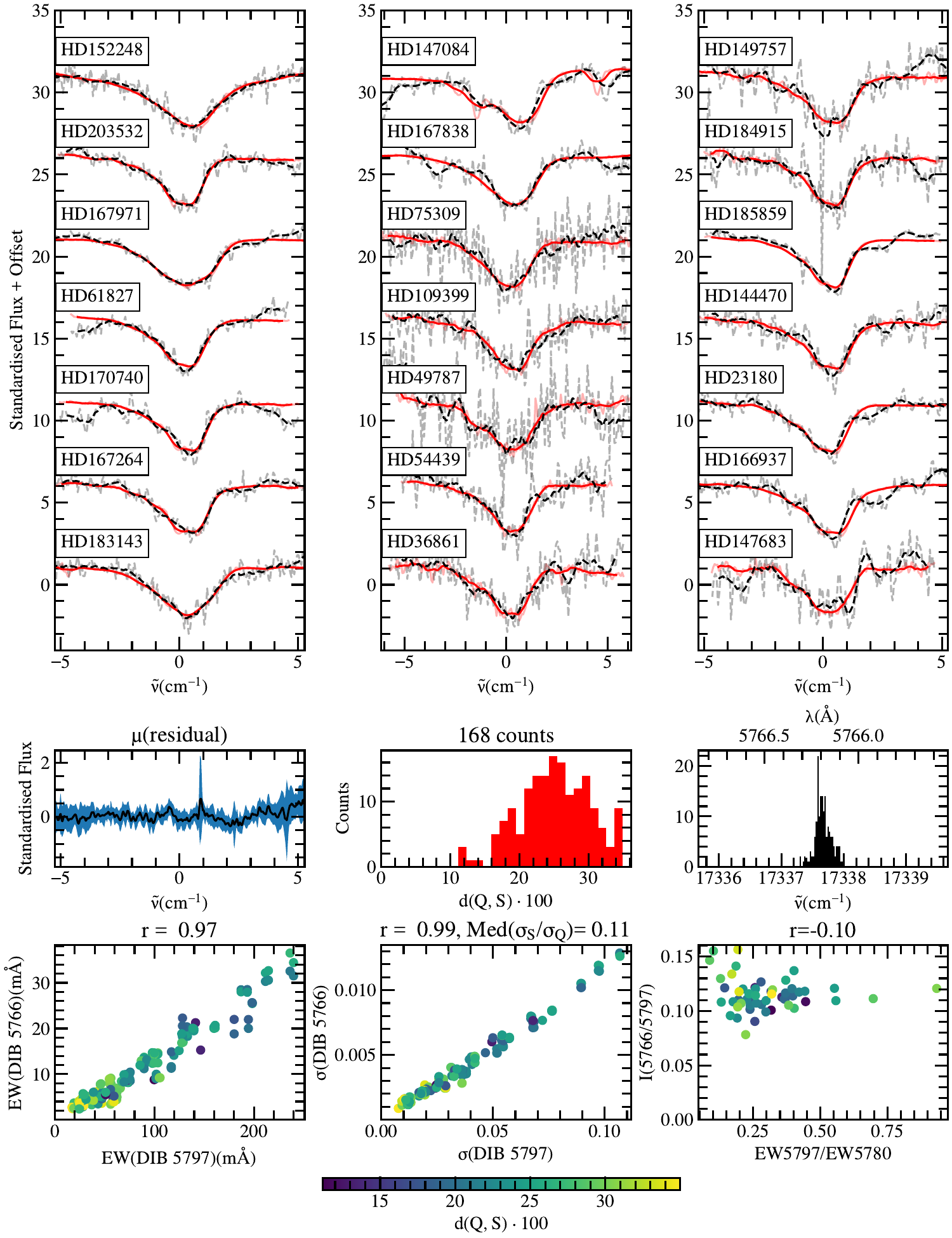}
\caption{Comparison between the 5797\,{\AA}~DIB and the 5766\,{\AA}~DIB.}
\label{fig:profiles_5797_5766}
\end{figure*}

\begin{figure*}
\includegraphics[width = .99\linewidth]{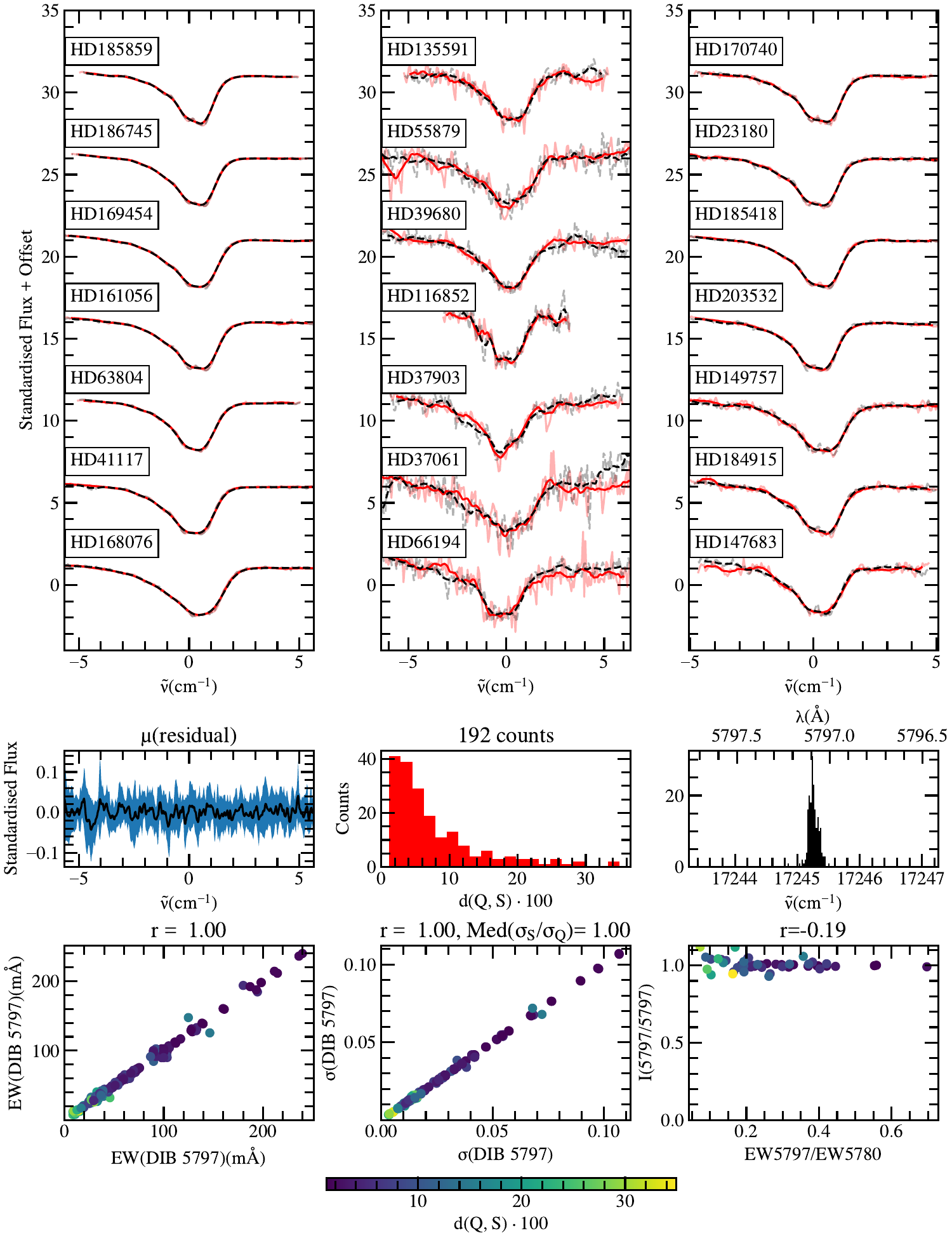}
\caption{Comparison between the 5797\,{\AA}~DIB and the 5797\,{\AA}~DIB.}
\label{fig:profiles_5797_5797}
\end{figure*}

\begin{figure*}
\includegraphics[width = .99\linewidth]{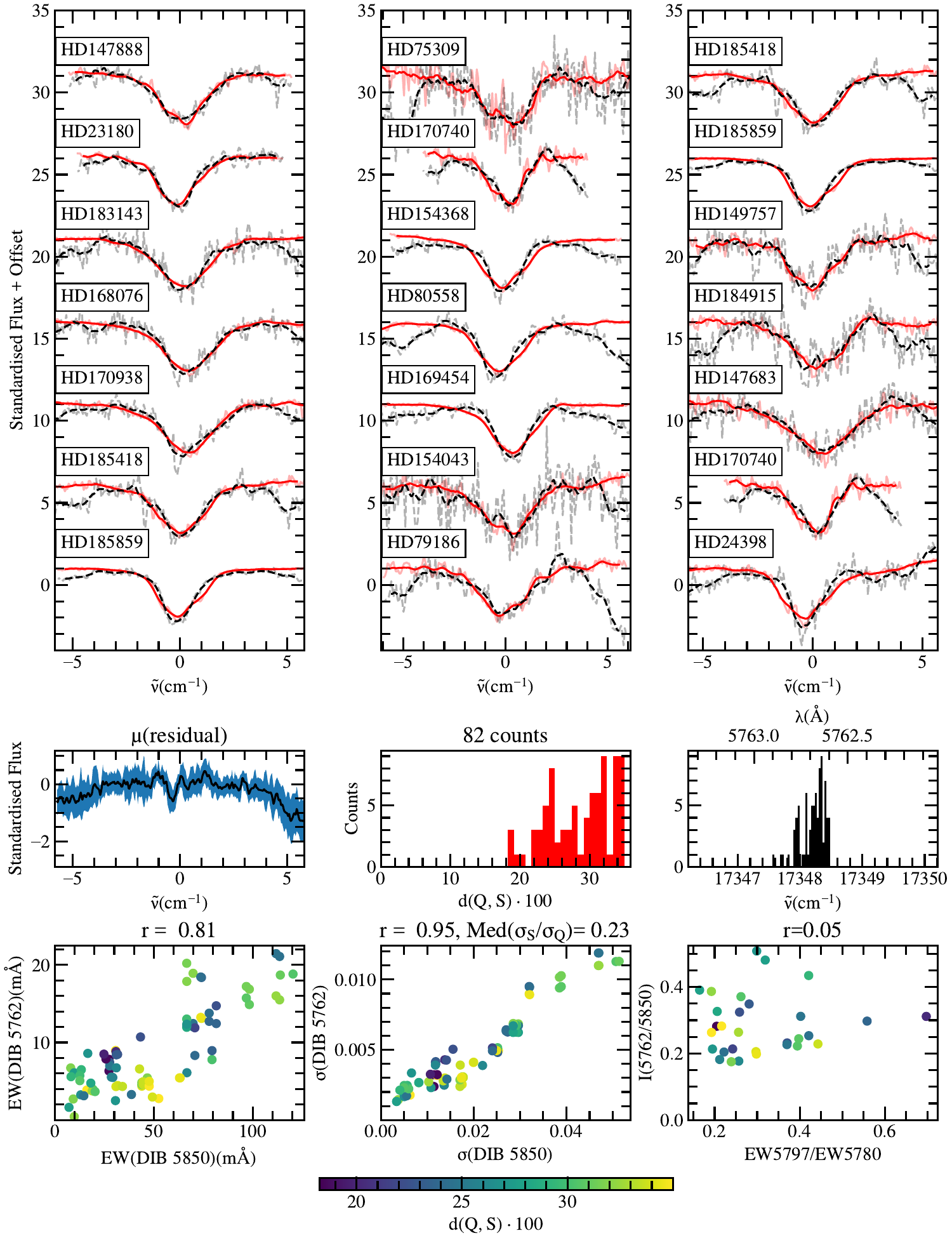}
\caption{Comparison between the 5850\,{\AA}~DIB and the 5762\,{\AA}~DIB.}
\label{fig:profiles_5850_5762}
\end{figure*}

\begin{figure*}
\includegraphics[width = .99\linewidth]{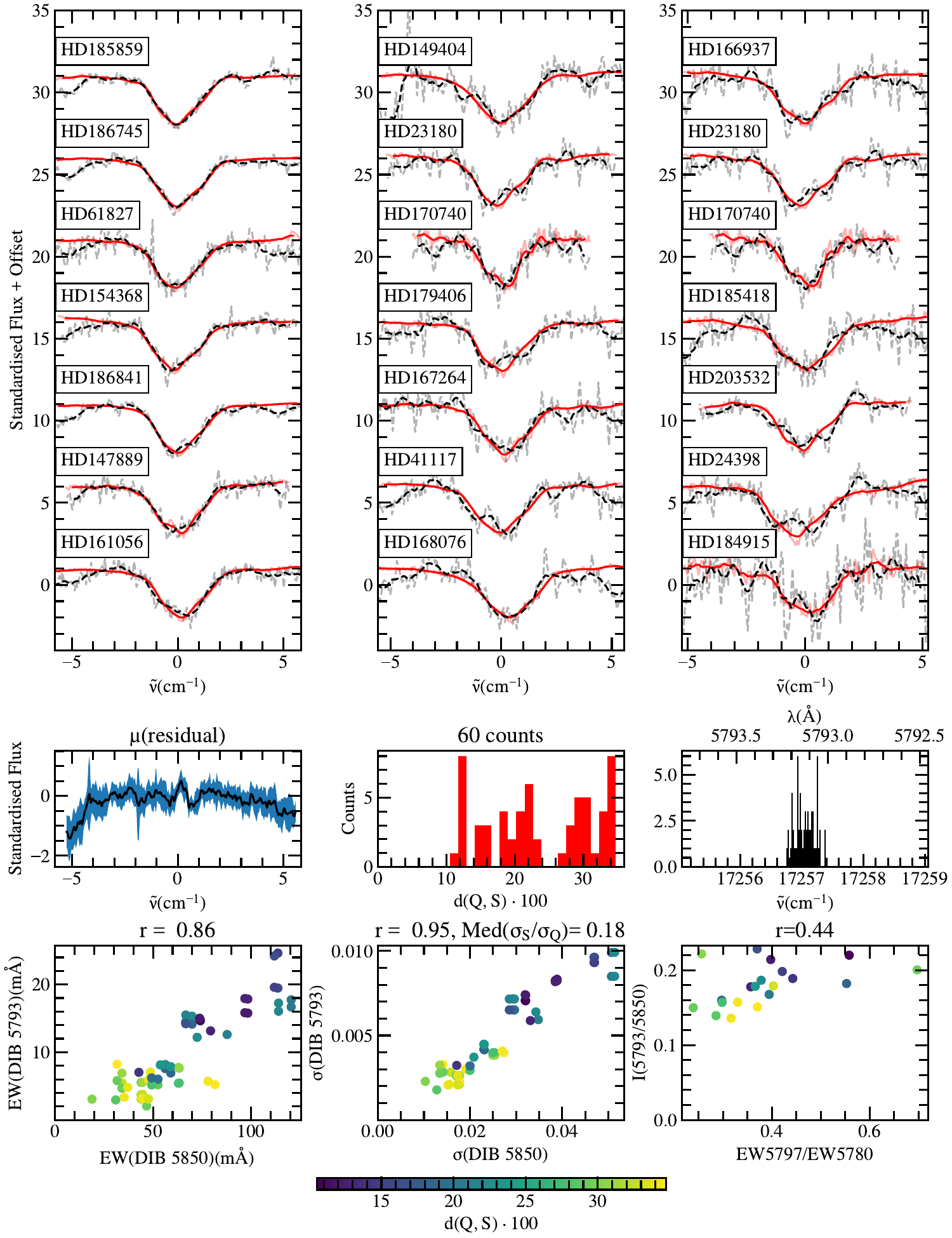}
\caption{Comparison between the 5850\,{\AA}~DIB and the 5793\,{\AA}~DIB.}
\label{fig:profiles_5850_5793}
\end{figure*}

\begin{figure*}
\includegraphics[width = .99\linewidth]{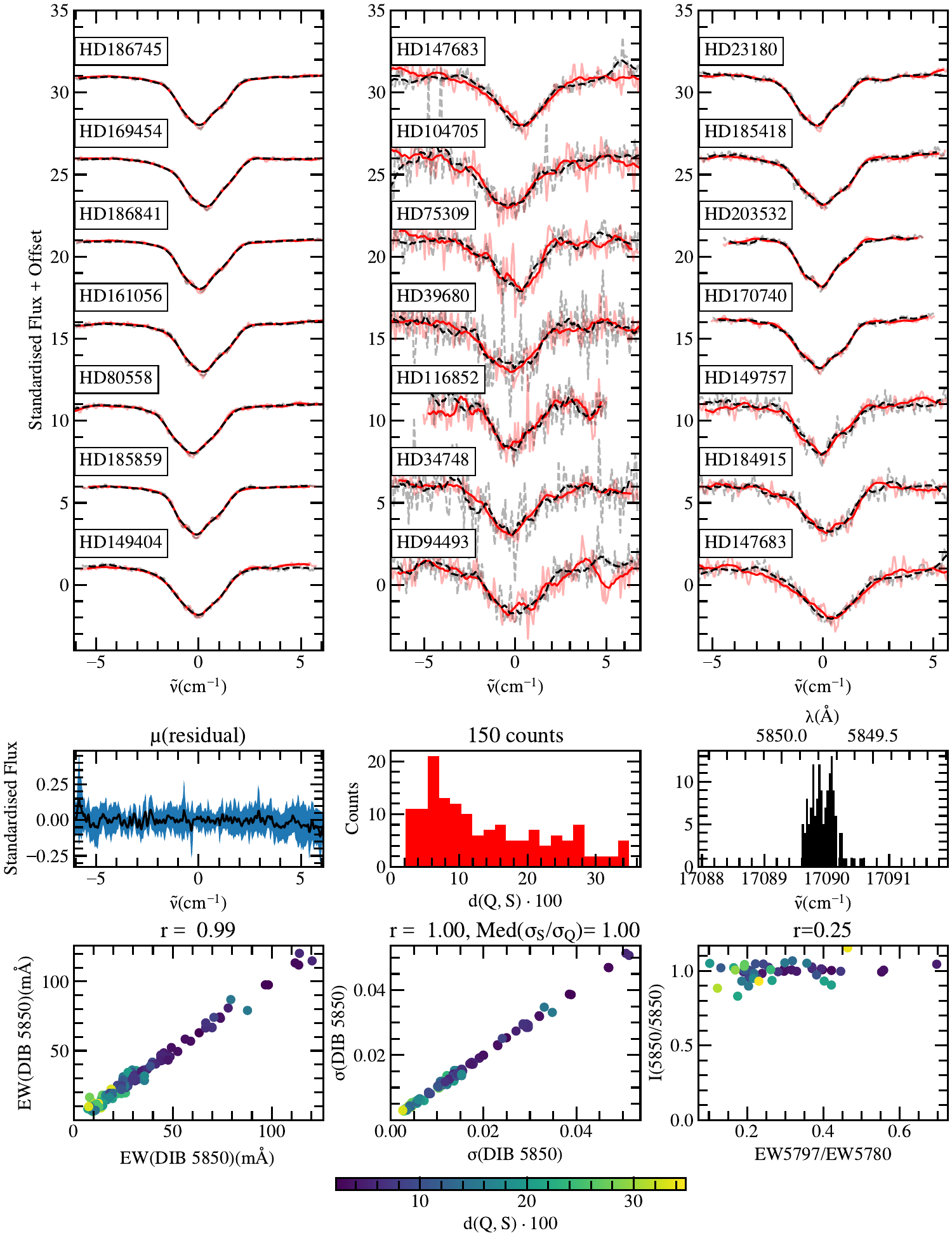}
\caption{Comparison between the 5850\,{\AA}~DIB and the 5850\,{\AA}~DIB.}
\label{fig:profiles_5850_5850}
\end{figure*}

\begin{figure*}
\includegraphics[width = .99\linewidth]{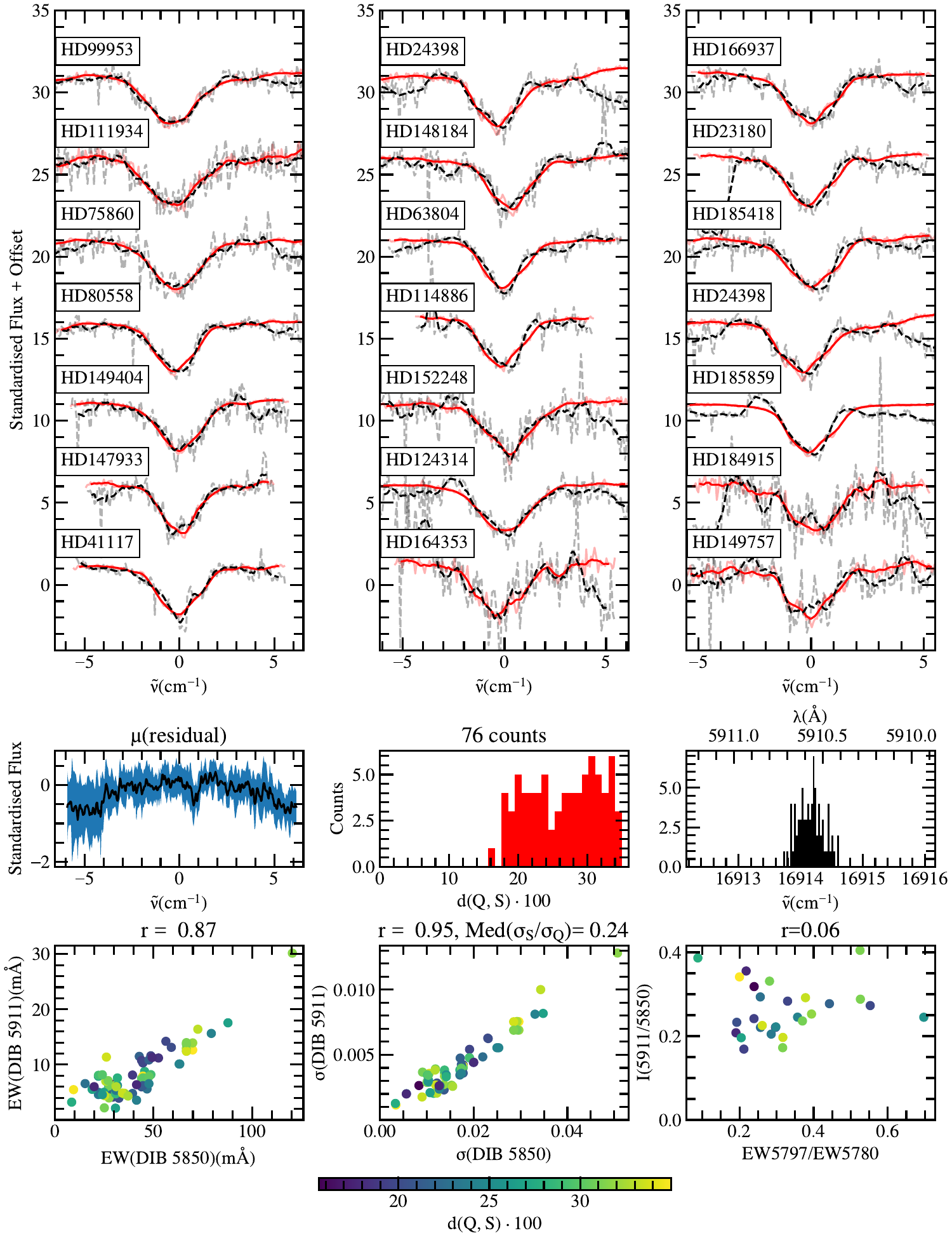}
\caption{Comparison between the 5850\,{\AA}~DIB and the 5911\,{\AA}~DIB.}
\label{fig:profiles_5850_5911}
\end{figure*}

\begin{figure*}
\includegraphics[width = .99\linewidth]{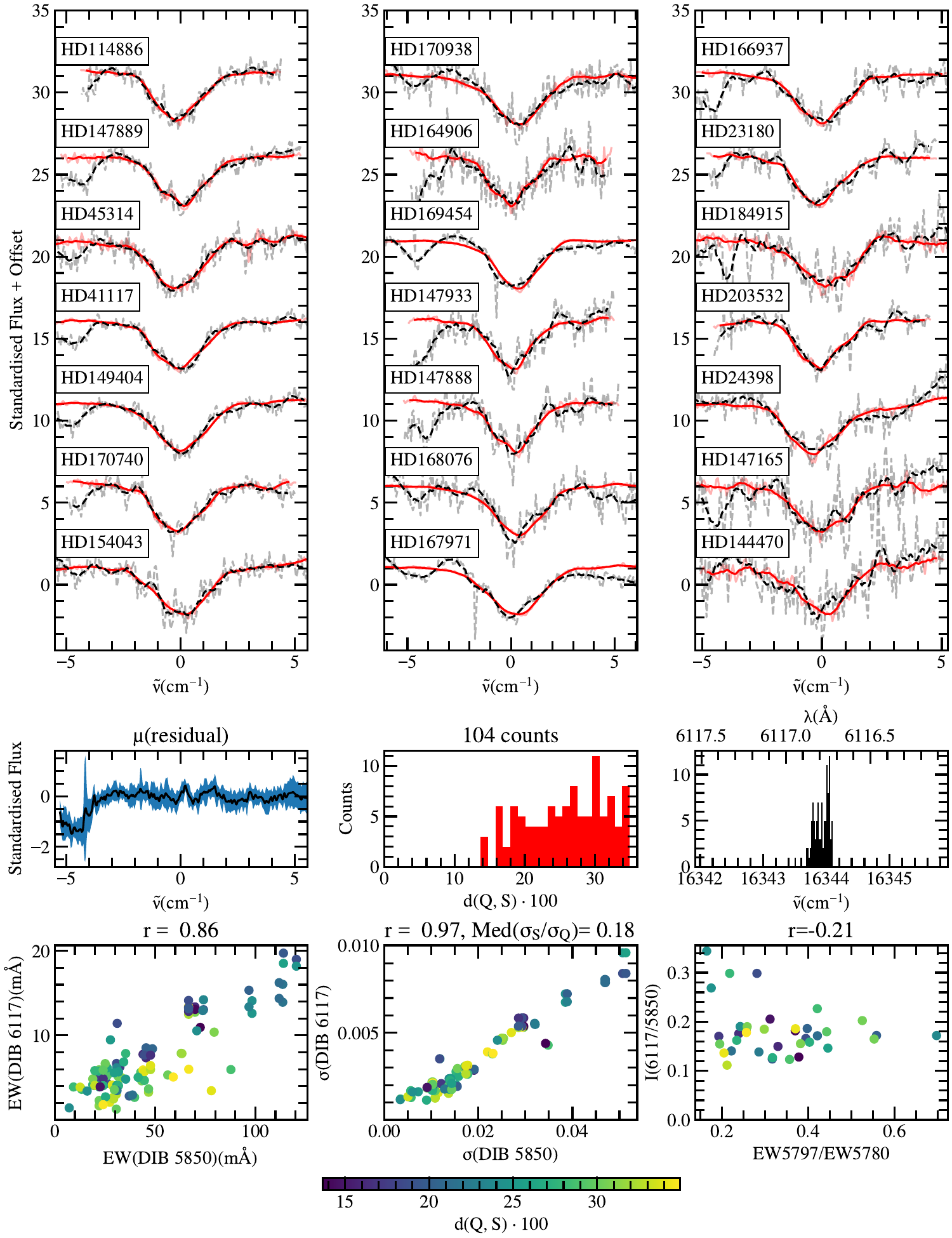}
\caption{Comparison between the 5850\,{\AA}~DIB and the 6117\,{\AA}~DIB.}
\label{fig:profiles_5850_6117}
\end{figure*}

\begin{figure*}
\includegraphics[width = .99\linewidth]{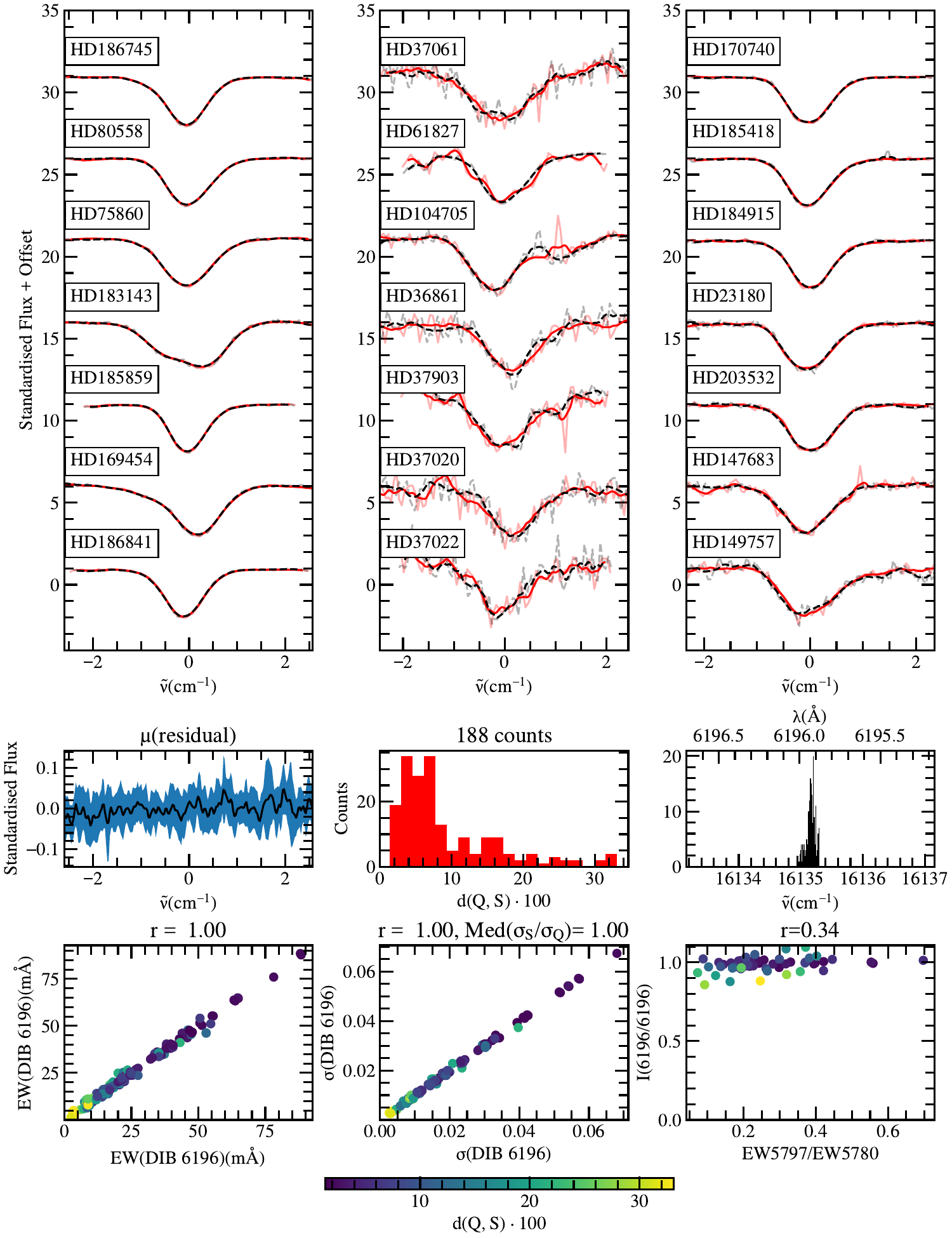}
\caption{Comparison between the 6196\,{\AA}~DIB and the 6196\,{\AA}~DIB.}
\label{fig:profiles_6196_6196}
\end{figure*}

\begin{figure*}
\includegraphics[width = .99\linewidth]{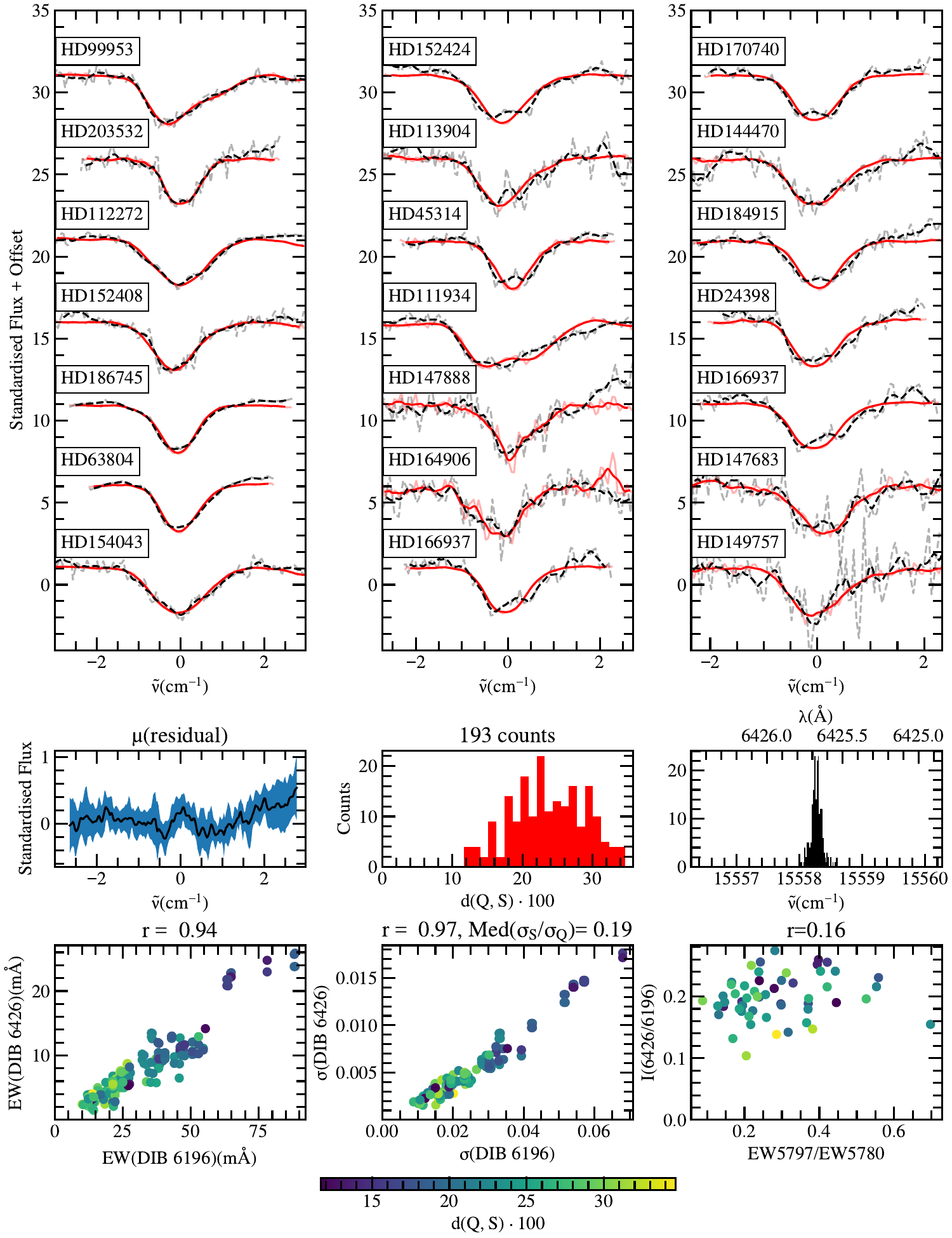}
\caption{Comparison between the 6196\,{\AA}~DIB and the 6426\,{\AA}~DIB.}
\label{fig:profiles_6196_6426}
\end{figure*}

\begin{figure*}
\includegraphics[width = .99\linewidth]{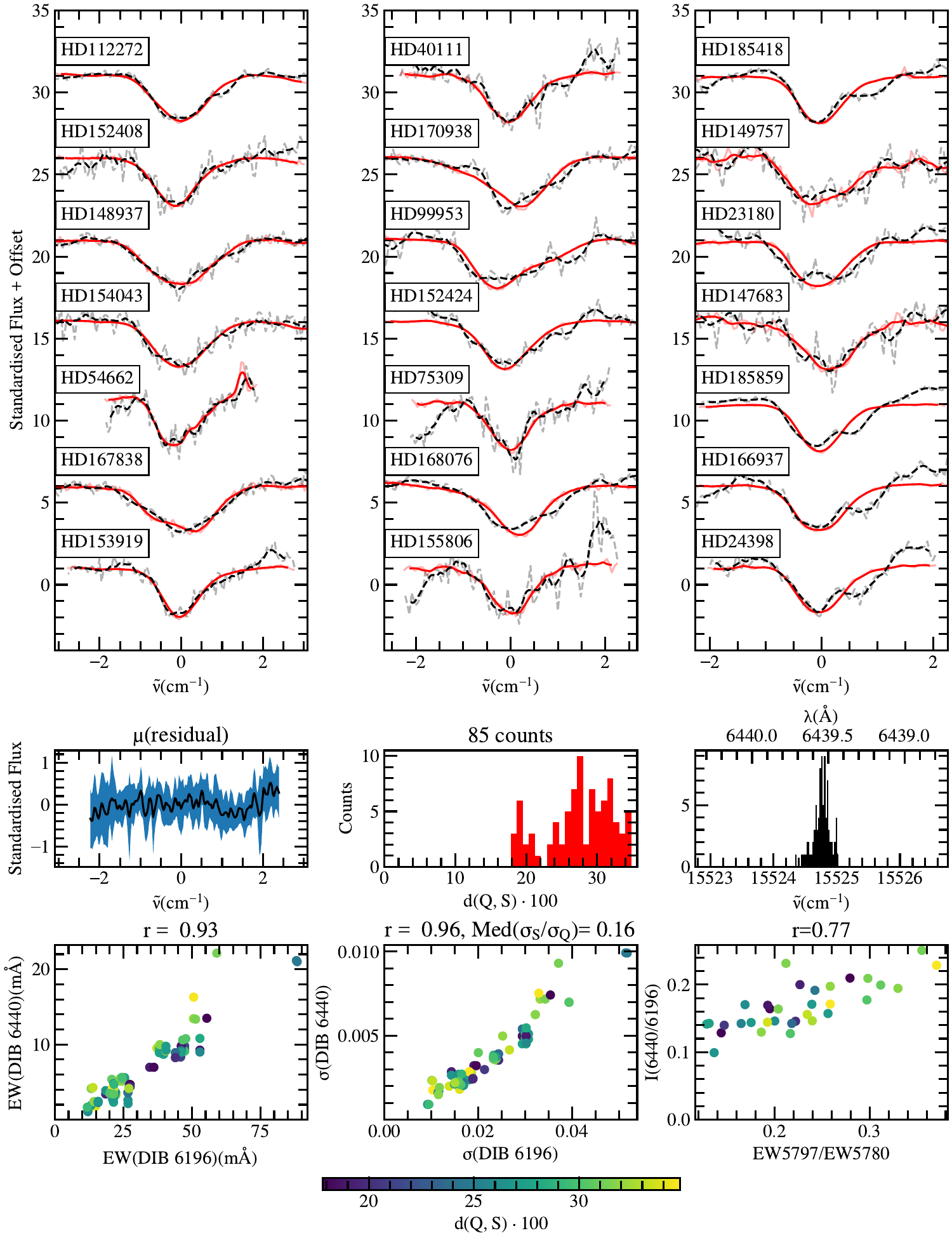}
\caption{Comparison between the 6196\,{\AA}~DIB and the 6440\,{\AA}~DIB.}
\label{fig:profiles_6196_6440}
\end{figure*}

\begin{figure*}
\includegraphics[width = .99\linewidth]{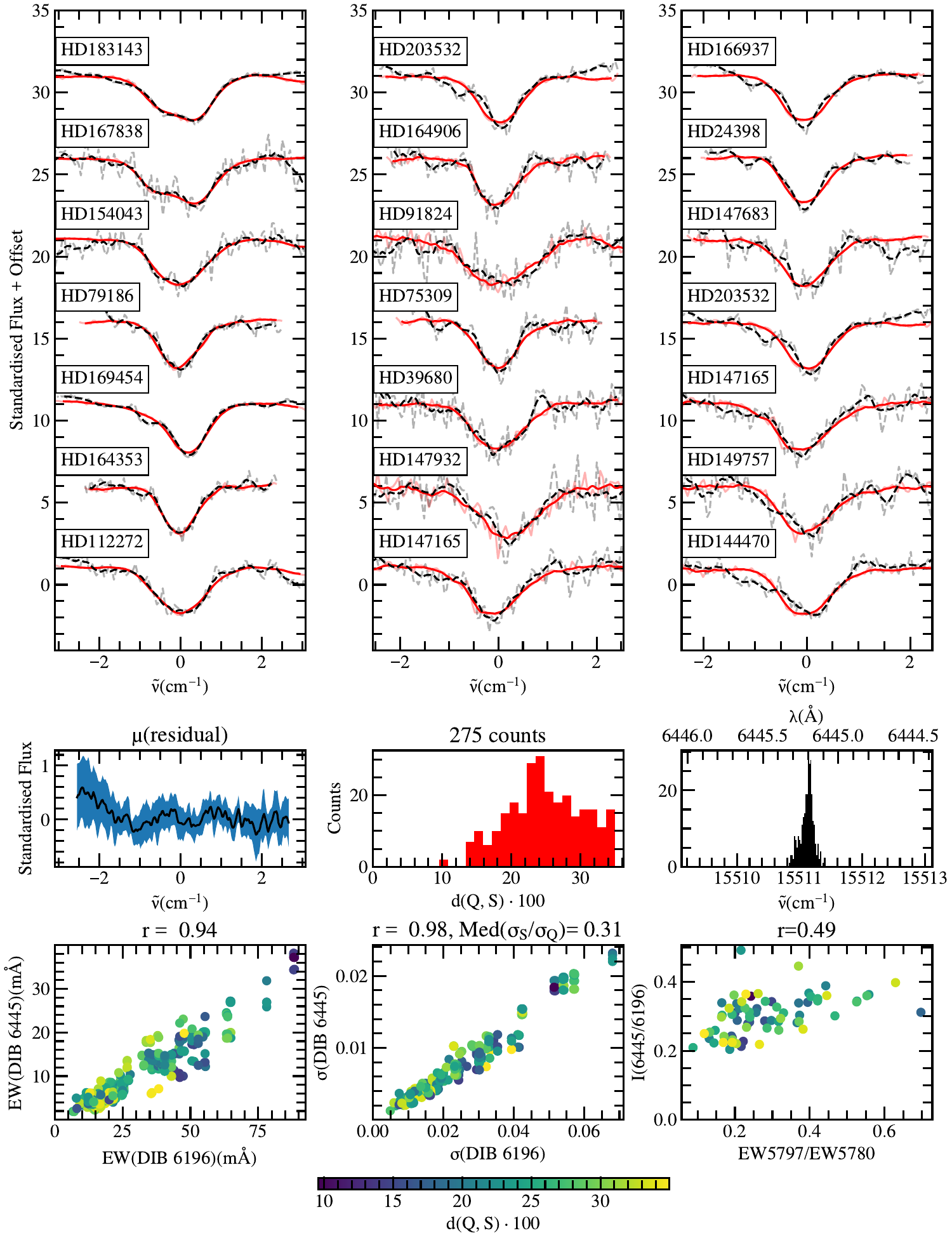}
\caption{Comparison between the 6196\,{\AA}~DIB and the 6445\,{\AA}~DIB.}
\label{fig:profiles_6196_6445}
\end{figure*}

\begin{figure*}
\includegraphics[width = .99\linewidth]{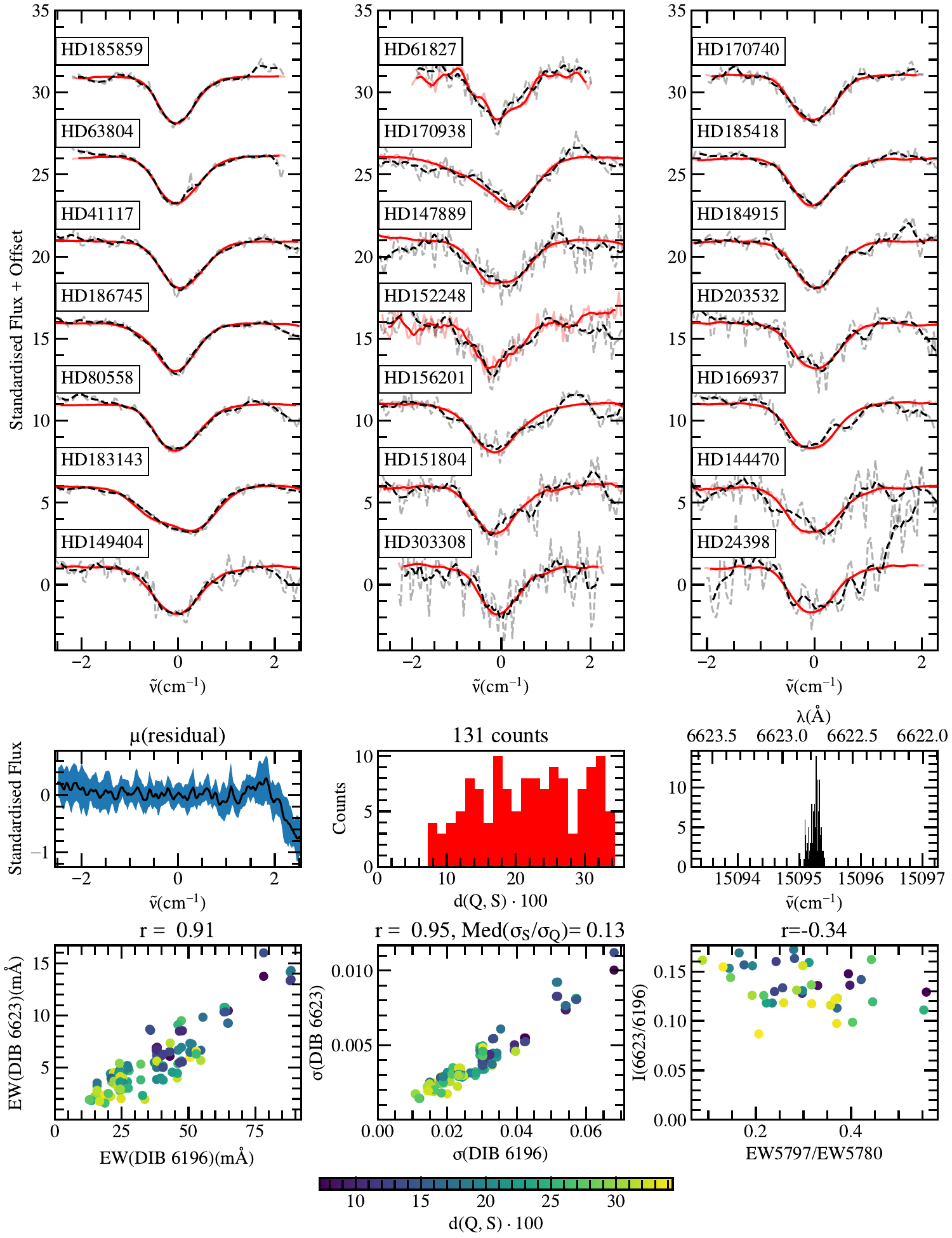}
\caption{Comparison between the 6196\,{\AA}~DIB and the 6623\,{\AA}~DIB.}
\label{fig:profiles_6196_6623}
\end{figure*}

\begin{figure*}
\includegraphics[width = .99\linewidth]{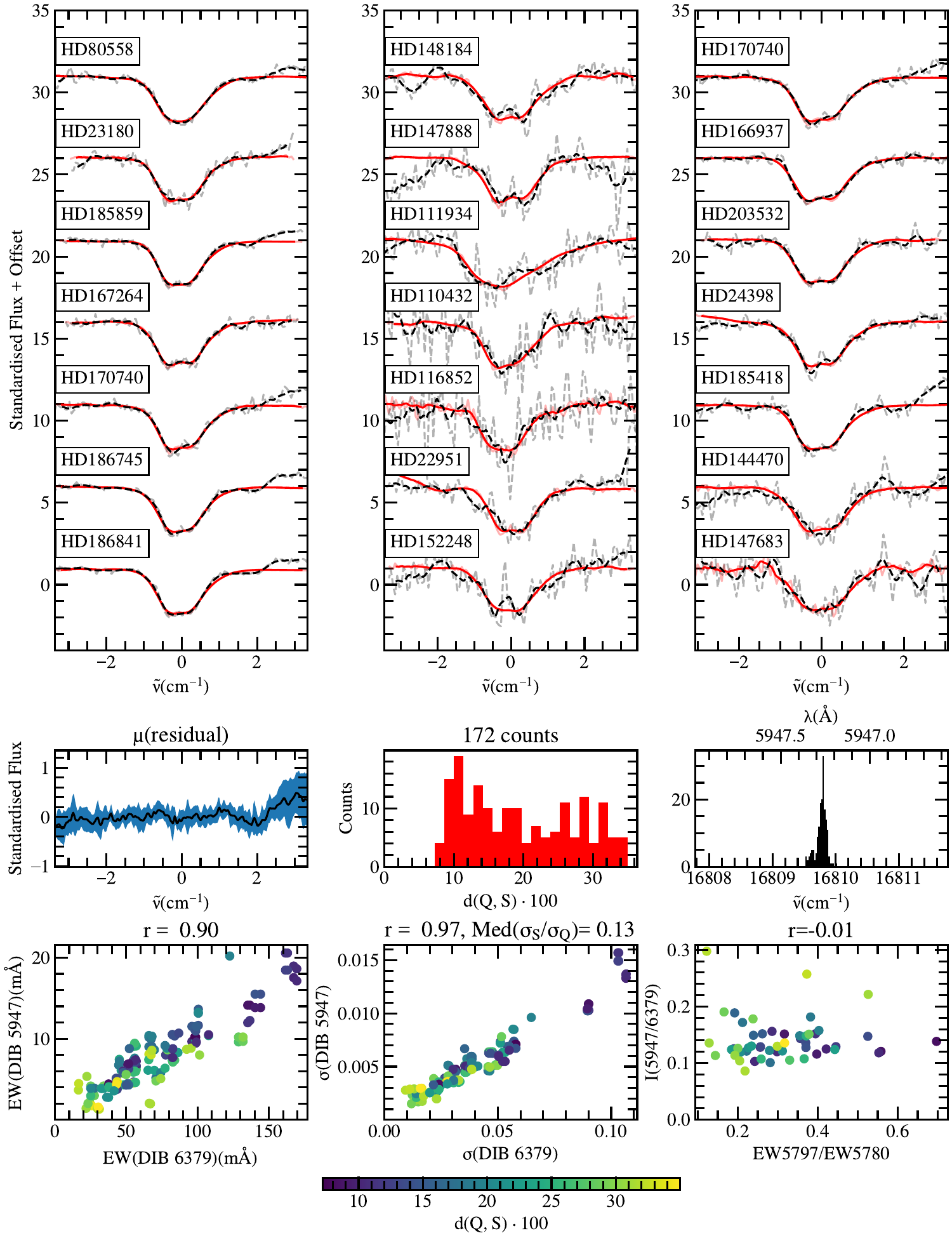}
\caption{Comparison between the 6379\,{\AA}~DIB and the 5947\,{\AA}~DIB.}
\label{fig:profiles_6379_5947}
\end{figure*}

\begin{figure*}
\includegraphics[width = .99\linewidth]{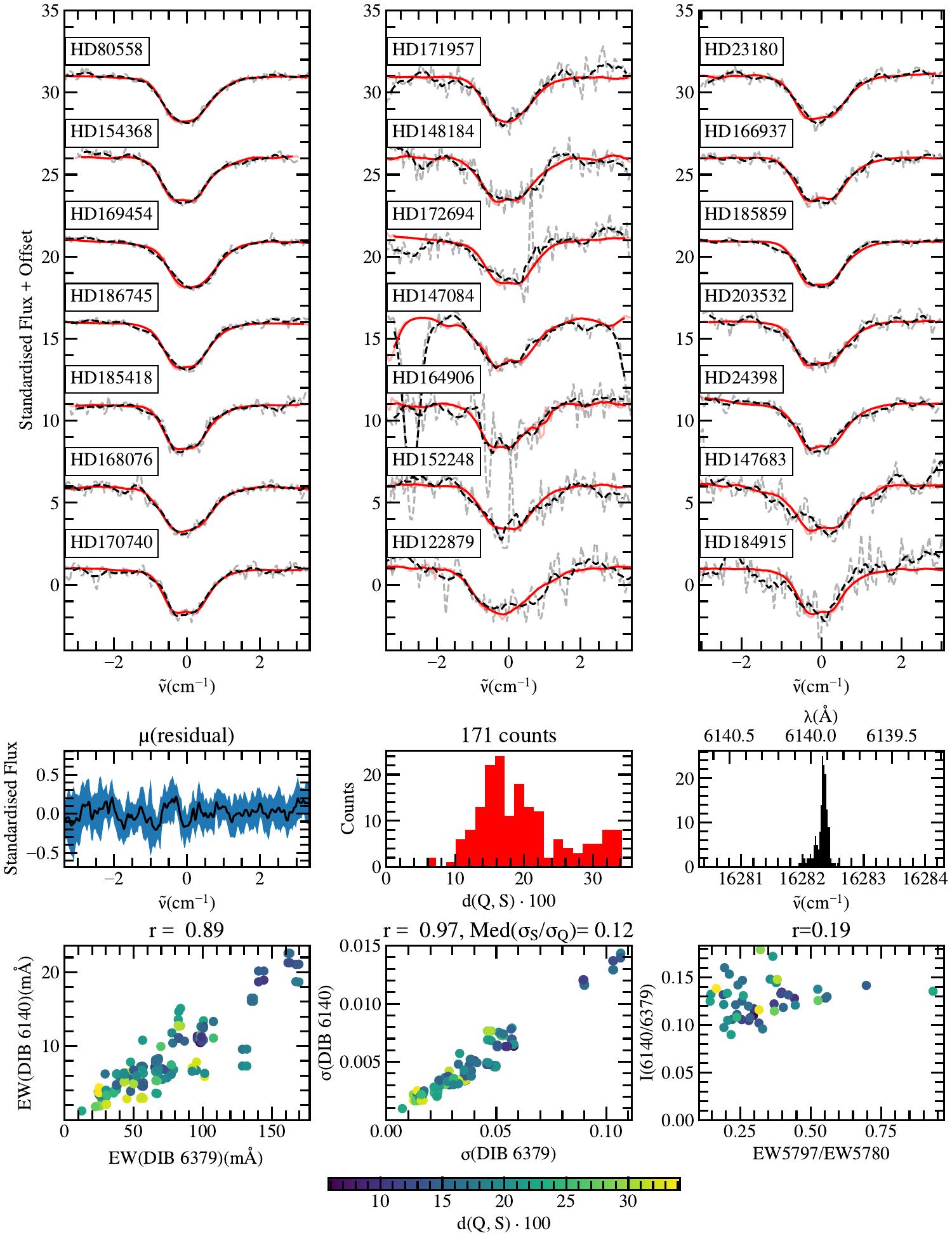}
\caption{Comparison between the 6379\,{\AA}~DIB and the 6140\,{\AA}~DIB.}
\label{fig:profiles_6379_6140}
\end{figure*}

\begin{figure*}
\includegraphics[width = .99\linewidth]{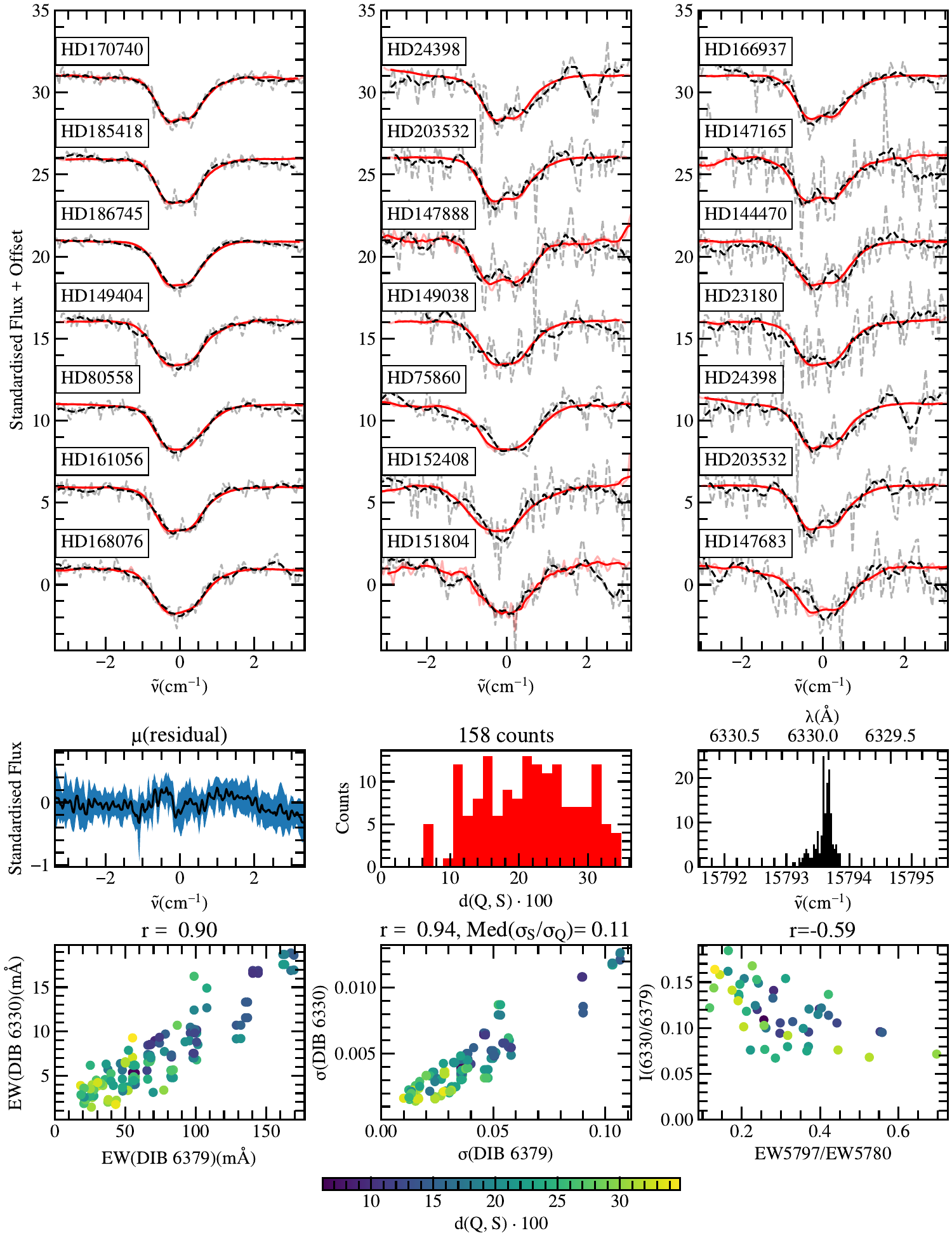}
\caption{Comparison between the 6379\,{\AA}~DIB and the 6330\,{\AA}~DIB.}
\label{fig:profiles_6379_6330}
\end{figure*}

\begin{figure*}
\includegraphics[width = .99\linewidth]{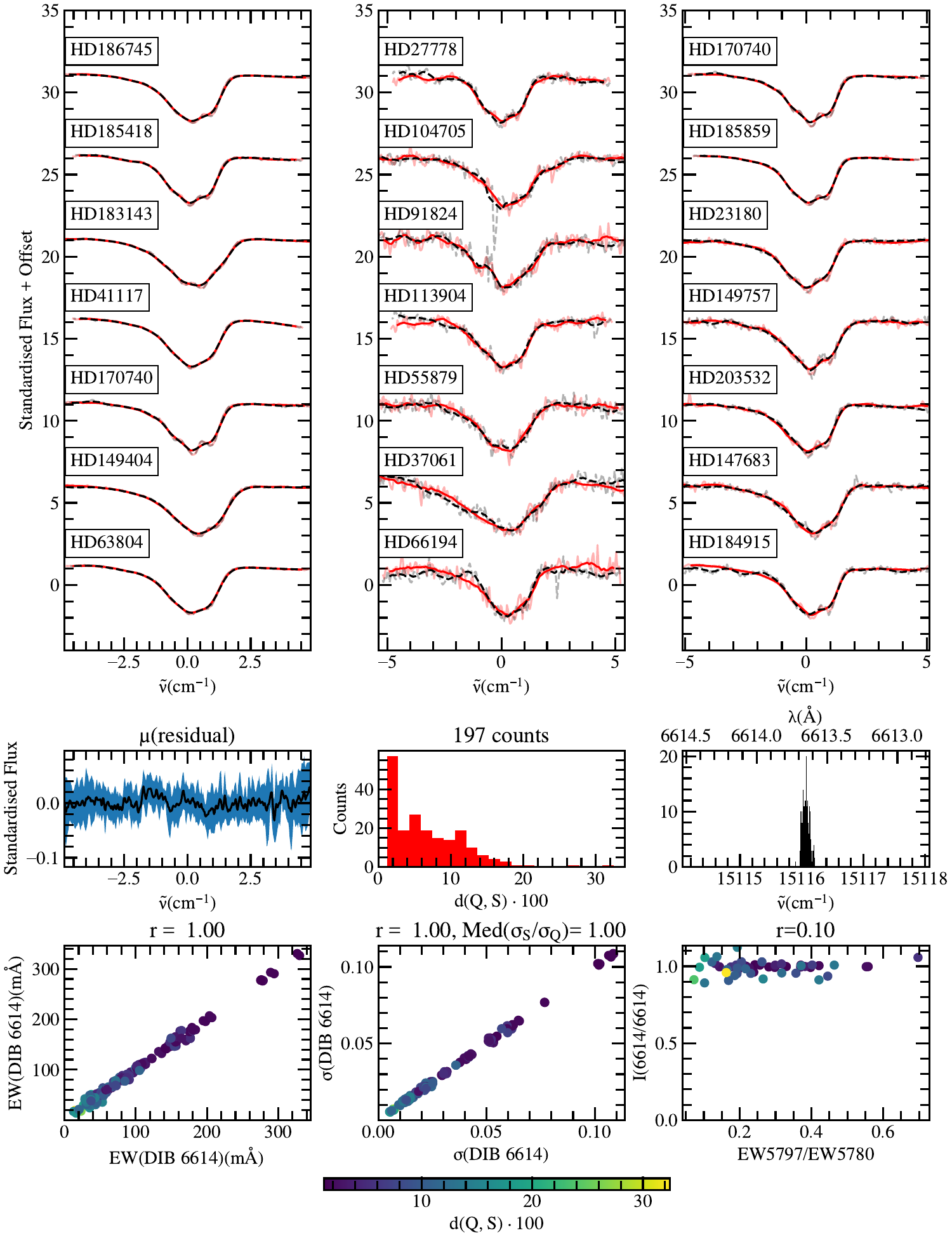}
\caption{Comparison between the 6614\,{\AA}~DIB and the 6614\,{\AA}~DIB.}
\label{fig:profiles_6614_6614}
\end{figure*}

\setcounter{section}{3}
\setcounter{figure}{0}

\begin{figure*}
\begin{flushleft}{\large\textsf{\textbf{Appendix C: New DIBs -- comparison plots}}}\end{flushleft}
\includegraphics[width = .97\linewidth]{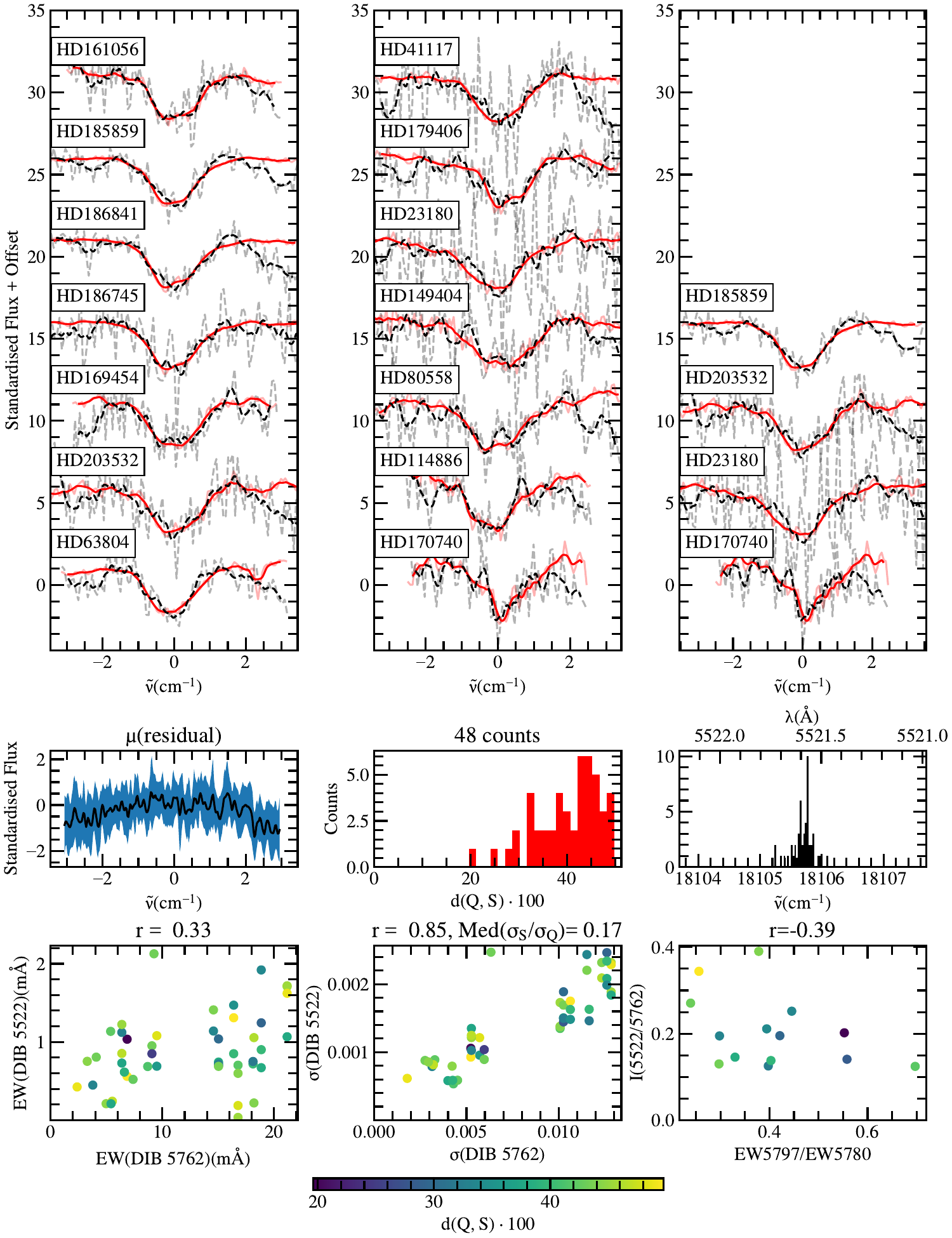}
\caption{Comparison between the 5762\,{\AA}~DIB and the 5522\,{\AA}~DIB.}
\label{fig:profiles_5762_5522}
\end{figure*}

\begin{figure*}
\includegraphics[width = .99\linewidth]{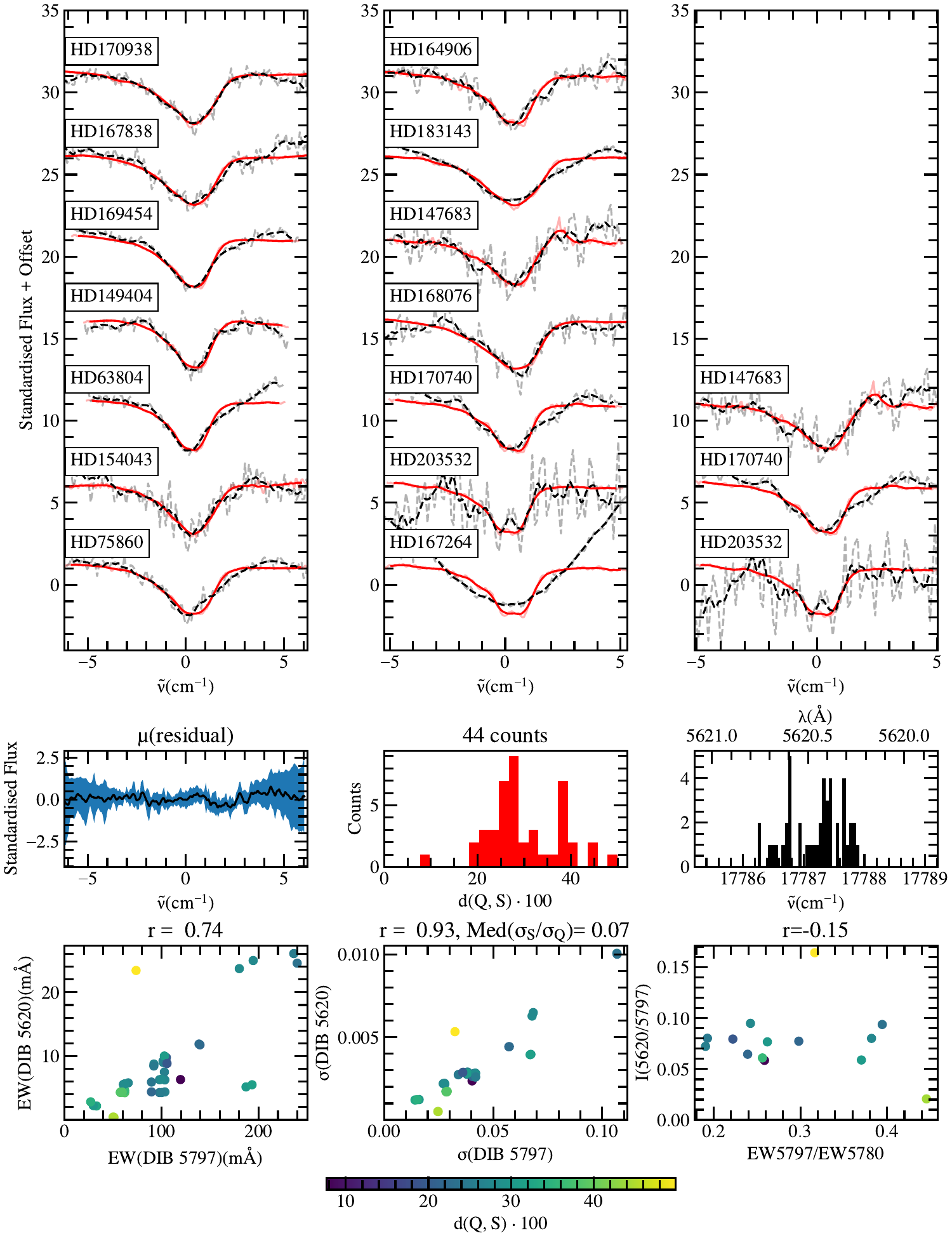}
\caption{Comparison between the 5797\,{\AA}~DIB and the 5620\,{\AA}~DIB.}
\label{fig:profiles_5797_5620}
\end{figure*}

\begin{figure*}
\includegraphics[width = .99\linewidth]{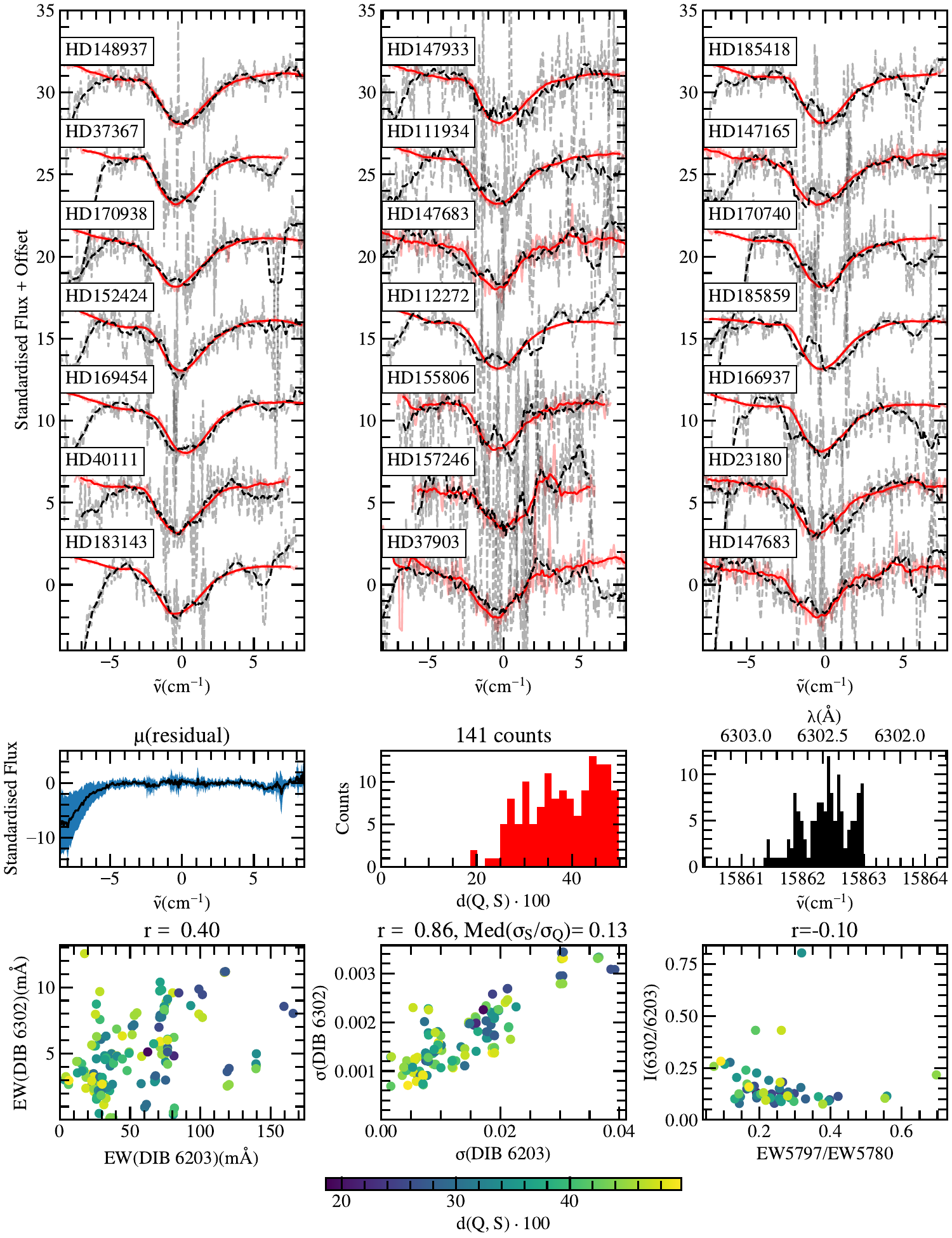}
\caption{Comparison between the 6203\,{\AA}~DIB and the 6302\,{\AA}~DIB.}
\label{fig:profiles_6203_6302}
\end{figure*}

\begin{figure*}
\includegraphics[width = .99\linewidth]{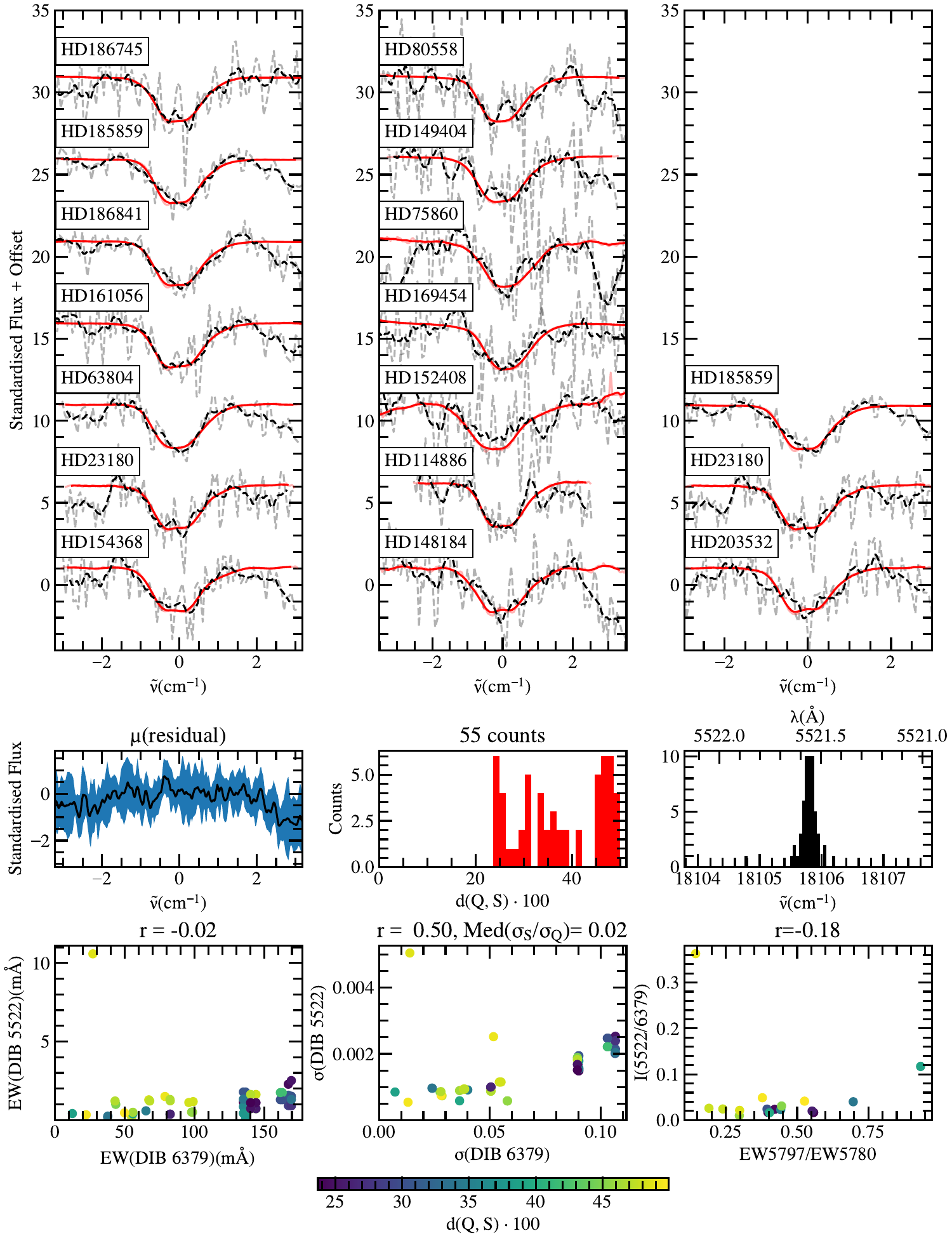}
\caption{Comparison between the 6379\,{\AA}~DIB and the 5522\,{\AA}~DIB.}
\label{fig:profiles_6379_5522}
\end{figure*}

\end{appendix}
\end{document}